\documentclass[%
reprint,
aps,nofootinbib,
notitlepage,
superscriptaddress,
twocolumn,
nobibnotes,
 amsmath,
 amssymb,
 prl,
]{revtex4-2}

\usepackage{graphicx}% Include figure files
\usepackage{dcolumn}% Align table columns on decimal point
\usepackage{bm}% bold math
\usepackage{braket}
\usepackage{float}
\usepackage{lipsum}
\usepackage{graphicx}% Include figure filesx
\usepackage{dcolumn}% Align table columns on decimal point
\usepackage{bm}% bold math
\usepackage{natbib} 
\usepackage{hyperref} 
\usepackage[caption=false]{subfig}
\usepackage{bm}

\hypersetup{
	colorlinks = true,
    linkcolor = Maroon,
    urlcolor  = Maroon,
    citecolor = Maroon
}

\usepackage{microtype}
\usepackage{verbatim}
\usepackage[amssymb]{SIunits}
\usepackage{tabularx}
\usepackage[dvipsnames]{xcolor}
\usepackage{tikz}
\usepackage{orcidlink}
\usepackage[normalem]{ulem}
\usepackage{xcolor}
\usepackage{soul}

\newcommand\aj{{\rm AJ}}%        % Astronomical Journal 
\renewcommand\apj{{\rm ApJ}}%    % Astrophysical Journal 
\newcommand\apjl{{\rm ApJL}}     % Astrophysical Journal, Letters 
\newcommand\mnras{{\rm MNRAS}}%   % Monthly Notices of the RAS 
\renewcommand\prd{{\rm PhRvD}}% % Physical Review D 
\newcommand\jcap{{\rm JCAP}}%  % Journal of Cosmology and Astroparticle Physics
\usepackage{verbatim}

\newcommand{\be}{\begin{equation}}
\newcommand{\ee}{\end{equation}}
\newcommand{\bea}{\begin{eqnarray}}
\newcommand{\eea}{\end{eqnarray}}
\newcommand{\beas}{\begin{eqnarray*}}
\newcommand{\eeas}{\end{eqnarray*}}

\newcommand{\bx}{{\bf x}}

\newcommand{\hatbn}{{\hat{\boldsymbol n}}}

\def\gsim{ \lower .75ex \hbox{$\sim$} \llap{\raise .27ex \hbox{$>$}} }
\def\lsim{ \lower .75ex \hbox{$\sim$} \llap{\raise .27ex \hbox{$<$}} }

\def\dalam{\hbox
{\vrule\vbox{\hrule\hbox to 1ex{ \hfill}\kern 1 ex\hrule}\vrule}}

\def\1/2{\hbox{$ {1 \over 2}$ }}

\newcommand{\columbia}{Department of Physics, Columbia University, New York, NY, USA 10027}
\newcommand{\perimeter}{Perimeter Institute for Theoretical Physics, 31 Caroline St N, Waterloo, ON N2L 2Y5, Canada}
\newcommand{\york}{Department of Physics and Astronomy, York University, Toronto, ON M3J 1P3, Canada}
\newcommand{\damtp}{DAMTP, Centre for Mathematical Sciences, Wilberforce Road, Cambridge CB3 0WA, UK}
\newcommand{\kavli}{Kavli Institute for Cosmology Cambridge, Madingley Road, Cambridge, CB3 0HA, UK}
\newcommand{\cca}{Center for Computational Astrophysics, Flatiron Institute, 162 5th Avenue, New York, NY 10010 USA}

\usepackage{color}

\begin{document}

\title{Constraints on axions from patchy screening of the cosmic microwave background}

\author{Samuel Goldstein}
\email{sjg2215@columbia.edu}
\affiliation{\columbia}

\author{Fiona McCarthy}
\affiliation{\damtp}
\affiliation{\kavli}
\affiliation{\cca}

\author{Cristina Mondino}
\affiliation{\perimeter}

\author{J.~Colin Hill}
\affiliation{\columbia}

\author{Junwu Huang}
\affiliation{\perimeter}

\author{Matthew~C.~Johnson}
\affiliation{\perimeter}
\affiliation{\york}

%\date{\today}

%TC:ignore
\begin{abstract}
The resonant conversion of cosmic microwave background (CMB) photons into axions within large-scale structure induces an anisotropic spectral distortion in CMB temperature maps. Applying state-of-the-art foreground cleaning techniques to \emph{Planck} CMB observations, we construct maps of axion-induced ``patchy screening" of the CMB. We cross-correlate these maps with data from the \emph{unWISE} galaxy survey and find no evidence of axions. We constrain the axion-photon coupling, $g_{a\gamma\gamma} \lesssim 2 \times 10^{-12}~{\rm GeV}^{-1}$, at the 95\% confidence level for axion masses in the range $10^{-13}~{\rm eV} \lesssim m_a \lesssim 10^{-12}~{\rm eV}$. These constraints are competitive with the tightest astrophysical axion limits in this mass range and are inferred from robust population-level statistics, which makes them complementary to existing searches that rely on modeling of individual systems. 
\end{abstract}

\maketitle
%TC:endignore

%TC:ignore
% \quickwordcount{main_supplemental_combined}
% \quickcharcount{main_supplemental_combined}
% \detailtexcount{main_supplemental_combined}
%TC:endignore

%%%%%%%%%%%%%%%%%%%%%%%%%%%%%%%%%%%%%%%%%%%%%%%%%%%%%%%%%%%%%%%%%%%%%%%%%%%%%%%%%%%%%%%%%%%%%%%%%%%%%%%%%%%%%%%%%%%%%
% Introduction
%%%%%%%%%%%%%%%%%%%%%%%%%%%%%%%%%%%%%%%%%%%%%%%%%%%%%%%%%%%%%%%%%%%%%%%%%%%%%%%%%%%%%%%%%%%%%%%%%%%%%%%%%%%%%%%%%%%%%

\paragraph{Introduction---} Axions are one of the most compelling extensions to the Standard Model of particle physics. Originally, the Quantum Chromodynamics (QCD) axion was proposed as a solution to the strong CP problem~\cite{Peccei:1977ur,Peccei:1977hh, Weinberg:1977ma, Wilczek:1977pj}. However, theoretical interest has expanded to more general ``axion-like particles," which couple to two photons via the following Lagrangian:
\begin{equation}\label{eq:lagrangian}
    \mathcal{L}\supset-\frac{1}{2}m_a^2a^2-\frac{1}{4}g_{a\gamma \gamma}F_{\mu\nu}\tilde{F}^{\mu\nu}a,
\end{equation}
where $a$ is the axion\footnote{Henceforth, we use the term ``axion" to refer to any pseudo-scalar field coupled to the electromagnetic sector via Eq.~\eqref{eq:lagrangian}.}, $m_a$ is the axion mass, $g_{a\gamma\gamma}$ is the axion-photon coupling, ${F}_{\mu\nu}$ is the electromagnetic field-strength tensor, and $\tilde{F}_{\mu\nu}$ is its dual. Such axions naturally emerge in many beyond Standard Model scenarios, including string theory, where they can span a wide range of masses~\cite{Witten:1984dg, Svrcek:2006yi, Arvanitaki:2009fg, Marsh:2017hbv}. Moreover, axions are a well-motivated dark matter candidate~\cite{Preskill:1982cy, Abbott:1982af, Dine:1982ah, Hui:2021tkt}. Consequently, there are many ongoing and upcoming lab experiments dedicated to searching for axions across a vast range of masses (see reviews~\cite{Graham:2015ouw, Irastorza:2018dyq, Choi:2020rgn} and references within).

Astrophysical axion searches typically leverage the interaction term in Eq.~\eqref{eq:lagrangian}, which induces photon-axion oscillations in the presence of an external magnetic field. 
Similar to the Mikheyev-Smirnov-Wolfenstein (MSW) effect in neutrino oscillations~\cite{Wolfenstein:1977ue, Mikheyev:1985zog}, the probability of photon-axion conversion depends on the properties of the medium through which the photon propagates. Notably, resonant (\emph{i.e.}, enhanced) conversion occurs when the axion mass equals the photon's plasma frequency, which is set by the free electron density of the medium~\cite{Mirizzi:2009nq}. Thus, cosmological environments, which typically have low free electron densities and large-scale magnetic fields, provide a powerful setting to search for resonant photon-axion conversions for light-axion models.

\begin{figure}[!t]
\centering
\includegraphics[width=0.995\linewidth]{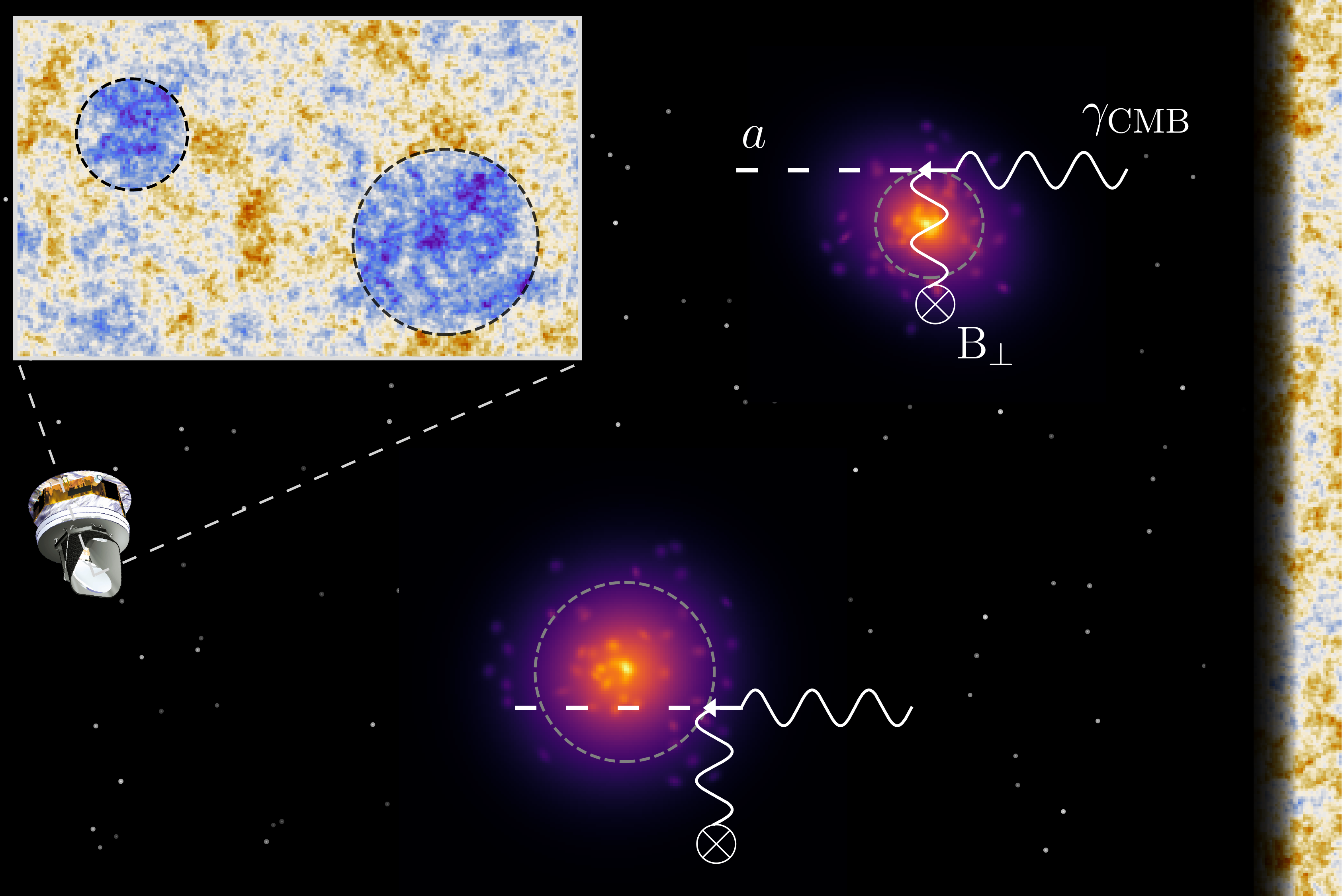}
\caption{Illustration of the physical process studied in this work: CMB photons ($\gamma_{\rm CMB})$ undergo resonant conversion into axions ($a$) within the magnetized plasma of galactic halos (${\rm B}_\perp)$. The dashed circles illustrate the resonance location for the halos, which is set by the free-electron density. The conversion of CMB photons into axions leads to an anisotropic screening of the CMB temperature that is correlated with the positions of galaxies. The inset panel shows the imprint of photon-axion conversion on a CMB temperature map. Note the effect has been greatly amplified for illustrative purposes.
}
\label{fig:axion_screening_diagram}
\end{figure}

The cosmic microwave background (CMB) acts as a natural backlight for exploring photon-axion conversion throughout the Universe. Specifically, the conversion of CMB photons into axions introduces secondary anisotropies in both the intensity and polarization of the CMB~\cite{Schlederer:2015jwa, DAmico:2015snf, Mukherjee:2018oeb, Mukherjee:2018zzg, Mukherjee:2019dsu, Mondino:2024rif, Mehta:2024pdz, Mehta:2024wfo}. In this work, we study the resonant conversion of CMB photons into axions as they traverse the ionized gas and magnetic fields within galactic halos. This conversion leads to anisotropic, \emph{i.e.}, ``patchy'', screening of the CMB temperature that is spatially correlated with the locations of galaxies~\cite{DAmico:2015snf, Mukherjee:2018oeb, Mondino:2024rif} (see Fig.~\ref{fig:axion_screening_diagram} for an illustration).

Recently, Ref.~\cite{Mondino:2024rif} derived theoretical predictions for the imprints of axion-induced screening by large-scale structure (LSS) in the CMB and forecasted that an analysis combining \emph{current} galaxy survey and CMB temperature data can provide competitive sensitivity to the axion-photon coupling for $m_a\approx 10^{-13}~\rm{eV}$ axions. In this \emph{Letter}, we perform the first such analysis. Specifically, we present the first measurement of the cross-correlation between axion-induced CMB screening and the galaxy number density field. We find no evidence for axion-induced screening of the CMB, and constrain the axion-photon coupling $g_{a\gamma\gamma}\lesssim 2\times 10^{-12}~{\rm GeV}^{-1}$ for axions with mass $10^{-13}~{\rm eV}\lesssim m_a\lesssim 10^{-12}~{\rm eV}$. The Supplemental Material (SM) contains further information validating our measurement and theoretical modeling.

\noindent\emph{Conventions:} We work in natural units, $\hbar= k_B=c = 1$. We assume a flat $\Lambda$CDM fiducial cosmology based on the \emph{Planck} 2018 analysis~\cite{Aghanim:2018eyx}: $\Omega_{\rm cdm}h^2=0.1201$, $\Omega_{\rm b}h^2=0.0224$, $h=0.6732$, $\ln\left(10^{10}A_s \right)=3.0448$, $n_s=0.9661$, and $\tau=0.0543$. Here, $\Omega_{\rm cdm}h^2$ and $\Omega_{\rm b}h^2$ are the present-day physical cold dark matter and baryon densities, respectively, $h\equiv H_0/100~{\rm km/s/Mpc}$ is the dimensionless Hubble constant, $A_s$ and $n_s$ are the amplitude and spectral index of primordial scalar perturbations, respectively, and $\tau$ is the optical depth to reionization.

\paragraph{Axion screening of the CMB---} We first review the theoretical formalism describing axion-induced screening of the CMB, and its cross-correlation with the galaxy distribution. For more details, see Ref.~\cite{Mondino:2024rif} and the SM of this work.

Consider a CMB photon with frequency $\nu$ propagating through a halo with free electron density $n_e(\bx)$ and magnetic field ${\bf{B}}(\bx)$. The interaction in Eq.~\eqref{eq:lagrangian} leads to a mixing between the component of the photon polarized along the direction of the transverse magnetic field and an ultra-relativistic axion with mass $m_a\ll \nu$. The mixing is most efficient if the axion mass is equal to the in-medium plasma mass, $m_\gamma(\bx)\equiv \sqrt{4\pi\,\alpha \, n_e(\bx)/m_e}\approx 3.7\times 10^{-11}~{\rm eV}\,\sqrt{n_e/{\rm cm}^{-3}}$, where $\alpha$ is the fine-structure constant.\footnote{For the halo masses considered here, the electron number density typically varies between $10^{-5}~{\rm cm}^{-3}\lesssim n_e \lesssim 10^{-3}~{\rm cm}^{-3}$, hence we are sensitive to axion masses between $10^{-13}~{\rm eV}$ and $10^{-12}~{\rm eV}$.} In the Landau-Zener approximation, the probability for the photon to resonantly convert into an axion is~\cite{Parke:1986, Mirizzi:2009nq, Tashiro:2013yea}
\begin{equation}\label{eq:photon_axion_conv_probability}
    P_{\gamma \rightarrow a}^{\rm res}(\bx(t_{\rm res}), \nu)\simeq \frac{2\pi^2\nu  g_{a\gamma\gamma}^2|{\bf{B}}_\perp^{\rm res}|^2}{m_a^2}\left\vert \frac{d \ln m_{\gamma}^2(\bx)}{dt} \right\vert^{-1}_{t_{\rm res}},
\end{equation}
where ${\bf{B}}_\perp^{\rm res}$ is the component of the magnetic field transverse to the propagation direction of the photon evaluated at the resonance location and $t_{\rm res}$ is the time of resonance. 

Eq.~\eqref{eq:photon_axion_conv_probability} leads to a spatially anisotropic screening of the CMB temperature which is conveniently described by the dark-screening optical depth~\cite{Pirvu:2023lch, Mondino:2024rif, McCarthy:2024ozh},
\begin{equation}\label{eq:optical_depth}
    \tau^{a}(\hatbn,\nu)=\sum_{t_{\rm res}} P_{\gamma \rightarrow a}^{\rm res}(\bx(t_{\rm res}), \nu),
\end{equation}
where the sum is taken over all resonances along the line of sight, $\hatbn$. Consequently, photon-axion conversion induces an anisotropic spectral distortion in the CMB temperature\footnote{We use the notation $\tilde{T}^a$ to denote the temperature fluctuation induced by photon-axion conversion. As described in~\cite{Mukherjee:2018oeb, Mukherjee:2019dsu, Mondino:2024rif, Mehta:2024pdz, Mehta:2024wfo}, photon-axion conversion also introduces polarized spectral distortions. However, this work focuses solely on the CMB temperature distortion, which is projected to provide the best sensitivity for current data.} $\tilde{T}^a(\nu, \hatbn)=T_{\rm CMB}\,\tau^{a}(\nu, \hatbn)\left(\frac{1-e^{-x}}{x}\right)$, where $x\equiv  2\pi \nu/T_{\rm CMB}$ %h\nu/k_BT_{\rm CMB} = 
is the dimensionless frequency, $T_{\rm CMB}\approx 2.726$ K is the CMB monopole temperature, and the term in parentheses accounts for the conversion from specific intensity to CMB thermodynamic temperature units. For convenience, we define the axion screening map at a reference frequency of $\nu=353$ GHz and label the map as $\tilde{T}^a_{353{\rm GHz}}(\hatbn)$ when the normalization is relevant.

In this \emph{Letter}, we investigate the cross-correlation between the axion-induced screened CMB temperature, $\tilde{T}^a(\hatbn)$, and the projected galaxy overdensity, $\delta_g(\hatbn)$. Since both $\tilde{T}^a$ and $\delta_g$ are defined on the sphere, it is convenient to work in the spherical harmonic basis. The cross-correlation is then described by the angular cross-power spectrum, $\langle \delta_{g,\,\ell m}\tilde{T}^{a\,*}_{\ell'm'}\rangle=\delta^K_{\ell\ell'}\delta_{mm'}^KC_\ell^{g\tilde{T}^a}$, where $\delta^K$ is the Kronecker delta, and we have assumed that $\tilde{T}^a$ and $\delta_g$ are statistically isotropic. Not only is $C_\ell^{g\tilde{T}^a}$ more sensitive to $g_{a\gamma\gamma}$ than $C_\ell^{\tilde{T}^a\tilde{T}^a}$ (see Ref.~\cite{Mondino:2024rif} and the SM), but the cross-correlation is immune to biases from foreground contaminants in the axion screening map that are uncorrelated with the galaxy overdensity map. 

We model $C_\ell^{g\tilde{T}^a}$ using the halo model formalism in Ref.~\cite{Mondino:2024rif}. We model the galaxy distribution using the halo occupation distribution (HOD) from Ref.~\cite{Kusiak:2022xkt}. We model the electron distribution using the ``AGN feedback" model from~\cite{Battaglia:2016xbi}. We model the halo magnetic field using the IllustrisTNG simulations~\cite{Nelson:2018uso}. The IllustrisTNG simulations are state-of-the-art cosmological magnetohydrodynamical simulations that have been shown to reproduce a wide range of observed galaxy properties~\cite{Pillepich:2017fcc, Springel:2017tpz, Nelson:2017cxy, 2018MNRAS.477.1206N, Marinacci:2017wew}. In this work, we compute the spherically-averaged, mass-weighted mean magnetic field profiles for halos in the TNG100-1 and TNG300-1 simulations --- the highest-resolution runs of the IllustrisTNG (110.7 Mpc)$^3$ and (302.6 Mpc)$^3$  comoving-volume simulations, respectively. We compute the profiles for $10^{11}~M_\odot-10^{15}~M_\odot$ halos between redshifts $0\leq z\leq 1.5$, covering the range of halo masses and redshifts in our galaxy sample. Due to the limited number of massive halos in the TNG100-1 simulation, we use the TNG300-1 simulation to estimate the magnetic field profiles for halos with $M_{\rm halo}\gtrsim 10^{14}~M_\odot$. Our improved magnetic field modeling in massive halos leads to a final constraint that is better than the forecast from Ref.~\cite{Mondino:2024rif}. Finally, we conservatively neglect conversions that occur deep within the halo or in its outermost regions, where theoretical modeling, particularly of the electron distribution and magnetic field, has more uncertainties. Specifically, we only consider conversions that occur between $0.1~R_{200c} < r < \min(R_{\rm vir}, 1.08\,R_{200c})$, where $r$ is the distance to the halo center, $R_{200c}$ is the radius at which the average density enclosed within the halo equals 200 times the critical density of the Universe at the halo redshift, and $R_{\rm vir}$ is the virial radius. The second upper bound comes from the truncation scale used in the best-fit HOD model for the galaxy sample considered here~\cite{Kusiak:2022xkt}. In the SM, we provide constraints for various models of the electron distribution, magnetic field, and galaxy distribution, and offer additional details regarding the halo model calculation.

\paragraph{Axion screening and galaxy data---} From Eq.~\eqref{eq:optical_depth}, the resonant conversion of CMB photons into axions causes a spectral distortion in the CMB temperature with the following spectral energy distribution (SED):\footnote{We normalize the SED to unity at 353 GHz.}
\begin{equation}\label{eq:SED}
    \frac{\tilde{T}^a(x)}{T_{\rm CMB}}\propto x\left(\frac{1-e^{-x}}{x} \right),
\end{equation}
where $x\equiv  2\pi \nu/T_{\rm CMB}.$ Consequently, we can leverage the constrained needlet internal linear combination (NILC) method~\cite{Delabrouille:2008qd,Remazeilles:2010hq,McCarthy:2023hpa} to construct maps of axion-induced screening from multi-frequency CMB temperature data. In general, the ILC procedure~\cite{1992ApJ...396L...7B, WMAP:2003cmr, Tegmark:2003ve, Eriksen:2004jg} constructs the minimum-variance linear combination of the observed frequency maps that ensures unit response to the signal of interest. In NILC, the weights are computed on a frame of needlets~\cite{doi:10.1137/040614359}, which have compact support in both real and harmonic space. NILC is particularly well-suited for CMB analyses since Galactic contributions are often localized in real space, whereas extragalactic contributions are often better described in harmonic space. We use \emph{constrained} NILC~\cite{Chen:2008gw, Remazeilles:2010hq}, which exactly removes contributions from unwanted signals in the map that have a known SED, albeit at the cost of increasing the noise of the map. We construct our NILC maps using \texttt{pyilc}~\cite{McCarthy:2023hpa}.\footnote{\href{https://github.com/jcolinhill/pyilc/}{https://github.com/jcolinhill/pyilc/}} 
  
For our fiducial analysis, we use the \emph{Planck} NPIPE (PR4)~\cite{Planck:2020olo} CMB temperature maps at 30, 44, 70, 100, 143, 217, 353, and 545 GHz. Following similar analyses in Refs.~\cite{McCarthy:2023hpa, McCarthy:2024ozh, McCarthy:2023cwg}, we pre-process the maps by masking the Galactic plane and bright point sources, and subsequently inpainting these masked regions. Since we are interested in cross-correlating our axion map with LSS tracers, we need to use the constrained NILC to ``deproject'' foreground contaminants that are correlated with the \emph{unWISE} galaxies, particularly the thermal Sunyaev-Zel'dovich (tSZ) effect~\cite{Sunyaev:1972eq, Sunyaev:1980nv} and the cosmic infrared background (CIB).\footnote{On large scales, the CMB is also correlated with \emph{unWISE} galaxies due to the integrated Sachs-Wolfe (ISW) effect~\cite{Sachs:1967er}. However, this contribution is negligible at the scales we analyze ($\ell>300$)~\cite{Krolewski:2021znk}, thus we do not explicitly deproject the CMB (note that it will be still partially removed due to the NILC minimum-variance criterion).} The tSZ effect has a known spectral response and can be deprojected exactly. Since there is no exact model for the CIB SED, we follow Refs.~\cite{McCarthy:2023hpa, McCarthy:2023cwg, McCarthy:2024ozh} and model the CIB as a modified blackbody. We fix the CIB temperature $T_{\rm CIB}=20$ K and spectral index $\beta_{\rm CIB}=1.6$. To account for deviations in the CIB SED from a modified blackbody, we also deproject the first derivative of the SED with respect to $\beta_{\rm CIB}$, following Ref.~\cite{Chluba:2017rtj}. The SM presents extensive validation of our map-making and foreground-cleaning procedure.

For our galaxy sample, we use infrared-selected galaxies from the \emph{unWISE} catalog~\cite{Lang_2014, Meisner_2017, Meisner_2017a, Schlafly_2019, Krolewski:2019yrv} constructed from data collected by the \emph{Wide-field Infrared Survey Explorer} (\emph{WISE}) mission~\cite{2010AJ....140.1868W,2011ApJ...731...53M}. The \emph{unWISE} catalog includes over 500 million galaxies across the full sky, making it particularly well-suited for cross-correlation analyses. In this work, we use the low-redshift \emph{unWISE} ``Blue" sample~\cite{Krolewski:2019yrv}, which has a mean redshift of $\bar{z}\approx0.6$. The \emph{unWISE} catalog also contains higher-redshift ``Green" ($\bar{z}\approx1.1$) and ``Red" ($\bar{z}\approx 1.5$) samples. However, these samples are more strongly correlated with the CIB, which is predominantly sourced by higher-redshift dusty star-forming galaxies~\cite{Maniyar:2018xfk, Yan:2023okq}. Indeed, in the SM we demonstrate that our fiducial analysis methods are insufficient for removing high-redshift CIB contributions that are correlated with the \emph{unWISE} Green sample, despite their robust performance for the Blue sample. Finally, even in the absence of CIB contamination, modeling the cross-correlation $C_\ell^{g\tilde{T}^a}$ for these high-redshift galaxy samples is challenging due to increased uncertainty in the magnetic field and electron number density models.

\paragraph{Cross-correlation measurement---} We compute the angular cross-power spectrum between the axion screening map and the \emph{unWISE} Blue galaxy catalog using \texttt{NaMaster}\footnote{\href{https://github.com/LSSTDESC/NaMaster}{https://github.com/LSSTDESC/NaMaster}}~\cite{Hivon:2001jp, Alonso:2018jzx}, which accounts for the mask-induced mode coupling. We mask the galaxy map using the union of the \emph{unWISE} mask~\cite{Krolewski:2019yrv} and the \emph{Planck} CMB lensing mask, leaving 51\% of the sky unmasked. We mask the axion map using the union of the pre-processing mask with the \emph{Planck} 70\% Galactic plane mask, leaving 62.5\% of the sky unmasked. We estimate the cross-power spectra in multipole bins between $5<\ell<3000$ with width $\Delta\ell=150.$ To estimate the covariance of $C_{\ell}^{g \tilde{T}^a_{\rm 353 GHz}}$, we compute the Gaussian covariance using the improved narrow kernel approximation~\cite{Nicola:2020lhi}. Our analysis includes only multipoles with bin centers between $300\leq \ell \leq 2000$. As described in the SM, we exclude the two bins below $\ell=300$ to avoid contributions from the ISW effect and to mitigate the impact of possible large-scale residual extra-galactic foregrounds present in our axion screening maps. We exclude bins with $\ell>2000$ to account for potential limitations in the HOD modeling on smaller scales. Detailed discussions of the cross-power spectrum measurement are provided in the SM.

\begin{figure}[!t]
\centering
\includegraphics[width=0.995\linewidth]{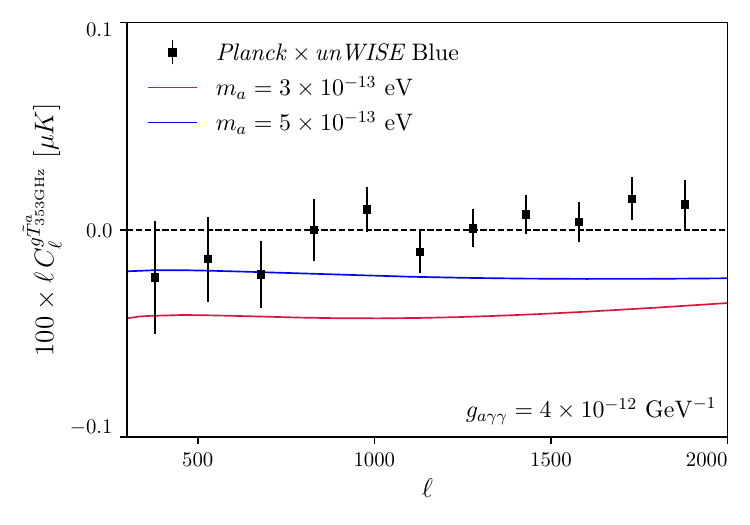}
\caption{Measured cross-correlation between our axion screening map and the \emph{unWISE} Blue galaxy sample. The measurement is consistent with zero (PTE of 0.61), thus we find no evidence of axion-induced screening. The solid lines show halo model predictions for the cross-correlation for two axion masses, assuming $g_{a\gamma\gamma}=4\times 10^{-12}~{\rm GeV}^{-1}.$ }
\label{fig:Cl_Tg_meas_and_fit}
\end{figure}

Fig.~\ref{fig:Cl_Tg_meas_and_fit} presents the measured cross-power spectrum between our axion-induced screening map and the \emph{unWISE} Blue galaxy sample,  $C_{\ell}^{g \tilde{T}^a_{\rm 353 GHz}}$. The measured cross-correlation is consistent with zero, with a $\chi^2$ value of 9.2 for 11 degrees of freedom. This corresponds to a probability-to-exceed (PTE) of 0.61. Thus, we do not detect axion-induced screening. For comparison, the solid lines in Fig.~\ref{fig:Cl_Tg_meas_and_fit} show the halo model predictions for two different axion masses assuming an axion-photon coupling $g_{a\gamma\gamma}=4\times 10^{-12}~{\rm GeV}^{-1}.$ The theoretical prediction for the cross-correlation is negative since photon-axion conversion \emph{decreases} the CMB intensity near galaxies.

\begin{figure}[!t]
\centering
\includegraphics[width=0.995\linewidth]{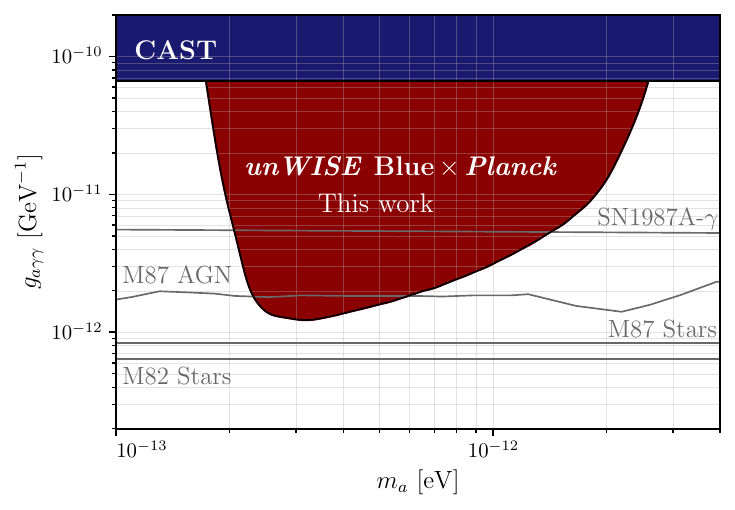}
    \caption{Constraints on the axion-photon coupling, $g_{a\gamma\gamma}$. The red region indicates the 95\% upper exclusion limit on $g_{a\gamma\gamma}$ derived in this work. The shaded blue region denotes the 95\% bound set by CAST~\cite{CAST:2017uph}. The solid grey lines show the 95\% confidence limits derived from various high-energy astrophysics searches for photon-axion or axion-photon conversion in a \emph{single} galaxy~\cite{AxionLimits}, including the absence of spectral distortions in the AGN spectrum of M87~\cite{Marsh:2017yvc}, the non-observation of $\gamma$-rays from axions produced in SN1987A and converting within the Galactic magnetic field~\cite{Hoof:2022xbe}, and the absence of X-ray signals from stellar axion production and conversion in M82 and M87~\cite{Ning:2024eky}. We present a more detailed comparison with existing constraints in the text.}
\label{fig:axion_constraints}
\end{figure}

\paragraph{Constraints on the axion-photon coupling---} Using the measurement and theoretical predictions for $C_{\ell}^{g \tilde{T}^a_{\rm 353 GHz}}$, we can exclude regions of the $m_a-g_{a\gamma \gamma}$ plane. We assume a Gaussian likelihood, which is given by, up to a constant, 
\begin{equation}
    \ln\mathcal{L}(\hat{C}_{\ell}^{g \tilde{T}^a}|\boldsymbol{\theta})=\left(\delta C_{\ell}^{g\tilde{T}^a}(\boldsymbol{\theta})\right)^T \mathcal{C}_{\ell \ell'}^{-1}\left(\delta C_{\ell'}^{g\tilde{T}^a}(\boldsymbol{\theta})\right),
\end{equation}
where $\boldsymbol{\theta}=(g_{a\gamma \gamma}, m_a)$ is the parameter vector, $\delta C_{\ell}^{g\tilde{T}^a}(\boldsymbol{\theta})$ is the difference between the measured cross-power spectrum and our theoretical model, and $\mathcal{C}_{\ell \ell'}^{-1}$ is the inverse of the $\hat{C}_{\ell}^{g \tilde{T}^a}$ covariance matrix, which accounts for the noise and mask-induced mode coupling in our maps. To obtain the posterior distribution of $\boldsymbol{\theta}$, we profile over axion masses between $10^{-13}$ eV and $4\times 10^{-12}$ eV, assuming a uniform prior on the axion-photon coupling, with $g_{a\gamma\gamma}\geq 0~{\rm GeV}^{-1}$.

Fig.~\ref{fig:axion_constraints} shows the constraints on the axion-photon coupling, $g_{a\gamma\gamma}$, derived in this work. The shaded red region shows the 95\% one-tailed upper confidence limit on $g_{a\gamma\gamma}$ derived from $C_\ell^{g \tilde{T}^a}$. The shaded blue region shows the laboratory constraints from the CERN Axion Solar Telescope (CAST)~\cite{CAST:2017uph}. The grey lines depict some of the most stringent astrophysical bounds in this region of parameter space, which are derived from Solar Maximum Mission $\gamma$-ray observations of SN1987A~\cite{Hoof:2022xbe}, \emph{Chandra} X-ray observations of active galactic nuclei (AGN) in M87~\cite{Marsh:2017yvc}, and \emph{NuSTAR} X-ray observations of M82 and M87~\cite{Ning:2024eky}. There are also tight bounds on the axion-photon coupling in this mass range from observations of the AGN spectra of NGC 1275~\cite{Reynolds:2019uqt} and H1821+643~\cite{Reynes:2021bpe}. We do not include these in Fig.~\ref{fig:axion_constraints} for visual clarity, as they overlap with the M82 stellar axion production constraint. We also note that constraints on axions in this mass range that are independent of the axion-photon coupling can be derived from black hole superradiance~\cite{Arvanitaki:2014wva, Cardoso:2018tly}. However, such constraints only apply if the axion self-interaction is small~\cite{Baryakhtar:2020gao}.

\paragraph{Conclusions---} In this work, we have derived new limits on the axion-photon coupling for light axions by analyzing the cross-correlation of galaxies with foreground-cleaned maps of resonant axion-induced screening in the CMB. Utilizing state-of-the-art CMB and LSS datasets, we have constrained the axion-photon coupling $g_{a\gamma\gamma}\lesssim 2\times 10^{-12}~{\rm GeV}^{-1}$ at the 95\% confidence level for $10^{-13}~{\rm eV}\lesssim m_a\lesssim 10^{-12}~{\rm eV}$. 

Our constraints are significantly tighter than the laboratory bounds from CAST and are competitive with the most stringent limits from previous astrophysical axion searches. Notably, unlike conventional astrophysical axion searches, which are typically based on photon-axion conversion within a single galaxy, our constraints are derived by searching for \emph{statistically averaged} photon-axion conversion in the circumgalactic medium (CGM) of tens of millions of galaxies. By focusing on photon-axion conversions outside of the halo center, our constraints are less sensitive to the complex modeling of the magnetic field within galaxies. Finally, the photon source in our analysis is the CMB, which has remarkably well-understood statistical and physical properties. Consequently, our constraints are highly complementary to existing astrophysical axion bounds.

There are several possible extensions to this work. Firstly, near-term CMB and LSS surveys are expected to improve our constraints by an order of magnitude~\cite{Mondino:2024rif}. Additionally, axion-photon conversion produces unique signatures in the CMB polarization that one could search for using similar methods to the analysis in this work. On the modeling side, improved measurements of the kinematic Sunyaev-Zel'dovich effect~\cite{Amodeo:2020mmu, AtacamaCosmologyTelescope:2020wtv, Kusiak:2021hai} and fast radio bursts~\cite{Madhavacheril:2019buy, Prochaska_2019} will help refine models for the electron distribution. Furthermore, future observations will lead to a better understanding of the magnetic field profile around halos~\cite{Heesen_2023, Muralidhara:2024ipg, Osinga:2024vzo}. These observational improvements will be supported by advancements in cosmological magnetohydrodynamical simulations~\cite{2024MNRAS.528.2308P}.
In conclusion, the axion search conducted in this work not only provides some of the most stringent constraints on $10^{-13}~{\rm eV}\lesssim m_a\lesssim 10^{-12}~{\rm eV}$ mass axions, but also lays the foundation for future astrophysical searches for these elusive particles using CMB and LSS datasets.

%TC:ignore

\paragraph{Acknowledgements---}  We thank Aleksandra Kusiak and Dalila P\^irvu for useful discussions. The authors acknowledge the Texas Advanced Computing Center (TACC)\footnote{\href{http://www.tacc.utexas.edu}{http://www.tacc.utexas.edu}} at the University of Texas at Austin for providing computational resources that have contributed to the research results reported within this paper. We acknowledge computing resources from Columbia University's Shared Research Computing Facility project, which is supported by NIH Research Facility Improvement Grant 1G20RR030893-01, and associated funds from the New York State Empire State Development, Division of Science Technology and Innovation (NYSTAR) Contract C090171, both awarded April 15, 2010.  SG and JCH acknowledge support from NSF grant AST-2307727 and DOE HEP grant DE-SC0011941.  JCH also acknowledges support from NSF grant AST-2108536, NASA grants 80NSSC22K0721 (ATP) and 80NSSC23K0463 (ADAP), the Sloan Foundation, and the Simons Foundation.  MCJ and JH are supported by the Natural Sciences and Engineering Research Council of Canada through a Discovery grant. Research at Perimeter Institute is supported in part by the Government of Canada through the Department of Innovation, Science and Economic Development Canada and by the Province of Ontario through the Ministry of Research, Innovation and Science.  FMcC acknowledges support from the European Research Council (ERC) under the European Union's Horizon 2020 research and innovation programme (Grant agreement No.~851274). The Flatiron Institute is a division of the Simons Foundation.

\bibliography{biblio.bib}

\providecommand{\noopsort}[1]{}\providecommand{\singleletter}[1]{#1}%
\begin{thebibliography}{109}%
\makeatletter
\providecommand \@ifxundefined [1]{%
 \@ifx{#1\undefined}
}%
\providecommand \@ifnum [1]{%
 \ifnum #1\expandafter \@firstoftwo
 \else \expandafter \@secondoftwo
 \fi
}%
\providecommand \@ifx [1]{%
 \ifx #1\expandafter \@firstoftwo
 \else \expandafter \@secondoftwo
 \fi
}%
\providecommand \natexlab [1]{#1}%
\providecommand \enquote  [1]{``#1''}%
\providecommand \bibnamefont  [1]{#1}%
\providecommand \bibfnamefont [1]{#1}%
\providecommand \citenamefont [1]{#1}%
\providecommand \href@noop [0]{\@secondoftwo}%
\providecommand \href [0]{\begingroup \@sanitize@url \@href}%
\providecommand \@href[1]{\@@startlink{#1}\@@href}%
\providecommand \@@href[1]{\endgroup#1\@@endlink}%
\providecommand \@sanitize@url [0]{\catcode `\\12\catcode `\$12\catcode `\&12\catcode `\#12\catcode `\^12\catcode `\_12\catcode `\%12\relax}%
\providecommand \@@startlink[1]{}%
\providecommand \@@endlink[0]{}%
\providecommand \url  [0]{\begingroup\@sanitize@url \@url }%
\providecommand \@url [1]{\endgroup\@href {#1}{\urlprefix }}%
\providecommand \urlprefix  [0]{URL }%
\providecommand \Eprint [0]{\href }%
\providecommand \doibase [0]{http://dx.doi.org/}%
\providecommand \selectlanguage [0]{\@gobble}%
\providecommand \bibinfo  [0]{\@secondoftwo}%
\providecommand \bibfield  [0]{\@secondoftwo}%
\providecommand \translation [1]{[#1]}%
\providecommand \BibitemOpen [0]{}%
\providecommand \bibitemStop [0]{}%
\providecommand \bibitemNoStop [0]{.\EOS\space}%
\providecommand \EOS [0]{\spacefactor3000\relax}%
\providecommand \BibitemShut  [1]{\csname bibitem#1\endcsname}%
\let\auto@bib@innerbib\@empty
%</preamble>
\bibitem [{\citenamefont {Peccei}\ and\ \citenamefont {Quinn}(1977{\natexlab{a}})}]{Peccei:1977ur}%
  \BibitemOpen
  \bibfield  {author} {\bibinfo {author} {\bibfnamefont {R.~D.}\ \bibnamefont {Peccei}}\ and\ \bibinfo {author} {\bibfnamefont {H.~R.}\ \bibnamefont {Quinn}},\ }\href {\doibase 10.1103/PhysRevD.16.1791} {\bibfield  {journal} {\bibinfo  {journal} {Phys. Rev. D}\ }\textbf {\bibinfo {volume} {16}},\ \bibinfo {pages} {1791} (\bibinfo {year} {1977}{\natexlab{a}})}\BibitemShut {NoStop}%
\bibitem [{\citenamefont {Peccei}\ and\ \citenamefont {Quinn}(1977{\natexlab{b}})}]{Peccei:1977hh}%
  \BibitemOpen
  \bibfield  {author} {\bibinfo {author} {\bibfnamefont {R.~D.}\ \bibnamefont {Peccei}}\ and\ \bibinfo {author} {\bibfnamefont {H.~R.}\ \bibnamefont {Quinn}},\ }\href {\doibase 10.1103/PhysRevLett.38.1440} {\bibfield  {journal} {\bibinfo  {journal} {Phys. Rev. Lett.}\ }\textbf {\bibinfo {volume} {38}},\ \bibinfo {pages} {1440} (\bibinfo {year} {1977}{\natexlab{b}})}\BibitemShut {NoStop}%
\bibitem [{\citenamefont {Weinberg}(1978)}]{Weinberg:1977ma}%
  \BibitemOpen
  \bibfield  {author} {\bibinfo {author} {\bibfnamefont {S.}~\bibnamefont {Weinberg}},\ }\href {\doibase 10.1103/PhysRevLett.40.223} {\bibfield  {journal} {\bibinfo  {journal} {Phys. Rev. Lett.}\ }\textbf {\bibinfo {volume} {40}},\ \bibinfo {pages} {223} (\bibinfo {year} {1978})}\BibitemShut {NoStop}%
\bibitem [{\citenamefont {Wilczek}(1978)}]{Wilczek:1977pj}%
  \BibitemOpen
  \bibfield  {author} {\bibinfo {author} {\bibfnamefont {F.}~\bibnamefont {Wilczek}},\ }\href {\doibase 10.1103/PhysRevLett.40.279} {\bibfield  {journal} {\bibinfo  {journal} {Phys. Rev. Lett.}\ }\textbf {\bibinfo {volume} {40}},\ \bibinfo {pages} {279} (\bibinfo {year} {1978})}\BibitemShut {NoStop}%
\bibitem [{\citenamefont {Witten}(1984)}]{Witten:1984dg}%
  \BibitemOpen
  \bibfield  {author} {\bibinfo {author} {\bibfnamefont {E.}~\bibnamefont {Witten}},\ }\href {\doibase 10.1016/0370-2693(84)90422-2} {\bibfield  {journal} {\bibinfo  {journal} {Phys. Lett. B}\ }\textbf {\bibinfo {volume} {149}},\ \bibinfo {pages} {351} (\bibinfo {year} {1984})}\BibitemShut {NoStop}%
\bibitem [{\citenamefont {Svrcek}\ and\ \citenamefont {Witten}(2006)}]{Svrcek:2006yi}%
  \BibitemOpen
  \bibfield  {author} {\bibinfo {author} {\bibfnamefont {P.}~\bibnamefont {Svrcek}}\ and\ \bibinfo {author} {\bibfnamefont {E.}~\bibnamefont {Witten}},\ }\href {\doibase 10.1088/1126-6708/2006/06/051} {\bibfield  {journal} {\bibinfo  {journal} {JHEP}\ }\textbf {\bibinfo {volume} {06}},\ \bibinfo {pages} {051} (\bibinfo {year} {2006})},\ \Eprint {http://arxiv.org/abs/hep-th/0605206} {arXiv:hep-th/0605206} \BibitemShut {NoStop}%
\bibitem [{\citenamefont {Arvanitaki}\ \emph {et~al.}(2010)\citenamefont {Arvanitaki}, \citenamefont {Dimopoulos}, \citenamefont {Dubovsky}, \citenamefont {Kaloper},\ and\ \citenamefont {March-Russell}}]{Arvanitaki:2009fg}%
  \BibitemOpen
  \bibfield  {author} {\bibinfo {author} {\bibfnamefont {A.}~\bibnamefont {Arvanitaki}}, \bibinfo {author} {\bibfnamefont {S.}~\bibnamefont {Dimopoulos}}, \bibinfo {author} {\bibfnamefont {S.}~\bibnamefont {Dubovsky}}, \bibinfo {author} {\bibfnamefont {N.}~\bibnamefont {Kaloper}}, \ and\ \bibinfo {author} {\bibfnamefont {J.}~\bibnamefont {March-Russell}},\ }\href {\doibase 10.1103/PhysRevD.81.123530} {\bibfield  {journal} {\bibinfo  {journal} {Phys. Rev. D}\ }\textbf {\bibinfo {volume} {81}},\ \bibinfo {pages} {123530} (\bibinfo {year} {2010})},\ \Eprint {http://arxiv.org/abs/0905.4720} {arXiv:0905.4720 [hep-th]} \BibitemShut {NoStop}%
\bibitem [{\citenamefont {Marsh}(2018)}]{Marsh:2017hbv}%
  \BibitemOpen
  \bibfield  {author} {\bibinfo {author} {\bibfnamefont {D.~J.~E.}\ \bibnamefont {Marsh}},\ }in\ \href {\doibase 10.3204/DESY-PROC-2017-02/marsh_david} {\emph {\bibinfo {booktitle} {{13th Patras Workshop on Axions, WIMPs and WISPs}}}}\ (\bibinfo {year} {2018})\ pp.\ \bibinfo {pages} {59--74},\ \Eprint {http://arxiv.org/abs/1712.03018} {arXiv:1712.03018 [hep-ph]} \BibitemShut {NoStop}%
\bibitem [{\citenamefont {Preskill}\ \emph {et~al.}(1983)\citenamefont {Preskill}, \citenamefont {Wise},\ and\ \citenamefont {Wilczek}}]{Preskill:1982cy}%
  \BibitemOpen
  \bibfield  {author} {\bibinfo {author} {\bibfnamefont {J.}~\bibnamefont {Preskill}}, \bibinfo {author} {\bibfnamefont {M.~B.}\ \bibnamefont {Wise}}, \ and\ \bibinfo {author} {\bibfnamefont {F.}~\bibnamefont {Wilczek}},\ }\href {\doibase 10.1016/0370-2693(83)90637-8} {\bibfield  {journal} {\bibinfo  {journal} {Phys. Lett. B}\ }\textbf {\bibinfo {volume} {120}},\ \bibinfo {pages} {127} (\bibinfo {year} {1983})}\BibitemShut {NoStop}%
\bibitem [{\citenamefont {Abbott}\ and\ \citenamefont {Sikivie}(1983)}]{Abbott:1982af}%
  \BibitemOpen
  \bibfield  {author} {\bibinfo {author} {\bibfnamefont {L.~F.}\ \bibnamefont {Abbott}}\ and\ \bibinfo {author} {\bibfnamefont {P.}~\bibnamefont {Sikivie}},\ }\href {\doibase 10.1016/0370-2693(83)90638-X} {\bibfield  {journal} {\bibinfo  {journal} {Phys. Lett. B}\ }\textbf {\bibinfo {volume} {120}},\ \bibinfo {pages} {133} (\bibinfo {year} {1983})}\BibitemShut {NoStop}%
\bibitem [{\citenamefont {Dine}\ and\ \citenamefont {Fischler}(1983)}]{Dine:1982ah}%
  \BibitemOpen
  \bibfield  {author} {\bibinfo {author} {\bibfnamefont {M.}~\bibnamefont {Dine}}\ and\ \bibinfo {author} {\bibfnamefont {W.}~\bibnamefont {Fischler}},\ }\href {\doibase 10.1016/0370-2693(83)90639-1} {\bibfield  {journal} {\bibinfo  {journal} {Phys. Lett. B}\ }\textbf {\bibinfo {volume} {120}},\ \bibinfo {pages} {137} (\bibinfo {year} {1983})}\BibitemShut {NoStop}%
\bibitem [{\citenamefont {Hui}(2021)}]{Hui:2021tkt}%
  \BibitemOpen
  \bibfield  {author} {\bibinfo {author} {\bibfnamefont {L.}~\bibnamefont {Hui}},\ }\href {\doibase 10.1146/annurev-astro-120920-010024} {\bibfield  {journal} {\bibinfo  {journal} {Ann. Rev. Astron. Astrophys.}\ }\textbf {\bibinfo {volume} {59}},\ \bibinfo {pages} {247} (\bibinfo {year} {2021})},\ \Eprint {http://arxiv.org/abs/2101.11735} {arXiv:2101.11735 [astro-ph.CO]} \BibitemShut {NoStop}%
\bibitem [{\citenamefont {Graham}\ \emph {et~al.}(2015)\citenamefont {Graham}, \citenamefont {Irastorza}, \citenamefont {Lamoreaux}, \citenamefont {Lindner},\ and\ \citenamefont {van Bibber}}]{Graham:2015ouw}%
  \BibitemOpen
  \bibfield  {author} {\bibinfo {author} {\bibfnamefont {P.~W.}\ \bibnamefont {Graham}}, \bibinfo {author} {\bibfnamefont {I.~G.}\ \bibnamefont {Irastorza}}, \bibinfo {author} {\bibfnamefont {S.~K.}\ \bibnamefont {Lamoreaux}}, \bibinfo {author} {\bibfnamefont {A.}~\bibnamefont {Lindner}}, \ and\ \bibinfo {author} {\bibfnamefont {K.~A.}\ \bibnamefont {van Bibber}},\ }\href {\doibase 10.1146/annurev-nucl-102014-022120} {\bibfield  {journal} {\bibinfo  {journal} {Ann. Rev. Nucl. Part. Sci.}\ }\textbf {\bibinfo {volume} {65}},\ \bibinfo {pages} {485} (\bibinfo {year} {2015})},\ \Eprint {http://arxiv.org/abs/1602.00039} {arXiv:1602.00039 [hep-ex]} \BibitemShut {NoStop}%
\bibitem [{\citenamefont {Irastorza}\ and\ \citenamefont {Redondo}(2018)}]{Irastorza:2018dyq}%
  \BibitemOpen
  \bibfield  {author} {\bibinfo {author} {\bibfnamefont {I.~G.}\ \bibnamefont {Irastorza}}\ and\ \bibinfo {author} {\bibfnamefont {J.}~\bibnamefont {Redondo}},\ }\href {\doibase 10.1016/j.ppnp.2018.05.003} {\bibfield  {journal} {\bibinfo  {journal} {Prog. Part. Nucl. Phys.}\ }\textbf {\bibinfo {volume} {102}},\ \bibinfo {pages} {89} (\bibinfo {year} {2018})},\ \Eprint {http://arxiv.org/abs/1801.08127} {arXiv:1801.08127 [hep-ph]} \BibitemShut {NoStop}%
\bibitem [{\citenamefont {Choi}\ \emph {et~al.}(2021)\citenamefont {Choi}, \citenamefont {Im},\ and\ \citenamefont {Sub~Shin}}]{Choi:2020rgn}%
  \BibitemOpen
  \bibfield  {author} {\bibinfo {author} {\bibfnamefont {K.}~\bibnamefont {Choi}}, \bibinfo {author} {\bibfnamefont {S.~H.}\ \bibnamefont {Im}}, \ and\ \bibinfo {author} {\bibfnamefont {C.}~\bibnamefont {Sub~Shin}},\ }\href {\doibase 10.1146/annurev-nucl-120720-031147} {\bibfield  {journal} {\bibinfo  {journal} {Ann. Rev. Nucl. Part. Sci.}\ }\textbf {\bibinfo {volume} {71}},\ \bibinfo {pages} {225} (\bibinfo {year} {2021})},\ \Eprint {http://arxiv.org/abs/2012.05029} {arXiv:2012.05029 [hep-ph]} \BibitemShut {NoStop}%
\bibitem [{\citenamefont {Wolfenstein}(1978)}]{Wolfenstein:1977ue}%
  \BibitemOpen
  \bibfield  {author} {\bibinfo {author} {\bibfnamefont {L.}~\bibnamefont {Wolfenstein}},\ }\href {\doibase 10.1103/PhysRevD.17.2369} {\bibfield  {journal} {\bibinfo  {journal} {\prd}\ }\textbf {\bibinfo {volume} {17}},\ \bibinfo {pages} {2369} (\bibinfo {year} {1978})}\BibitemShut {NoStop}%
\bibitem [{\citenamefont {Mikheyev}\ and\ \citenamefont {Smirnov}(1985)}]{Mikheyev:1985zog}%
  \BibitemOpen
  \bibfield  {author} {\bibinfo {author} {\bibfnamefont {S.~P.}\ \bibnamefont {Mikheyev}}\ and\ \bibinfo {author} {\bibfnamefont {A.~Y.}\ \bibnamefont {Smirnov}},\ }\href@noop {} {\bibfield  {journal} {\bibinfo  {journal} {Sov. J. Nucl. Phys.}\ }\textbf {\bibinfo {volume} {42}},\ \bibinfo {pages} {913} (\bibinfo {year} {1985})}\BibitemShut {NoStop}%
\bibitem [{\citenamefont {Mirizzi}\ \emph {et~al.}(2009)\citenamefont {Mirizzi}, \citenamefont {Redondo},\ and\ \citenamefont {Sigl}}]{Mirizzi:2009nq}%
  \BibitemOpen
  \bibfield  {author} {\bibinfo {author} {\bibfnamefont {A.}~\bibnamefont {Mirizzi}}, \bibinfo {author} {\bibfnamefont {J.}~\bibnamefont {Redondo}}, \ and\ \bibinfo {author} {\bibfnamefont {G.}~\bibnamefont {Sigl}},\ }\href {\doibase 10.1088/1475-7516/2009/08/001} {\bibfield  {journal} {\bibinfo  {journal} {JCAP}\ }\textbf {\bibinfo {volume} {08}},\ \bibinfo {pages} {001} (\bibinfo {year} {2009})},\ \Eprint {http://arxiv.org/abs/0905.4865} {arXiv:0905.4865 [hep-ph]} \BibitemShut {NoStop}%
\bibitem [{\citenamefont {Schlederer}\ and\ \citenamefont {Sigl}(2016)}]{Schlederer:2015jwa}%
  \BibitemOpen
  \bibfield  {author} {\bibinfo {author} {\bibfnamefont {M.}~\bibnamefont {Schlederer}}\ and\ \bibinfo {author} {\bibfnamefont {G.}~\bibnamefont {Sigl}},\ }\href {\doibase 10.1088/1475-7516/2016/01/038} {\bibfield  {journal} {\bibinfo  {journal} {JCAP}\ }\textbf {\bibinfo {volume} {01}},\ \bibinfo {pages} {038} (\bibinfo {year} {2016})},\ \Eprint {http://arxiv.org/abs/1507.02855} {arXiv:1507.02855 [hep-ph]} \BibitemShut {NoStop}%
\bibitem [{\citenamefont {D'Amico}\ and\ \citenamefont {Kaloper}(2015)}]{DAmico:2015snf}%
  \BibitemOpen
  \bibfield  {author} {\bibinfo {author} {\bibfnamefont {G.}~\bibnamefont {D'Amico}}\ and\ \bibinfo {author} {\bibfnamefont {N.}~\bibnamefont {Kaloper}},\ }\href {\doibase 10.1103/PhysRevD.91.085015} {\bibfield  {journal} {\bibinfo  {journal} {Phys. Rev. D}\ }\textbf {\bibinfo {volume} {91}},\ \bibinfo {pages} {085015} (\bibinfo {year} {2015})},\ \Eprint {http://arxiv.org/abs/1501.01642} {arXiv:1501.01642 [astro-ph.CO]} \BibitemShut {NoStop}%
\bibitem [{\citenamefont {Mukherjee}\ \emph {et~al.}(2018)\citenamefont {Mukherjee}, \citenamefont {Khatri},\ and\ \citenamefont {Wandelt}}]{Mukherjee:2018oeb}%
  \BibitemOpen
  \bibfield  {author} {\bibinfo {author} {\bibfnamefont {S.}~\bibnamefont {Mukherjee}}, \bibinfo {author} {\bibfnamefont {R.}~\bibnamefont {Khatri}}, \ and\ \bibinfo {author} {\bibfnamefont {B.~D.}\ \bibnamefont {Wandelt}},\ }\href {\doibase 10.1088/1475-7516/2018/04/045} {\bibfield  {journal} {\bibinfo  {journal} {JCAP}\ }\textbf {\bibinfo {volume} {04}},\ \bibinfo {pages} {045} (\bibinfo {year} {2018})},\ \Eprint {http://arxiv.org/abs/1801.09701} {arXiv:1801.09701 [astro-ph.CO]} \BibitemShut {NoStop}%
\bibitem [{\citenamefont {Mukherjee}\ \emph {et~al.}(2019)\citenamefont {Mukherjee}, \citenamefont {Khatri},\ and\ \citenamefont {Wandelt}}]{Mukherjee:2018zzg}%
  \BibitemOpen
  \bibfield  {author} {\bibinfo {author} {\bibfnamefont {S.}~\bibnamefont {Mukherjee}}, \bibinfo {author} {\bibfnamefont {R.}~\bibnamefont {Khatri}}, \ and\ \bibinfo {author} {\bibfnamefont {B.~D.}\ \bibnamefont {Wandelt}},\ }\href {\doibase 10.1088/1475-7516/2019/06/031} {\bibfield  {journal} {\bibinfo  {journal} {JCAP}\ }\textbf {\bibinfo {volume} {06}},\ \bibinfo {pages} {031} (\bibinfo {year} {2019})},\ \Eprint {http://arxiv.org/abs/1811.11177} {arXiv:1811.11177 [astro-ph.CO]} \BibitemShut {NoStop}%
\bibitem [{\citenamefont {Mukherjee}\ \emph {et~al.}(2020)\citenamefont {Mukherjee}, \citenamefont {Spergel}, \citenamefont {Khatri},\ and\ \citenamefont {Wandelt}}]{Mukherjee:2019dsu}%
  \BibitemOpen
  \bibfield  {author} {\bibinfo {author} {\bibfnamefont {S.}~\bibnamefont {Mukherjee}}, \bibinfo {author} {\bibfnamefont {D.~N.}\ \bibnamefont {Spergel}}, \bibinfo {author} {\bibfnamefont {R.}~\bibnamefont {Khatri}}, \ and\ \bibinfo {author} {\bibfnamefont {B.~D.}\ \bibnamefont {Wandelt}},\ }\href {\doibase 10.1088/1475-7516/2020/02/032} {\bibfield  {journal} {\bibinfo  {journal} {JCAP}\ }\textbf {\bibinfo {volume} {02}},\ \bibinfo {pages} {032} (\bibinfo {year} {2020})},\ \Eprint {http://arxiv.org/abs/1908.07534} {arXiv:1908.07534 [astro-ph.CO]} \BibitemShut {NoStop}%
\bibitem [{\citenamefont {Mondino}\ \emph {et~al.}(2024)\citenamefont {Mondino}, \citenamefont {Pîrvu}, \citenamefont {Huang},\ and\ \citenamefont {Johnson}}]{Mondino:2024rif}%
  \BibitemOpen
  \bibfield  {author} {\bibinfo {author} {\bibfnamefont {C.}~\bibnamefont {Mondino}}, \bibinfo {author} {\bibfnamefont {D.}~\bibnamefont {Pîrvu}}, \bibinfo {author} {\bibfnamefont {J.}~\bibnamefont {Huang}}, \ and\ \bibinfo {author} {\bibfnamefont {M.~C.}\ \bibnamefont {Johnson}},\ }\href@noop {} {\  (\bibinfo {year} {2024})},\ \Eprint {http://arxiv.org/abs/2405.08059} {arXiv:2405.08059 [hep-ph]} \BibitemShut {NoStop}%
\bibitem [{\citenamefont {Mehta}\ and\ \citenamefont {Mukherjee}(2024{\natexlab{a}})}]{Mehta:2024pdz}%
  \BibitemOpen
  \bibfield  {author} {\bibinfo {author} {\bibfnamefont {H.}~\bibnamefont {Mehta}}\ and\ \bibinfo {author} {\bibfnamefont {S.}~\bibnamefont {Mukherjee}},\ }\href {\doibase 10.1088/1475-7516/2024/07/084} {\bibfield  {journal} {\bibinfo  {journal} {JCAP}\ }\textbf {\bibinfo {volume} {07}},\ \bibinfo {pages} {084} (\bibinfo {year} {2024}{\natexlab{a}})},\ \Eprint {http://arxiv.org/abs/2405.08879} {arXiv:2405.08879 [astro-ph.CO]} \BibitemShut {NoStop}%
\bibitem [{\citenamefont {Mehta}\ and\ \citenamefont {Mukherjee}(2024{\natexlab{b}})}]{Mehta:2024wfo}%
  \BibitemOpen
  \bibfield  {author} {\bibinfo {author} {\bibfnamefont {H.}~\bibnamefont {Mehta}}\ and\ \bibinfo {author} {\bibfnamefont {S.}~\bibnamefont {Mukherjee}},\ }\href@noop {} {\  (\bibinfo {year} {2024}{\natexlab{b}})},\ \Eprint {http://arxiv.org/abs/2405.08878} {arXiv:2405.08878 [astro-ph.CO]} \BibitemShut {NoStop}%
\bibitem [{\citenamefont {Aghanim}\ \emph {et~al.}(2020)\citenamefont {Aghanim} \emph {et~al.}}]{Aghanim:2018eyx}%
  \BibitemOpen
  \bibfield  {author} {\bibinfo {author} {\bibfnamefont {N.}~\bibnamefont {Aghanim}} \emph {et~al.} (\bibinfo {collaboration} {Planck}),\ }\href {\doibase 10.1051/0004-6361/201833910} {\bibfield  {journal} {\bibinfo  {journal} {Astron. Astrophys.}\ }\textbf {\bibinfo {volume} {641}},\ \bibinfo {pages} {A6} (\bibinfo {year} {2020})},\ \bibinfo {note} {[Erratum: Astron.Astrophys. 652, C4 (2021)]},\ \Eprint {http://arxiv.org/abs/1807.06209} {arXiv:1807.06209 [astro-ph.CO]} \BibitemShut {NoStop}%
\bibitem [{\citenamefont {Parke}(1986)}]{Parke:1986}%
  \BibitemOpen
  \bibfield  {author} {\bibinfo {author} {\bibfnamefont {S.~J.}\ \bibnamefont {Parke}},\ }\href {\doibase 10.1103/PhysRevLett.57.1275} {\bibfield  {journal} {\bibinfo  {journal} {Phys. Rev. Lett.}\ }\textbf {\bibinfo {volume} {57}},\ \bibinfo {pages} {1275} (\bibinfo {year} {1986})}\BibitemShut {NoStop}%
\bibitem [{\citenamefont {Tashiro}\ \emph {et~al.}(2013)\citenamefont {Tashiro}, \citenamefont {Silk},\ and\ \citenamefont {Marsh}}]{Tashiro:2013yea}%
  \BibitemOpen
  \bibfield  {author} {\bibinfo {author} {\bibfnamefont {H.}~\bibnamefont {Tashiro}}, \bibinfo {author} {\bibfnamefont {J.}~\bibnamefont {Silk}}, \ and\ \bibinfo {author} {\bibfnamefont {D.~J.~E.}\ \bibnamefont {Marsh}},\ }\href {\doibase 10.1103/PhysRevD.88.125024} {\bibfield  {journal} {\bibinfo  {journal} {Phys. Rev. D}\ }\textbf {\bibinfo {volume} {88}},\ \bibinfo {pages} {125024} (\bibinfo {year} {2013})},\ \Eprint {http://arxiv.org/abs/1308.0314} {arXiv:1308.0314 [astro-ph.CO]} \BibitemShut {NoStop}%
\bibitem [{\citenamefont {Pîrvu}\ \emph {et~al.}(2024)\citenamefont {Pîrvu}, \citenamefont {Huang},\ and\ \citenamefont {Johnson}}]{Pirvu:2023lch}%
  \BibitemOpen
  \bibfield  {author} {\bibinfo {author} {\bibfnamefont {D.}~\bibnamefont {Pîrvu}}, \bibinfo {author} {\bibfnamefont {J.}~\bibnamefont {Huang}}, \ and\ \bibinfo {author} {\bibfnamefont {M.~C.}\ \bibnamefont {Johnson}},\ }\href {\doibase 10.1088/1475-7516/2024/01/019} {\bibfield  {journal} {\bibinfo  {journal} {\jcap}\ }\textbf {\bibinfo {volume} {2024}},\ \bibinfo {pages} {019} (\bibinfo {year} {2024})}\BibitemShut {NoStop}%
\bibitem [{\citenamefont {McCarthy}\ \emph {et~al.}(2024)\citenamefont {McCarthy}, \citenamefont {Pirvu}, \citenamefont {Hill}, \citenamefont {Huang}, \citenamefont {Johnson},\ and\ \citenamefont {Rogers}}]{McCarthy:2024ozh}%
  \BibitemOpen
  \bibfield  {author} {\bibinfo {author} {\bibfnamefont {F.}~\bibnamefont {McCarthy}}, \bibinfo {author} {\bibfnamefont {D.}~\bibnamefont {Pirvu}}, \bibinfo {author} {\bibfnamefont {J.~C.}\ \bibnamefont {Hill}}, \bibinfo {author} {\bibfnamefont {J.}~\bibnamefont {Huang}}, \bibinfo {author} {\bibfnamefont {M.~C.}\ \bibnamefont {Johnson}}, \ and\ \bibinfo {author} {\bibfnamefont {K.~K.}\ \bibnamefont {Rogers}},\ }\href@noop {} {\  (\bibinfo {year} {2024})},\ \Eprint {http://arxiv.org/abs/2406.02546} {arXiv:2406.02546 [hep-ph]} \BibitemShut {NoStop}%
\bibitem [{\citenamefont {Kusiak}\ \emph {et~al.}(2022)\citenamefont {Kusiak}, \citenamefont {Bolliet}, \citenamefont {Krolewski},\ and\ \citenamefont {Hill}}]{Kusiak:2022xkt}%
  \BibitemOpen
  \bibfield  {author} {\bibinfo {author} {\bibfnamefont {A.}~\bibnamefont {Kusiak}}, \bibinfo {author} {\bibfnamefont {B.}~\bibnamefont {Bolliet}}, \bibinfo {author} {\bibfnamefont {A.}~\bibnamefont {Krolewski}}, \ and\ \bibinfo {author} {\bibfnamefont {J.~C.}\ \bibnamefont {Hill}},\ }\href {\doibase 10.1103/PhysRevD.106.123517} {\bibfield  {journal} {\bibinfo  {journal} {Phys. Rev. D}\ }\textbf {\bibinfo {volume} {106}},\ \bibinfo {pages} {123517} (\bibinfo {year} {2022})},\ \Eprint {http://arxiv.org/abs/2203.12583} {arXiv:2203.12583 [astro-ph.CO]} \BibitemShut {NoStop}%
\bibitem [{\citenamefont {Battaglia}(2016)}]{Battaglia:2016xbi}%
  \BibitemOpen
  \bibfield  {author} {\bibinfo {author} {\bibfnamefont {N.}~\bibnamefont {Battaglia}},\ }\href {\doibase 10.1088/1475-7516/2016/08/058} {\bibfield  {journal} {\bibinfo  {journal} {JCAP}\ }\textbf {\bibinfo {volume} {08}},\ \bibinfo {pages} {058} (\bibinfo {year} {2016})},\ \Eprint {http://arxiv.org/abs/1607.02442} {arXiv:1607.02442 [astro-ph.CO]} \BibitemShut {NoStop}%
\bibitem [{\citenamefont {Nelson}\ \emph {et~al.}(2019)\citenamefont {Nelson} \emph {et~al.}}]{Nelson:2018uso}%
  \BibitemOpen
  \bibfield  {author} {\bibinfo {author} {\bibfnamefont {D.}~\bibnamefont {Nelson}} \emph {et~al.},\ }\href {\doibase 10.1186/s40668-019-0028-x} {\bibfield  {journal} {\bibinfo  {journal} {Comput. Astrophys. Cosmol.}\ }\textbf {\bibinfo {volume} {6}},\ \bibinfo {pages} {2} (\bibinfo {year} {2019})},\ \Eprint {http://arxiv.org/abs/1812.05609} {arXiv:1812.05609 [astro-ph.GA]} \BibitemShut {NoStop}%
\bibitem [{\citenamefont {Pillepich}\ \emph {et~al.}(2018)\citenamefont {Pillepich} \emph {et~al.}}]{Pillepich:2017fcc}%
  \BibitemOpen
  \bibfield  {author} {\bibinfo {author} {\bibfnamefont {A.}~\bibnamefont {Pillepich}} \emph {et~al.},\ }\href {\doibase 10.1093/mnras/stx3112} {\bibfield  {journal} {\bibinfo  {journal} {Mon. Not. Roy. Astron. Soc.}\ }\textbf {\bibinfo {volume} {475}},\ \bibinfo {pages} {648} (\bibinfo {year} {2018})},\ \Eprint {http://arxiv.org/abs/1707.03406} {arXiv:1707.03406 [astro-ph.GA]} \BibitemShut {NoStop}%
\bibitem [{\citenamefont {Springel}\ \emph {et~al.}(2018)\citenamefont {Springel} \emph {et~al.}}]{Springel:2017tpz}%
  \BibitemOpen
  \bibfield  {author} {\bibinfo {author} {\bibfnamefont {V.}~\bibnamefont {Springel}} \emph {et~al.},\ }\href {\doibase 10.1093/mnras/stx3304} {\bibfield  {journal} {\bibinfo  {journal} {Mon. Not. Roy. Astron. Soc.}\ }\textbf {\bibinfo {volume} {475}},\ \bibinfo {pages} {676} (\bibinfo {year} {2018})},\ \Eprint {http://arxiv.org/abs/1707.03397} {arXiv:1707.03397 [astro-ph.GA]} \BibitemShut {NoStop}%
\bibitem [{\citenamefont {Nelson}\ \emph {et~al.}(2018)\citenamefont {Nelson} \emph {et~al.}}]{Nelson:2017cxy}%
  \BibitemOpen
  \bibfield  {author} {\bibinfo {author} {\bibfnamefont {D.}~\bibnamefont {Nelson}} \emph {et~al.},\ }\href {\doibase 10.1093/mnras/stx3040} {\bibfield  {journal} {\bibinfo  {journal} {Mon. Not. Roy. Astron. Soc.}\ }\textbf {\bibinfo {volume} {475}},\ \bibinfo {pages} {624} (\bibinfo {year} {2018})},\ \Eprint {http://arxiv.org/abs/1707.03395} {arXiv:1707.03395 [astro-ph.GA]} \BibitemShut {NoStop}%
\bibitem [{\citenamefont {{Naiman}}\ \emph {et~al.}(2018)\citenamefont {{Naiman}}, \citenamefont {{Pillepich}}, \citenamefont {{Springel}}, \citenamefont {{Ramirez-Ruiz}}, \citenamefont {{Torrey}}, \citenamefont {{Vogelsberger}}, \citenamefont {{Pakmor}}, \citenamefont {{Nelson}}, \citenamefont {{Marinacci}}, \citenamefont {{Hernquist}}, \citenamefont {{Weinberger}},\ and\ \citenamefont {{Genel}}}]{2018MNRAS.477.1206N}%
  \BibitemOpen
  \bibfield  {author} {\bibinfo {author} {\bibfnamefont {J.~P.}\ \bibnamefont {{Naiman}}}, \bibinfo {author} {\bibfnamefont {A.}~\bibnamefont {{Pillepich}}}, \bibinfo {author} {\bibfnamefont {V.}~\bibnamefont {{Springel}}}, \bibinfo {author} {\bibfnamefont {E.}~\bibnamefont {{Ramirez-Ruiz}}}, \bibinfo {author} {\bibfnamefont {P.}~\bibnamefont {{Torrey}}}, \bibinfo {author} {\bibfnamefont {M.}~\bibnamefont {{Vogelsberger}}}, \bibinfo {author} {\bibfnamefont {R.}~\bibnamefont {{Pakmor}}}, \bibinfo {author} {\bibfnamefont {D.}~\bibnamefont {{Nelson}}}, \bibinfo {author} {\bibfnamefont {F.}~\bibnamefont {{Marinacci}}}, \bibinfo {author} {\bibfnamefont {L.}~\bibnamefont {{Hernquist}}}, \bibinfo {author} {\bibfnamefont {R.}~\bibnamefont {{Weinberger}}}, \ and\ \bibinfo {author} {\bibfnamefont {S.}~\bibnamefont {{Genel}}},\ }\href {\doibase 10.1093/mnras/sty618} {\bibfield  {journal} {\bibinfo  {journal} {\mnras}\ }\textbf {\bibinfo {volume} {477}},\ \bibinfo {pages} {1206} (\bibinfo {year} {2018})},\ \Eprint
  {http://arxiv.org/abs/1707.03401} {arXiv:1707.03401 [astro-ph.GA]} \BibitemShut {NoStop}%
\bibitem [{\citenamefont {Marinacci}\ \emph {et~al.}(2018)\citenamefont {Marinacci} \emph {et~al.}}]{Marinacci:2017wew}%
  \BibitemOpen
  \bibfield  {author} {\bibinfo {author} {\bibfnamefont {F.}~\bibnamefont {Marinacci}} \emph {et~al.},\ }\href {\doibase 10.1093/mnras/sty2206} {\bibfield  {journal} {\bibinfo  {journal} {Mon. Not. Roy. Astron. Soc.}\ }\textbf {\bibinfo {volume} {480}},\ \bibinfo {pages} {5113} (\bibinfo {year} {2018})},\ \Eprint {http://arxiv.org/abs/1707.03396} {arXiv:1707.03396 [astro-ph.CO]} \BibitemShut {NoStop}%
\bibitem [{\citenamefont {Delabrouille}\ \emph {et~al.}(2009)\citenamefont {Delabrouille}, \citenamefont {Cardoso}, \citenamefont {Jeune}, \citenamefont {Betoule}, \citenamefont {Fay},\ and\ \citenamefont {Guilloux}}]{Delabrouille:2008qd}%
  \BibitemOpen
  \bibfield  {author} {\bibinfo {author} {\bibfnamefont {J.}~\bibnamefont {Delabrouille}}, \bibinfo {author} {\bibfnamefont {J.~F.}\ \bibnamefont {Cardoso}}, \bibinfo {author} {\bibfnamefont {M.~L.}\ \bibnamefont {Jeune}}, \bibinfo {author} {\bibfnamefont {M.}~\bibnamefont {Betoule}}, \bibinfo {author} {\bibfnamefont {G.}~\bibnamefont {Fay}}, \ and\ \bibinfo {author} {\bibfnamefont {F.}~\bibnamefont {Guilloux}},\ }\href {\doibase 10.1051/0004-6361:200810514} {\bibfield  {journal} {\bibinfo  {journal} {Astron. Astrophys.}\ }\textbf {\bibinfo {volume} {493}},\ \bibinfo {pages} {835} (\bibinfo {year} {2009})},\ \Eprint {http://arxiv.org/abs/0807.0773} {arXiv:0807.0773 [astro-ph]} \BibitemShut {NoStop}%
\bibitem [{\citenamefont {Remazeilles}\ \emph {et~al.}(2011)\citenamefont {Remazeilles}, \citenamefont {Delabrouille},\ and\ \citenamefont {Cardoso}}]{Remazeilles:2010hq}%
  \BibitemOpen
  \bibfield  {author} {\bibinfo {author} {\bibfnamefont {M.}~\bibnamefont {Remazeilles}}, \bibinfo {author} {\bibfnamefont {J.}~\bibnamefont {Delabrouille}}, \ and\ \bibinfo {author} {\bibfnamefont {J.-F.}\ \bibnamefont {Cardoso}},\ }\href {\doibase 10.1111/j.1365-2966.2010.17624.x} {\bibfield  {journal} {\bibinfo  {journal} {Mon. Not. Roy. Astron. Soc.}\ }\textbf {\bibinfo {volume} {410}},\ \bibinfo {pages} {2481} (\bibinfo {year} {2011})},\ \Eprint {http://arxiv.org/abs/1006.5599} {arXiv:1006.5599 [astro-ph.CO]} \BibitemShut {NoStop}%
\bibitem [{\citenamefont {McCarthy}\ and\ \citenamefont {Hill}(2024{\natexlab{a}})}]{McCarthy:2023hpa}%
  \BibitemOpen
  \bibfield  {author} {\bibinfo {author} {\bibfnamefont {F.}~\bibnamefont {McCarthy}}\ and\ \bibinfo {author} {\bibfnamefont {J.~C.}\ \bibnamefont {Hill}},\ }\href {\doibase 10.1103/PhysRevD.109.023528} {\bibfield  {journal} {\bibinfo  {journal} {Phys. Rev. D}\ }\textbf {\bibinfo {volume} {109}},\ \bibinfo {pages} {023528} (\bibinfo {year} {2024}{\natexlab{a}})},\ \Eprint {http://arxiv.org/abs/2307.01043} {arXiv:2307.01043 [astro-ph.CO]} \BibitemShut {NoStop}%
\bibitem [{\citenamefont {{Bennett}}\ \emph {et~al.}(1992)\citenamefont {{Bennett}}, \citenamefont {{Smoot}}, \citenamefont {{Hinshaw}}, \citenamefont {{Wright}}, \citenamefont {{Kogut}}, \citenamefont {{de Amici}}, \citenamefont {{Meyer}}, \citenamefont {{Weiss}}, \citenamefont {{Wilkinson}}, \citenamefont {{Gulkis}}, \citenamefont {{Janssen}}, \citenamefont {{Boggess}}, \citenamefont {{Cheng}}, \citenamefont {{Hauser}}, \citenamefont {{Kelsall}}, \citenamefont {{Mather}}, \citenamefont {{Moseley}}, \citenamefont {{Murdock}},\ and\ \citenamefont {{Silverberg}}}]{1992ApJ...396L...7B}%
  \BibitemOpen
  \bibfield  {author} {\bibinfo {author} {\bibfnamefont {C.~L.}\ \bibnamefont {{Bennett}}}, \bibinfo {author} {\bibfnamefont {G.~F.}\ \bibnamefont {{Smoot}}}, \bibinfo {author} {\bibfnamefont {G.}~\bibnamefont {{Hinshaw}}}, \bibinfo {author} {\bibfnamefont {E.~L.}\ \bibnamefont {{Wright}}}, \bibinfo {author} {\bibfnamefont {A.}~\bibnamefont {{Kogut}}}, \bibinfo {author} {\bibfnamefont {G.}~\bibnamefont {{de Amici}}}, \bibinfo {author} {\bibfnamefont {S.~S.}\ \bibnamefont {{Meyer}}}, \bibinfo {author} {\bibfnamefont {R.}~\bibnamefont {{Weiss}}}, \bibinfo {author} {\bibfnamefont {D.~T.}\ \bibnamefont {{Wilkinson}}}, \bibinfo {author} {\bibfnamefont {S.}~\bibnamefont {{Gulkis}}}, \bibinfo {author} {\bibfnamefont {M.}~\bibnamefont {{Janssen}}}, \bibinfo {author} {\bibfnamefont {N.~W.}\ \bibnamefont {{Boggess}}}, \bibinfo {author} {\bibfnamefont {E.~S.}\ \bibnamefont {{Cheng}}}, \bibinfo {author} {\bibfnamefont {M.~G.}\ \bibnamefont {{Hauser}}}, \bibinfo {author} {\bibfnamefont {T.}~\bibnamefont {{Kelsall}}},
  \bibinfo {author} {\bibfnamefont {J.~C.}\ \bibnamefont {{Mather}}}, \bibinfo {author} {\bibfnamefont {J.}~\bibnamefont {{Moseley}}, \bibfnamefont {S.~H.}}, \bibinfo {author} {\bibfnamefont {T.~L.}\ \bibnamefont {{Murdock}}}, \ and\ \bibinfo {author} {\bibfnamefont {R.~F.}\ \bibnamefont {{Silverberg}}},\ }\href {\doibase 10.1086/186505} {\bibfield  {journal} {\bibinfo  {journal} {\apjl}\ }\textbf {\bibinfo {volume} {396}},\ \bibinfo {pages} {L7} (\bibinfo {year} {1992})}\BibitemShut {NoStop}%
\bibitem [{\citenamefont {Bennett}\ \emph {et~al.}(2003)\citenamefont {Bennett} \emph {et~al.}}]{WMAP:2003cmr}%
  \BibitemOpen
  \bibfield  {author} {\bibinfo {author} {\bibfnamefont {C.}~\bibnamefont {Bennett}} \emph {et~al.} (\bibinfo {collaboration} {WMAP}),\ }\href {\doibase 10.1086/377252} {\bibfield  {journal} {\bibinfo  {journal} {Astrophys. J. Suppl.}\ }\textbf {\bibinfo {volume} {148}},\ \bibinfo {pages} {97} (\bibinfo {year} {2003})},\ \Eprint {http://arxiv.org/abs/astro-ph/0302208} {arXiv:astro-ph/0302208} \BibitemShut {NoStop}%
\bibitem [{\citenamefont {Tegmark}\ \emph {et~al.}(2003)\citenamefont {Tegmark}, \citenamefont {de~Oliveira-Costa},\ and\ \citenamefont {Hamilton}}]{Tegmark:2003ve}%
  \BibitemOpen
  \bibfield  {author} {\bibinfo {author} {\bibfnamefont {M.}~\bibnamefont {Tegmark}}, \bibinfo {author} {\bibfnamefont {A.}~\bibnamefont {de~Oliveira-Costa}}, \ and\ \bibinfo {author} {\bibfnamefont {A.}~\bibnamefont {Hamilton}},\ }\href {\doibase 10.1103/PhysRevD.68.123523} {\bibfield  {journal} {\bibinfo  {journal} {Phys. Rev. D}\ }\textbf {\bibinfo {volume} {68}},\ \bibinfo {pages} {123523} (\bibinfo {year} {2003})},\ \Eprint {http://arxiv.org/abs/astro-ph/0302496} {arXiv:astro-ph/0302496} \BibitemShut {NoStop}%
\bibitem [{\citenamefont {Eriksen}\ \emph {et~al.}(2004)\citenamefont {Eriksen}, \citenamefont {Banday}, \citenamefont {Gorski},\ and\ \citenamefont {Lilje}}]{Eriksen:2004jg}%
  \BibitemOpen
  \bibfield  {author} {\bibinfo {author} {\bibfnamefont {H.~K.}\ \bibnamefont {Eriksen}}, \bibinfo {author} {\bibfnamefont {A.~J.}\ \bibnamefont {Banday}}, \bibinfo {author} {\bibfnamefont {K.~M.}\ \bibnamefont {Gorski}}, \ and\ \bibinfo {author} {\bibfnamefont {P.~B.}\ \bibnamefont {Lilje}},\ }\href {\doibase 10.1086/422807} {\bibfield  {journal} {\bibinfo  {journal} {Astrophys. J.}\ }\textbf {\bibinfo {volume} {612}},\ \bibinfo {pages} {633} (\bibinfo {year} {2004})},\ \Eprint {http://arxiv.org/abs/astro-ph/0403098} {arXiv:astro-ph/0403098} \BibitemShut {NoStop}%
\bibitem [{\citenamefont {Narcowich}\ \emph {et~al.}(2006)\citenamefont {Narcowich}, \citenamefont {Petrushev},\ and\ \citenamefont {Ward}}]{doi:10.1137/040614359}%
  \BibitemOpen
  \bibfield  {author} {\bibinfo {author} {\bibfnamefont {F.~J.}\ \bibnamefont {Narcowich}}, \bibinfo {author} {\bibfnamefont {P.}~\bibnamefont {Petrushev}}, \ and\ \bibinfo {author} {\bibfnamefont {J.~D.}\ \bibnamefont {Ward}},\ }\href {\doibase 10.1137/040614359} {\bibfield  {journal} {\bibinfo  {journal} {SIAM Journal on Mathematical Analysis}\ }\textbf {\bibinfo {volume} {38}},\ \bibinfo {pages} {574} (\bibinfo {year} {2006})},\ \Eprint {http://arxiv.org/abs/https://doi.org/10.1137/040614359} {https://doi.org/10.1137/040614359} \BibitemShut {NoStop}%
\bibitem [{\citenamefont {Chen}\ and\ \citenamefont {Wright}(2009)}]{Chen:2008gw}%
  \BibitemOpen
  \bibfield  {author} {\bibinfo {author} {\bibfnamefont {X.}~\bibnamefont {Chen}}\ and\ \bibinfo {author} {\bibfnamefont {E.~L.}\ \bibnamefont {Wright}},\ }\href {\doibase 10.1088/0004-637X/694/1/222} {\bibfield  {journal} {\bibinfo  {journal} {Astrophys. J.}\ }\textbf {\bibinfo {volume} {694}},\ \bibinfo {pages} {222} (\bibinfo {year} {2009})},\ \Eprint {http://arxiv.org/abs/0809.4025} {arXiv:0809.4025 [astro-ph]} \BibitemShut {NoStop}%
\bibitem [{\citenamefont {Akrami}\ \emph {et~al.}(2020)\citenamefont {Akrami} \emph {et~al.}}]{Planck:2020olo}%
  \BibitemOpen
  \bibfield  {author} {\bibinfo {author} {\bibfnamefont {Y.}~\bibnamefont {Akrami}} \emph {et~al.} (\bibinfo {collaboration} {Planck}),\ }\href {\doibase 10.1051/0004-6361/202038073} {\bibfield  {journal} {\bibinfo  {journal} {Astron. Astrophys.}\ }\textbf {\bibinfo {volume} {643}},\ \bibinfo {pages} {A42} (\bibinfo {year} {2020})},\ \Eprint {http://arxiv.org/abs/2007.04997} {arXiv:2007.04997 [astro-ph.CO]} \BibitemShut {NoStop}%
\bibitem [{\citenamefont {McCarthy}\ and\ \citenamefont {Hill}(2024{\natexlab{b}})}]{McCarthy:2023cwg}%
  \BibitemOpen
  \bibfield  {author} {\bibinfo {author} {\bibfnamefont {F.}~\bibnamefont {McCarthy}}\ and\ \bibinfo {author} {\bibfnamefont {J.~C.}\ \bibnamefont {Hill}},\ }\href {\doibase 10.1103/PhysRevD.109.023529} {\bibfield  {journal} {\bibinfo  {journal} {Phys. Rev. D}\ }\textbf {\bibinfo {volume} {109}},\ \bibinfo {pages} {023529} (\bibinfo {year} {2024}{\natexlab{b}})},\ \Eprint {http://arxiv.org/abs/2308.16260} {arXiv:2308.16260 [astro-ph.CO]} \BibitemShut {NoStop}%
\bibitem [{\citenamefont {Sunyaev}\ and\ \citenamefont {Zeldovich}(1972)}]{Sunyaev:1972eq}%
  \BibitemOpen
  \bibfield  {author} {\bibinfo {author} {\bibfnamefont {R.~A.}\ \bibnamefont {Sunyaev}}\ and\ \bibinfo {author} {\bibfnamefont {Y.~B.}\ \bibnamefont {Zeldovich}},\ }\href@noop {} {\bibfield  {journal} {\bibinfo  {journal} {Comments Astrophys. Space Phys.}\ }\textbf {\bibinfo {volume} {4}},\ \bibinfo {pages} {173} (\bibinfo {year} {1972})}\BibitemShut {NoStop}%
\bibitem [{\citenamefont {Sunyaev}\ and\ \citenamefont {Zeldovich}(1980)}]{Sunyaev:1980nv}%
  \BibitemOpen
  \bibfield  {author} {\bibinfo {author} {\bibfnamefont {R.~A.}\ \bibnamefont {Sunyaev}}\ and\ \bibinfo {author} {\bibfnamefont {Y.~B.}\ \bibnamefont {Zeldovich}},\ }\href@noop {} {\bibfield  {journal} {\bibinfo  {journal} {\mnras}\ }\textbf {\bibinfo {volume} {190}},\ \bibinfo {pages} {413} (\bibinfo {year} {1980})}\BibitemShut {NoStop}%
\bibitem [{\citenamefont {Sachs}\ and\ \citenamefont {Wolfe}(1967)}]{Sachs:1967er}%
  \BibitemOpen
  \bibfield  {author} {\bibinfo {author} {\bibfnamefont {R.~K.}\ \bibnamefont {Sachs}}\ and\ \bibinfo {author} {\bibfnamefont {A.~M.}\ \bibnamefont {Wolfe}},\ }\href {\doibase 10.1007/s10714-007-0448-9} {\bibfield  {journal} {\bibinfo  {journal} {Astrophys. J.}\ }\textbf {\bibinfo {volume} {147}},\ \bibinfo {pages} {73} (\bibinfo {year} {1967})}\BibitemShut {NoStop}%
\bibitem [{\citenamefont {Krolewski}\ and\ \citenamefont {Ferraro}(2022)}]{Krolewski:2021znk}%
  \BibitemOpen
  \bibfield  {author} {\bibinfo {author} {\bibfnamefont {A.}~\bibnamefont {Krolewski}}\ and\ \bibinfo {author} {\bibfnamefont {S.}~\bibnamefont {Ferraro}},\ }\href {\doibase 10.1088/1475-7516/2022/04/033} {\bibfield  {journal} {\bibinfo  {journal} {JCAP}\ }\textbf {\bibinfo {volume} {04}},\ \bibinfo {pages} {033} (\bibinfo {year} {2022})},\ \Eprint {http://arxiv.org/abs/2110.13959} {arXiv:2110.13959 [astro-ph.CO]} \BibitemShut {NoStop}%
\bibitem [{\citenamefont {Chluba}\ \emph {et~al.}(2017)\citenamefont {Chluba}, \citenamefont {Hill},\ and\ \citenamefont {Abitbol}}]{Chluba:2017rtj}%
  \BibitemOpen
  \bibfield  {author} {\bibinfo {author} {\bibfnamefont {J.}~\bibnamefont {Chluba}}, \bibinfo {author} {\bibfnamefont {J.~C.}\ \bibnamefont {Hill}}, \ and\ \bibinfo {author} {\bibfnamefont {M.~H.}\ \bibnamefont {Abitbol}},\ }\href {\doibase 10.1093/mnras/stx1982} {\bibfield  {journal} {\bibinfo  {journal} {Mon. Not. Roy. Astron. Soc.}\ }\textbf {\bibinfo {volume} {472}},\ \bibinfo {pages} {1195} (\bibinfo {year} {2017})},\ \Eprint {http://arxiv.org/abs/1701.00274} {arXiv:1701.00274 [astro-ph.CO]} \BibitemShut {NoStop}%
\bibitem [{\citenamefont {Lang}(2014)}]{Lang_2014}%
  \BibitemOpen
  \bibfield  {author} {\bibinfo {author} {\bibfnamefont {D.}~\bibnamefont {Lang}},\ }\href {\doibase 10.1088/0004-6256/147/5/108} {\bibfield  {journal} {\bibinfo  {journal} {\aj}\ }\textbf {\bibinfo {volume} {147}},\ \bibinfo {pages} {108} (\bibinfo {year} {2014})}\BibitemShut {NoStop}%
\bibitem [{\citenamefont {Meisner}\ \emph {et~al.}(2017{\natexlab{a}})\citenamefont {Meisner}, \citenamefont {Lang},\ and\ \citenamefont {Schlegel}}]{Meisner_2017}%
  \BibitemOpen
  \bibfield  {author} {\bibinfo {author} {\bibfnamefont {A.~M.}\ \bibnamefont {Meisner}}, \bibinfo {author} {\bibfnamefont {D.}~\bibnamefont {Lang}}, \ and\ \bibinfo {author} {\bibfnamefont {D.~J.}\ \bibnamefont {Schlegel}},\ }\href {\doibase 10.3847/1538-3881/153/1/38} {\bibfield  {journal} {\bibinfo  {journal} {\aj}\ }\textbf {\bibinfo {volume} {153}},\ \bibinfo {pages} {38} (\bibinfo {year} {2017}{\natexlab{a}})}\BibitemShut {NoStop}%
\bibitem [{\citenamefont {Meisner}\ \emph {et~al.}(2017{\natexlab{b}})\citenamefont {Meisner}, \citenamefont {Lang},\ and\ \citenamefont {Schlegel}}]{Meisner_2017a}%
  \BibitemOpen
  \bibfield  {author} {\bibinfo {author} {\bibfnamefont {A.~M.}\ \bibnamefont {Meisner}}, \bibinfo {author} {\bibfnamefont {D.}~\bibnamefont {Lang}}, \ and\ \bibinfo {author} {\bibfnamefont {D.~J.}\ \bibnamefont {Schlegel}},\ }\href {\doibase 10.3847/1538-3881/aa894e} {\bibfield  {journal} {\bibinfo  {journal} {\aj}\ }\textbf {\bibinfo {volume} {154}},\ \bibinfo {pages} {161} (\bibinfo {year} {2017}{\natexlab{b}})}\BibitemShut {NoStop}%
\bibitem [{\citenamefont {Schlafly}\ \emph {et~al.}(2019)\citenamefont {Schlafly}, \citenamefont {Meisner},\ and\ \citenamefont {Green}}]{Schlafly_2019}%
  \BibitemOpen
  \bibfield  {author} {\bibinfo {author} {\bibfnamefont {E.~F.}\ \bibnamefont {Schlafly}}, \bibinfo {author} {\bibfnamefont {A.~M.}\ \bibnamefont {Meisner}}, \ and\ \bibinfo {author} {\bibfnamefont {G.~M.}\ \bibnamefont {Green}},\ }\href {\doibase 10.3847/1538-4365/aafbea} {\bibfield  {journal} {\bibinfo  {journal} {The Astrophysical Journal Supplement Series}\ }\textbf {\bibinfo {volume} {240}},\ \bibinfo {pages} {30} (\bibinfo {year} {2019})}\BibitemShut {NoStop}%
\bibitem [{\citenamefont {Krolewski}\ \emph {et~al.}(2020)\citenamefont {Krolewski}, \citenamefont {Ferraro}, \citenamefont {Schlafly},\ and\ \citenamefont {White}}]{Krolewski:2019yrv}%
  \BibitemOpen
  \bibfield  {author} {\bibinfo {author} {\bibfnamefont {A.}~\bibnamefont {Krolewski}}, \bibinfo {author} {\bibfnamefont {S.}~\bibnamefont {Ferraro}}, \bibinfo {author} {\bibfnamefont {E.~F.}\ \bibnamefont {Schlafly}}, \ and\ \bibinfo {author} {\bibfnamefont {M.}~\bibnamefont {White}},\ }\href {\doibase 10.1088/1475-7516/2020/05/047} {\bibfield  {journal} {\bibinfo  {journal} {JCAP}\ }\textbf {\bibinfo {volume} {05}},\ \bibinfo {pages} {047} (\bibinfo {year} {2020})},\ \Eprint {http://arxiv.org/abs/1909.07412} {arXiv:1909.07412 [astro-ph.CO]} \BibitemShut {NoStop}%
\bibitem [{\citenamefont {{Wright}}\ \emph {et~al.}(2010)\citenamefont {{Wright}} \emph {et~al.}}]{2010AJ....140.1868W}%
  \BibitemOpen
  \bibfield  {author} {\bibinfo {author} {\bibfnamefont {E.~L.}\ \bibnamefont {{Wright}}} \emph {et~al.},\ }\href {\doibase 10.1088/0004-6256/140/6/1868} {\bibfield  {journal} {\bibinfo  {journal} {\aj}\ }\textbf {\bibinfo {volume} {140}},\ \bibinfo {pages} {1868} (\bibinfo {year} {2010})}\BibitemShut {NoStop}%
\bibitem [{\citenamefont {{Mainzer}}\ \emph {et~al.}(2011)\citenamefont {{Mainzer}} \emph {et~al.}}]{2011ApJ...731...53M}%
  \BibitemOpen
  \bibfield  {author} {\bibinfo {author} {\bibfnamefont {A.}~\bibnamefont {{Mainzer}}} \emph {et~al.},\ }\href {\doibase 10.1088/0004-637X/731/1/53} {\bibfield  {journal} {\bibinfo  {journal} {\apj}\ }\textbf {\bibinfo {volume} {731}},\ \bibinfo {pages} {53} (\bibinfo {year} {2011})}\BibitemShut {NoStop}%
\bibitem [{\citenamefont {Maniyar}\ \emph {et~al.}(2018)\citenamefont {Maniyar}, \citenamefont {B\'ethermin},\ and\ \citenamefont {Lagache}}]{Maniyar:2018xfk}%
  \BibitemOpen
  \bibfield  {author} {\bibinfo {author} {\bibfnamefont {A.~S.}\ \bibnamefont {Maniyar}}, \bibinfo {author} {\bibfnamefont {M.}~\bibnamefont {B\'ethermin}}, \ and\ \bibinfo {author} {\bibfnamefont {G.}~\bibnamefont {Lagache}},\ }\href {\doibase 10.1051/0004-6361/201732499} {\bibfield  {journal} {\bibinfo  {journal} {Astron. Astrophys.}\ }\textbf {\bibinfo {volume} {614}},\ \bibinfo {pages} {A39} (\bibinfo {year} {2018})},\ \Eprint {http://arxiv.org/abs/1801.10146} {arXiv:1801.10146 [astro-ph.CO]} \BibitemShut {NoStop}%
\bibitem [{\citenamefont {Yan}\ \emph {et~al.}(2024)\citenamefont {Yan}, \citenamefont {Maniyar},\ and\ \citenamefont {van Waerbeke}}]{Yan:2023okq}%
  \BibitemOpen
  \bibfield  {author} {\bibinfo {author} {\bibfnamefont {Z.}~\bibnamefont {Yan}}, \bibinfo {author} {\bibfnamefont {A.~S.}\ \bibnamefont {Maniyar}}, \ and\ \bibinfo {author} {\bibfnamefont {L.}~\bibnamefont {van Waerbeke}},\ }\href {\doibase 10.1088/1475-7516/2024/05/058} {\bibfield  {journal} {\bibinfo  {journal} {JCAP}\ }\textbf {\bibinfo {volume} {05}},\ \bibinfo {pages} {058} (\bibinfo {year} {2024})},\ \Eprint {http://arxiv.org/abs/2310.10848} {arXiv:2310.10848 [astro-ph.CO]} \BibitemShut {NoStop}%
\bibitem [{\citenamefont {Hivon}\ \emph {et~al.}(2002)\citenamefont {Hivon}, \citenamefont {Gorski}, \citenamefont {Netterfield}, \citenamefont {Crill}, \citenamefont {Prunet},\ and\ \citenamefont {Hansen}}]{Hivon:2001jp}%
  \BibitemOpen
  \bibfield  {author} {\bibinfo {author} {\bibfnamefont {E.}~\bibnamefont {Hivon}}, \bibinfo {author} {\bibfnamefont {K.~M.}\ \bibnamefont {Gorski}}, \bibinfo {author} {\bibfnamefont {C.~B.}\ \bibnamefont {Netterfield}}, \bibinfo {author} {\bibfnamefont {B.~P.}\ \bibnamefont {Crill}}, \bibinfo {author} {\bibfnamefont {S.}~\bibnamefont {Prunet}}, \ and\ \bibinfo {author} {\bibfnamefont {F.}~\bibnamefont {Hansen}},\ }\href {\doibase 10.1086/338126} {\bibfield  {journal} {\bibinfo  {journal} {Astrophys. J.}\ }\textbf {\bibinfo {volume} {567}},\ \bibinfo {pages} {2} (\bibinfo {year} {2002})},\ \Eprint {http://arxiv.org/abs/astro-ph/0105302} {arXiv:astro-ph/0105302} \BibitemShut {NoStop}%
\bibitem [{\citenamefont {Alonso}\ \emph {et~al.}(2019)\citenamefont {Alonso}, \citenamefont {Sanchez},\ and\ \citenamefont {Slosar}}]{Alonso:2018jzx}%
  \BibitemOpen
  \bibfield  {author} {\bibinfo {author} {\bibfnamefont {D.}~\bibnamefont {Alonso}}, \bibinfo {author} {\bibfnamefont {J.}~\bibnamefont {Sanchez}}, \ and\ \bibinfo {author} {\bibfnamefont {A.}~\bibnamefont {Slosar}} (\bibinfo {collaboration} {LSST Dark Energy Science}),\ }\href {\doibase 10.1093/mnras/stz093} {\bibfield  {journal} {\bibinfo  {journal} {Mon. Not. Roy. Astron. Soc.}\ }\textbf {\bibinfo {volume} {484}},\ \bibinfo {pages} {4127} (\bibinfo {year} {2019})},\ \Eprint {http://arxiv.org/abs/1809.09603} {arXiv:1809.09603 [astro-ph.CO]} \BibitemShut {NoStop}%
\bibitem [{\citenamefont {Nicola}\ \emph {et~al.}(2021)\citenamefont {Nicola}, \citenamefont {Garcia-Garcia}, \citenamefont {Alonso}, \citenamefont {Dunkley}, \citenamefont {Ferreira}, \citenamefont {Slosar},\ and\ \citenamefont {Spergel}}]{Nicola:2020lhi}%
  \BibitemOpen
  \bibfield  {author} {\bibinfo {author} {\bibfnamefont {A.}~\bibnamefont {Nicola}}, \bibinfo {author} {\bibfnamefont {C.}~\bibnamefont {Garcia-Garcia}}, \bibinfo {author} {\bibfnamefont {D.}~\bibnamefont {Alonso}}, \bibinfo {author} {\bibfnamefont {J.}~\bibnamefont {Dunkley}}, \bibinfo {author} {\bibfnamefont {P.~G.}\ \bibnamefont {Ferreira}}, \bibinfo {author} {\bibfnamefont {A.}~\bibnamefont {Slosar}}, \ and\ \bibinfo {author} {\bibfnamefont {D.~N.}\ \bibnamefont {Spergel}},\ }\href {\doibase 10.1088/1475-7516/2021/03/067} {\bibfield  {journal} {\bibinfo  {journal} {JCAP}\ }\textbf {\bibinfo {volume} {03}},\ \bibinfo {pages} {067} (\bibinfo {year} {2021})},\ \Eprint {http://arxiv.org/abs/2010.09717} {arXiv:2010.09717 [astro-ph.CO]} \BibitemShut {NoStop}%
\bibitem [{\citenamefont {Anastassopoulos}\ \emph {et~al.}(2017)\citenamefont {Anastassopoulos} \emph {et~al.}}]{CAST:2017uph}%
  \BibitemOpen
  \bibfield  {author} {\bibinfo {author} {\bibfnamefont {V.}~\bibnamefont {Anastassopoulos}} \emph {et~al.} (\bibinfo {collaboration} {CAST}),\ }\href {\doibase 10.1038/nphys4109} {\bibfield  {journal} {\bibinfo  {journal} {Nature Phys.}\ }\textbf {\bibinfo {volume} {13}},\ \bibinfo {pages} {584} (\bibinfo {year} {2017})},\ \Eprint {http://arxiv.org/abs/1705.02290} {arXiv:1705.02290 [hep-ex]} \BibitemShut {NoStop}%
\bibitem [{\citenamefont {O'Hare}(2020)}]{AxionLimits}%
  \BibitemOpen
  \bibfield  {author} {\bibinfo {author} {\bibfnamefont {C.}~\bibnamefont {O'Hare}},\ }\href {\doibase 10.5281/zenodo.3932430} {\enquote {\bibinfo {title} {cajohare/axionlimits: Axionlimits},}\ }\bibinfo {howpublished} {\url{https://cajohare.github.io/AxionLimits/}} (\bibinfo {year} {2020})\BibitemShut {NoStop}%
\bibitem [{\citenamefont {Marsh}\ \emph {et~al.}(2017)\citenamefont {Marsh}, \citenamefont {Russell}, \citenamefont {Fabian}, \citenamefont {McNamara}, \citenamefont {Nulsen},\ and\ \citenamefont {Reynolds}}]{Marsh:2017yvc}%
  \BibitemOpen
  \bibfield  {author} {\bibinfo {author} {\bibfnamefont {M.~C.~D.}\ \bibnamefont {Marsh}}, \bibinfo {author} {\bibfnamefont {H.~R.}\ \bibnamefont {Russell}}, \bibinfo {author} {\bibfnamefont {A.~C.}\ \bibnamefont {Fabian}}, \bibinfo {author} {\bibfnamefont {B.~P.}\ \bibnamefont {McNamara}}, \bibinfo {author} {\bibfnamefont {P.}~\bibnamefont {Nulsen}}, \ and\ \bibinfo {author} {\bibfnamefont {C.~S.}\ \bibnamefont {Reynolds}},\ }\href {\doibase 10.1088/1475-7516/2017/12/036} {\bibfield  {journal} {\bibinfo  {journal} {JCAP}\ }\textbf {\bibinfo {volume} {12}},\ \bibinfo {pages} {036} (\bibinfo {year} {2017})},\ \Eprint {http://arxiv.org/abs/1703.07354} {arXiv:1703.07354 [hep-ph]} \BibitemShut {NoStop}%
\bibitem [{\citenamefont {Hoof}\ and\ \citenamefont {Schulz}(2023)}]{Hoof:2022xbe}%
  \BibitemOpen
  \bibfield  {author} {\bibinfo {author} {\bibfnamefont {S.}~\bibnamefont {Hoof}}\ and\ \bibinfo {author} {\bibfnamefont {L.}~\bibnamefont {Schulz}},\ }\href {\doibase 10.1088/1475-7516/2023/03/054} {\bibfield  {journal} {\bibinfo  {journal} {JCAP}\ }\textbf {\bibinfo {volume} {03}},\ \bibinfo {pages} {054} (\bibinfo {year} {2023})},\ \Eprint {http://arxiv.org/abs/2212.09764} {arXiv:2212.09764 [hep-ph]} \BibitemShut {NoStop}%
\bibitem [{\citenamefont {Ning}\ and\ \citenamefont {Safdi}(2024)}]{Ning:2024eky}%
  \BibitemOpen
  \bibfield  {author} {\bibinfo {author} {\bibfnamefont {O.}~\bibnamefont {Ning}}\ and\ \bibinfo {author} {\bibfnamefont {B.~R.}\ \bibnamefont {Safdi}},\ }\href@noop {} {\  (\bibinfo {year} {2024})},\ \Eprint {http://arxiv.org/abs/2404.14476} {arXiv:2404.14476 [hep-ph]} \BibitemShut {NoStop}%
\bibitem [{\citenamefont {Reynolds}\ \emph {et~al.}(2020)\citenamefont {Reynolds}, \citenamefont {Marsh}, \citenamefont {Russell}, \citenamefont {Fabian}, \citenamefont {Smith}, \citenamefont {Tombesi},\ and\ \citenamefont {Veilleux}}]{Reynolds:2019uqt}%
  \BibitemOpen
  \bibfield  {author} {\bibinfo {author} {\bibfnamefont {C.~S.}\ \bibnamefont {Reynolds}}, \bibinfo {author} {\bibfnamefont {M.~C.~D.}\ \bibnamefont {Marsh}}, \bibinfo {author} {\bibfnamefont {H.~R.}\ \bibnamefont {Russell}}, \bibinfo {author} {\bibfnamefont {A.~C.}\ \bibnamefont {Fabian}}, \bibinfo {author} {\bibfnamefont {R.}~\bibnamefont {Smith}}, \bibinfo {author} {\bibfnamefont {F.}~\bibnamefont {Tombesi}}, \ and\ \bibinfo {author} {\bibfnamefont {S.}~\bibnamefont {Veilleux}},\ }\href {\doibase 10.3847/1538-4357/ab6a0c} {\bibfield  {journal} {\bibinfo  {journal} {Astrophys. J.}\ }\textbf {\bibinfo {volume} {890}},\ \bibinfo {pages} {59} (\bibinfo {year} {2020})},\ \Eprint {http://arxiv.org/abs/1907.05475} {arXiv:1907.05475 [hep-ph]} \BibitemShut {NoStop}%
\bibitem [{\citenamefont {Reyn\'es}\ \emph {et~al.}(2021)\citenamefont {Reyn\'es}, \citenamefont {Matthews}, \citenamefont {Reynolds}, \citenamefont {Russell}, \citenamefont {Smith},\ and\ \citenamefont {Marsh}}]{Reynes:2021bpe}%
  \BibitemOpen
  \bibfield  {author} {\bibinfo {author} {\bibfnamefont {J.~S.}\ \bibnamefont {Reyn\'es}}, \bibinfo {author} {\bibfnamefont {J.~H.}\ \bibnamefont {Matthews}}, \bibinfo {author} {\bibfnamefont {C.~S.}\ \bibnamefont {Reynolds}}, \bibinfo {author} {\bibfnamefont {H.~R.}\ \bibnamefont {Russell}}, \bibinfo {author} {\bibfnamefont {R.~N.}\ \bibnamefont {Smith}}, \ and\ \bibinfo {author} {\bibfnamefont {M.~C.~D.}\ \bibnamefont {Marsh}},\ }\href {\doibase 10.1093/mnras/stab3464} {\bibfield  {journal} {\bibinfo  {journal} {Mon. Not. Roy. Astron. Soc.}\ }\textbf {\bibinfo {volume} {510}},\ \bibinfo {pages} {1264} (\bibinfo {year} {2021})},\ \Eprint {http://arxiv.org/abs/2109.03261} {arXiv:2109.03261 [astro-ph.HE]} \BibitemShut {NoStop}%
\bibitem [{\citenamefont {Arvanitaki}\ \emph {et~al.}(2015)\citenamefont {Arvanitaki}, \citenamefont {Baryakhtar},\ and\ \citenamefont {Huang}}]{Arvanitaki:2014wva}%
  \BibitemOpen
  \bibfield  {author} {\bibinfo {author} {\bibfnamefont {A.}~\bibnamefont {Arvanitaki}}, \bibinfo {author} {\bibfnamefont {M.}~\bibnamefont {Baryakhtar}}, \ and\ \bibinfo {author} {\bibfnamefont {X.}~\bibnamefont {Huang}},\ }\href {\doibase 10.1103/PhysRevD.91.084011} {\bibfield  {journal} {\bibinfo  {journal} {Phys. Rev. D}\ }\textbf {\bibinfo {volume} {91}},\ \bibinfo {pages} {084011} (\bibinfo {year} {2015})},\ \Eprint {http://arxiv.org/abs/1411.2263} {arXiv:1411.2263 [hep-ph]} \BibitemShut {NoStop}%
\bibitem [{\citenamefont {Cardoso}\ \emph {et~al.}(2018)\citenamefont {Cardoso}, \citenamefont {Dias}, \citenamefont {Hartnett}, \citenamefont {Middleton}, \citenamefont {Pani},\ and\ \citenamefont {Santos}}]{Cardoso:2018tly}%
  \BibitemOpen
  \bibfield  {author} {\bibinfo {author} {\bibfnamefont {V.}~\bibnamefont {Cardoso}}, \bibinfo {author} {\bibfnamefont {O.~J.~C.}\ \bibnamefont {Dias}}, \bibinfo {author} {\bibfnamefont {G.~S.}\ \bibnamefont {Hartnett}}, \bibinfo {author} {\bibfnamefont {M.}~\bibnamefont {Middleton}}, \bibinfo {author} {\bibfnamefont {P.}~\bibnamefont {Pani}}, \ and\ \bibinfo {author} {\bibfnamefont {J.~E.}\ \bibnamefont {Santos}},\ }\href {\doibase 10.1088/1475-7516/2018/03/043} {\bibfield  {journal} {\bibinfo  {journal} {JCAP}\ }\textbf {\bibinfo {volume} {03}},\ \bibinfo {pages} {043} (\bibinfo {year} {2018})},\ \Eprint {http://arxiv.org/abs/1801.01420} {arXiv:1801.01420 [gr-qc]} \BibitemShut {NoStop}%
\bibitem [{\citenamefont {Baryakhtar}\ \emph {et~al.}(2021)\citenamefont {Baryakhtar}, \citenamefont {Galanis}, \citenamefont {Lasenby},\ and\ \citenamefont {Simon}}]{Baryakhtar:2020gao}%
  \BibitemOpen
  \bibfield  {author} {\bibinfo {author} {\bibfnamefont {M.}~\bibnamefont {Baryakhtar}}, \bibinfo {author} {\bibfnamefont {M.}~\bibnamefont {Galanis}}, \bibinfo {author} {\bibfnamefont {R.}~\bibnamefont {Lasenby}}, \ and\ \bibinfo {author} {\bibfnamefont {O.}~\bibnamefont {Simon}},\ }\href {\doibase 10.1103/PhysRevD.103.095019} {\bibfield  {journal} {\bibinfo  {journal} {Phys. Rev. D}\ }\textbf {\bibinfo {volume} {103}},\ \bibinfo {pages} {095019} (\bibinfo {year} {2021})},\ \Eprint {http://arxiv.org/abs/2011.11646} {arXiv:2011.11646 [hep-ph]} \BibitemShut {NoStop}%
\bibitem [{\citenamefont {Amodeo}\ \emph {et~al.}(2021)\citenamefont {Amodeo} \emph {et~al.}}]{Amodeo:2020mmu}%
  \BibitemOpen
  \bibfield  {author} {\bibinfo {author} {\bibfnamefont {S.}~\bibnamefont {Amodeo}} \emph {et~al.},\ }\href {\doibase 10.1103/PhysRevD.103.063514} {\bibfield  {journal} {\bibinfo  {journal} {Phys. Rev. D}\ }\textbf {\bibinfo {volume} {103}},\ \bibinfo {pages} {063514} (\bibinfo {year} {2021})},\ \bibinfo {note} {[Erratum: Phys.Rev.D 107, 063514 (2023)]},\ \Eprint {http://arxiv.org/abs/2009.05558} {arXiv:2009.05558 [astro-ph.CO]} \BibitemShut {NoStop}%
\bibitem [{\citenamefont {Schaan}\ \emph {et~al.}(2021)\citenamefont {Schaan} \emph {et~al.}}]{AtacamaCosmologyTelescope:2020wtv}%
  \BibitemOpen
  \bibfield  {author} {\bibinfo {author} {\bibfnamefont {E.}~\bibnamefont {Schaan}} \emph {et~al.} (\bibinfo {collaboration} {Atacama Cosmology Telescope}),\ }\href {\doibase 10.1103/PhysRevD.103.063513} {\bibfield  {journal} {\bibinfo  {journal} {Phys. Rev. D}\ }\textbf {\bibinfo {volume} {103}},\ \bibinfo {pages} {063513} (\bibinfo {year} {2021})},\ \Eprint {http://arxiv.org/abs/2009.05557} {arXiv:2009.05557 [astro-ph.CO]} \BibitemShut {NoStop}%
\bibitem [{\citenamefont {Kusiak}\ \emph {et~al.}(2021)\citenamefont {Kusiak}, \citenamefont {Bolliet}, \citenamefont {Ferraro}, \citenamefont {Hill},\ and\ \citenamefont {Krolewski}}]{Kusiak:2021hai}%
  \BibitemOpen
  \bibfield  {author} {\bibinfo {author} {\bibfnamefont {A.}~\bibnamefont {Kusiak}}, \bibinfo {author} {\bibfnamefont {B.}~\bibnamefont {Bolliet}}, \bibinfo {author} {\bibfnamefont {S.}~\bibnamefont {Ferraro}}, \bibinfo {author} {\bibfnamefont {J.~C.}\ \bibnamefont {Hill}}, \ and\ \bibinfo {author} {\bibfnamefont {A.}~\bibnamefont {Krolewski}},\ }\href {\doibase 10.1103/PhysRevD.104.043518} {\bibfield  {journal} {\bibinfo  {journal} {Phys. Rev. D}\ }\textbf {\bibinfo {volume} {104}},\ \bibinfo {pages} {043518} (\bibinfo {year} {2021})},\ \Eprint {http://arxiv.org/abs/2102.01068} {arXiv:2102.01068 [astro-ph.CO]} \BibitemShut {NoStop}%
\bibitem [{\citenamefont {Madhavacheril}\ \emph {et~al.}(2019)\citenamefont {Madhavacheril}, \citenamefont {Battaglia}, \citenamefont {Smith},\ and\ \citenamefont {Sievers}}]{Madhavacheril:2019buy}%
  \BibitemOpen
  \bibfield  {author} {\bibinfo {author} {\bibfnamefont {M.~S.}\ \bibnamefont {Madhavacheril}}, \bibinfo {author} {\bibfnamefont {N.}~\bibnamefont {Battaglia}}, \bibinfo {author} {\bibfnamefont {K.~M.}\ \bibnamefont {Smith}}, \ and\ \bibinfo {author} {\bibfnamefont {J.~L.}\ \bibnamefont {Sievers}},\ }\href {\doibase 10.1103/PhysRevD.100.103532} {\bibfield  {journal} {\bibinfo  {journal} {Phys. Rev. D}\ }\textbf {\bibinfo {volume} {100}},\ \bibinfo {pages} {103532} (\bibinfo {year} {2019})},\ \Eprint {http://arxiv.org/abs/1901.02418} {arXiv:1901.02418 [astro-ph.CO]} \BibitemShut {NoStop}%
\bibitem [{\citenamefont {Prochaska}\ and\ \citenamefont {Zheng}(2019)}]{Prochaska_2019}%
  \BibitemOpen
  \bibfield  {author} {\bibinfo {author} {\bibfnamefont {J.~X.}\ \bibnamefont {Prochaska}}\ and\ \bibinfo {author} {\bibfnamefont {Y.}~\bibnamefont {Zheng}},\ }\href {\doibase 10.1093/mnras/stz261} {\bibfield  {journal} {\bibinfo  {journal} {Monthly Notices of the Royal Astronomical Society}\ } (\bibinfo {year} {2019}),\ 10.1093/mnras/stz261}\BibitemShut {NoStop}%
\bibitem [{\citenamefont {Heesen}\ \emph {et~al.}(2023)\citenamefont {Heesen}, \citenamefont {O’Sullivan}, \citenamefont {Brüggen}, \citenamefont {Basu}, \citenamefont {Beck}, \citenamefont {Seta}, \citenamefont {Carretti}, \citenamefont {Krause}, \citenamefont {Haverkorn}, \citenamefont {Hutschenreuter}, \citenamefont {Bracco}, \citenamefont {Stein}, \citenamefont {Bomans}, \citenamefont {Dettmar}, \citenamefont {Chyży}, \citenamefont {Heald}, \citenamefont {Paladino},\ and\ \citenamefont {Horellou}}]{Heesen_2023}%
  \BibitemOpen
  \bibfield  {author} {\bibinfo {author} {\bibfnamefont {V.}~\bibnamefont {Heesen}}, \bibinfo {author} {\bibfnamefont {S.~P.}\ \bibnamefont {O’Sullivan}}, \bibinfo {author} {\bibfnamefont {M.}~\bibnamefont {Brüggen}}, \bibinfo {author} {\bibfnamefont {A.}~\bibnamefont {Basu}}, \bibinfo {author} {\bibfnamefont {R.}~\bibnamefont {Beck}}, \bibinfo {author} {\bibfnamefont {A.}~\bibnamefont {Seta}}, \bibinfo {author} {\bibfnamefont {E.}~\bibnamefont {Carretti}}, \bibinfo {author} {\bibfnamefont {M.~G.~H.}\ \bibnamefont {Krause}}, \bibinfo {author} {\bibfnamefont {M.}~\bibnamefont {Haverkorn}}, \bibinfo {author} {\bibfnamefont {S.}~\bibnamefont {Hutschenreuter}}, \bibinfo {author} {\bibfnamefont {A.}~\bibnamefont {Bracco}}, \bibinfo {author} {\bibfnamefont {M.}~\bibnamefont {Stein}}, \bibinfo {author} {\bibfnamefont {D.~J.}\ \bibnamefont {Bomans}}, \bibinfo {author} {\bibfnamefont {R.-J.}\ \bibnamefont {Dettmar}}, \bibinfo {author} {\bibfnamefont {K.~T.}\ \bibnamefont {Chyży}}, \bibinfo {author} {\bibfnamefont
  {G.~H.}\ \bibnamefont {Heald}}, \bibinfo {author} {\bibfnamefont {R.}~\bibnamefont {Paladino}}, \ and\ \bibinfo {author} {\bibfnamefont {C.}~\bibnamefont {Horellou}},\ }\href {\doibase 10.1051/0004-6361/202346008} {\bibfield  {journal} {\bibinfo  {journal} {Astronomy \&; Astrophysics}\ }\textbf {\bibinfo {volume} {670}},\ \bibinfo {pages} {L23} (\bibinfo {year} {2023})}\BibitemShut {NoStop}%
\bibitem [{\citenamefont {Muralidhara}\ and\ \citenamefont {Basu}(2024)}]{Muralidhara:2024ipg}%
  \BibitemOpen
  \bibfield  {author} {\bibinfo {author} {\bibfnamefont {V.}~\bibnamefont {Muralidhara}}\ and\ \bibinfo {author} {\bibfnamefont {K.}~\bibnamefont {Basu}},\ }\href@noop {} {\  (\bibinfo {year} {2024})},\ \Eprint {http://arxiv.org/abs/2402.17445} {arXiv:2402.17445 [astro-ph.CO]} \BibitemShut {NoStop}%
\bibitem [{\citenamefont {Osinga}\ \emph {et~al.}(2024)\citenamefont {Osinga}, \citenamefont {van Weeren}, \citenamefont {Rudnick}, \citenamefont {Andrade-Santos}, \citenamefont {Bonafede}, \citenamefont {Clarke}, \citenamefont {Duncan}, \citenamefont {Giacintucci},\ and\ \citenamefont {R\"ottgering}}]{Osinga:2024vzo}%
  \BibitemOpen
  \bibfield  {author} {\bibinfo {author} {\bibfnamefont {E.}~\bibnamefont {Osinga}}, \bibinfo {author} {\bibfnamefont {R.~J.}\ \bibnamefont {van Weeren}}, \bibinfo {author} {\bibfnamefont {L.}~\bibnamefont {Rudnick}}, \bibinfo {author} {\bibfnamefont {F.}~\bibnamefont {Andrade-Santos}}, \bibinfo {author} {\bibfnamefont {A.}~\bibnamefont {Bonafede}}, \bibinfo {author} {\bibfnamefont {T.}~\bibnamefont {Clarke}}, \bibinfo {author} {\bibfnamefont {K.}~\bibnamefont {Duncan}}, \bibinfo {author} {\bibfnamefont {S.}~\bibnamefont {Giacintucci}}, \ and\ \bibinfo {author} {\bibfnamefont {H.~J.~A.}\ \bibnamefont {R\"ottgering}},\ }\href@noop {} {\  (\bibinfo {year} {2024})},\ \Eprint {http://arxiv.org/abs/2408.07178} {arXiv:2408.07178 [astro-ph.CO]} \BibitemShut {NoStop}%
\bibitem [{\citenamefont {{Pakmor}}\ \emph {et~al.}(2024)\citenamefont {{Pakmor}}, \citenamefont {{Bieri}}, \citenamefont {{van de Voort}}, \citenamefont {{Werhahn}}, \citenamefont {{Fattahi}}, \citenamefont {{Guillet}}, \citenamefont {{Pfrommer}}, \citenamefont {{Springel}},\ and\ \citenamefont {{Talbot}}}]{2024MNRAS.528.2308P}%
  \BibitemOpen
  \bibfield  {author} {\bibinfo {author} {\bibfnamefont {R.}~\bibnamefont {{Pakmor}}}, \bibinfo {author} {\bibfnamefont {R.}~\bibnamefont {{Bieri}}}, \bibinfo {author} {\bibfnamefont {F.}~\bibnamefont {{van de Voort}}}, \bibinfo {author} {\bibfnamefont {M.}~\bibnamefont {{Werhahn}}}, \bibinfo {author} {\bibfnamefont {A.}~\bibnamefont {{Fattahi}}}, \bibinfo {author} {\bibfnamefont {T.}~\bibnamefont {{Guillet}}}, \bibinfo {author} {\bibfnamefont {C.}~\bibnamefont {{Pfrommer}}}, \bibinfo {author} {\bibfnamefont {V.}~\bibnamefont {{Springel}}}, \ and\ \bibinfo {author} {\bibfnamefont {R.~Y.}\ \bibnamefont {{Talbot}}},\ }\href {\doibase 10.1093/mnras/stae112} {\bibfield  {journal} {\bibinfo  {journal} {\mnras}\ }\textbf {\bibinfo {volume} {528}},\ \bibinfo {pages} {2308} (\bibinfo {year} {2024})},\ \Eprint {http://arxiv.org/abs/2309.13104} {arXiv:2309.13104 [astro-ph.GA]} \BibitemShut {NoStop}%
\bibitem [{\citenamefont {Rees}\ and\ \citenamefont {Sciama}(1968)}]{Rees:1968zza}%
  \BibitemOpen
  \bibfield  {author} {\bibinfo {author} {\bibfnamefont {M.~J.}\ \bibnamefont {Rees}}\ and\ \bibinfo {author} {\bibfnamefont {D.~W.}\ \bibnamefont {Sciama}},\ }\href {\doibase 10.1038/217511a0} {\bibfield  {journal} {\bibinfo  {journal} {Nature}\ }\textbf {\bibinfo {volume} {217}},\ \bibinfo {pages} {511} (\bibinfo {year} {1968})}\BibitemShut {NoStop}%
\bibitem [{\citenamefont {Ferraro}\ \emph {et~al.}(2022)\citenamefont {Ferraro}, \citenamefont {Schaan},\ and\ \citenamefont {Pierpaoli}}]{Ferraro:2022twg}%
  \BibitemOpen
  \bibfield  {author} {\bibinfo {author} {\bibfnamefont {S.}~\bibnamefont {Ferraro}}, \bibinfo {author} {\bibfnamefont {E.}~\bibnamefont {Schaan}}, \ and\ \bibinfo {author} {\bibfnamefont {E.}~\bibnamefont {Pierpaoli}},\ }\href@noop {} {\  (\bibinfo {year} {2022})},\ \Eprint {http://arxiv.org/abs/2205.10332} {arXiv:2205.10332 [astro-ph.CO]} \BibitemShut {NoStop}%
\bibitem [{\citenamefont {Dore}\ \emph {et~al.}(2004)\citenamefont {Dore}, \citenamefont {Hennawi},\ and\ \citenamefont {Spergel}}]{Dore:2003ex}%
  \BibitemOpen
  \bibfield  {author} {\bibinfo {author} {\bibfnamefont {O.}~\bibnamefont {Dore}}, \bibinfo {author} {\bibfnamefont {J.~F.}\ \bibnamefont {Hennawi}}, \ and\ \bibinfo {author} {\bibfnamefont {D.~N.}\ \bibnamefont {Spergel}},\ }\href {\doibase 10.1086/382946} {\bibfield  {journal} {\bibinfo  {journal} {Astrophys. J.}\ }\textbf {\bibinfo {volume} {606}},\ \bibinfo {pages} {46} (\bibinfo {year} {2004})},\ \Eprint {http://arxiv.org/abs/astro-ph/0309337} {arXiv:astro-ph/0309337} \BibitemShut {NoStop}%
\bibitem [{\citenamefont {Ade}\ \emph {et~al.}(2016)\citenamefont {Ade} \emph {et~al.}}]{Planck:2015wtm}%
  \BibitemOpen
  \bibfield  {author} {\bibinfo {author} {\bibfnamefont {P.~A.~R.}\ \bibnamefont {Ade}} \emph {et~al.} (\bibinfo {collaboration} {Planck}),\ }\href {\doibase 10.1051/0004-6361/201525809} {\bibfield  {journal} {\bibinfo  {journal} {Astron. Astrophys.}\ }\textbf {\bibinfo {volume} {594}},\ \bibinfo {pages} {A4} (\bibinfo {year} {2016})},\ \Eprint {http://arxiv.org/abs/1502.01584} {arXiv:1502.01584 [astro-ph.CO]} \BibitemShut {NoStop}%
\bibitem [{\citenamefont {Adam}\ \emph {et~al.}(2016)\citenamefont {Adam} \emph {et~al.}}]{Planck:2015aiq}%
  \BibitemOpen
  \bibfield  {author} {\bibinfo {author} {\bibfnamefont {R.}~\bibnamefont {Adam}} \emph {et~al.} (\bibinfo {collaboration} {Planck}),\ }\href {\doibase 10.1051/0004-6361/201525844} {\bibfield  {journal} {\bibinfo  {journal} {Astron. Astrophys.}\ }\textbf {\bibinfo {volume} {594}},\ \bibinfo {pages} {A7} (\bibinfo {year} {2016})},\ \Eprint {http://arxiv.org/abs/1502.01586} {arXiv:1502.01586 [astro-ph.IM]} \BibitemShut {NoStop}%
\bibitem [{\citenamefont {Kusiak}\ \emph {et~al.}(2023)\citenamefont {Kusiak}, \citenamefont {Surrao},\ and\ \citenamefont {Hill}}]{Kusiak:2023hrz}%
  \BibitemOpen
  \bibfield  {author} {\bibinfo {author} {\bibfnamefont {A.}~\bibnamefont {Kusiak}}, \bibinfo {author} {\bibfnamefont {K.~M.}\ \bibnamefont {Surrao}}, \ and\ \bibinfo {author} {\bibfnamefont {J.~C.}\ \bibnamefont {Hill}},\ }\href {\doibase 10.1103/PhysRevD.108.123501} {\bibfield  {journal} {\bibinfo  {journal} {Phys. Rev. D}\ }\textbf {\bibinfo {volume} {108}},\ \bibinfo {pages} {123501} (\bibinfo {year} {2023})},\ \Eprint {http://arxiv.org/abs/2303.08121} {arXiv:2303.08121 [astro-ph.CO]} \BibitemShut {NoStop}%
\bibitem [{\citenamefont {Peacock}\ and\ \citenamefont {Smith}(2000)}]{Peacock:2000qk}%
  \BibitemOpen
  \bibfield  {author} {\bibinfo {author} {\bibfnamefont {J.~A.}\ \bibnamefont {Peacock}}\ and\ \bibinfo {author} {\bibfnamefont {R.~E.}\ \bibnamefont {Smith}},\ }\href {\doibase 10.1046/j.1365-8711.2000.03779.x} {\bibfield  {journal} {\bibinfo  {journal} {Mon. Not. Roy. Astron. Soc.}\ }\textbf {\bibinfo {volume} {318}},\ \bibinfo {pages} {1144} (\bibinfo {year} {2000})},\ \Eprint {http://arxiv.org/abs/astro-ph/0005010} {arXiv:astro-ph/0005010} \BibitemShut {NoStop}%
\bibitem [{\citenamefont {Zheng}\ \emph {et~al.}(2007)\citenamefont {Zheng}, \citenamefont {Coil},\ and\ \citenamefont {Zehavi}}]{Zheng:2007zg}%
  \BibitemOpen
  \bibfield  {author} {\bibinfo {author} {\bibfnamefont {Z.}~\bibnamefont {Zheng}}, \bibinfo {author} {\bibfnamefont {A.~L.}\ \bibnamefont {Coil}}, \ and\ \bibinfo {author} {\bibfnamefont {I.}~\bibnamefont {Zehavi}},\ }\href {\doibase 10.1086/521074} {\bibfield  {journal} {\bibinfo  {journal} {Astrophys. J.}\ }\textbf {\bibinfo {volume} {667}},\ \bibinfo {pages} {760} (\bibinfo {year} {2007})},\ \Eprint {http://arxiv.org/abs/astro-ph/0703457} {arXiv:astro-ph/0703457} \BibitemShut {NoStop}%
\bibitem [{\citenamefont {Bolliet}\ \emph {et~al.}(2023)\citenamefont {Bolliet}, \citenamefont {Hill}, \citenamefont {Ferraro}, \citenamefont {Kusiak},\ and\ \citenamefont {Krolewski}}]{Bolliet:2022pze}%
  \BibitemOpen
  \bibfield  {author} {\bibinfo {author} {\bibfnamefont {B.}~\bibnamefont {Bolliet}}, \bibinfo {author} {\bibfnamefont {J.~C.}\ \bibnamefont {Hill}}, \bibinfo {author} {\bibfnamefont {S.}~\bibnamefont {Ferraro}}, \bibinfo {author} {\bibfnamefont {A.}~\bibnamefont {Kusiak}}, \ and\ \bibinfo {author} {\bibfnamefont {A.}~\bibnamefont {Krolewski}},\ }\href {\doibase 10.1088/1475-7516/2023/03/039} {\bibfield  {journal} {\bibinfo  {journal} {JCAP}\ }\textbf {\bibinfo {volume} {03}},\ \bibinfo {pages} {039} (\bibinfo {year} {2023})},\ \Eprint {http://arxiv.org/abs/2208.07847} {arXiv:2208.07847 [astro-ph.CO]} \BibitemShut {NoStop}%
\bibitem [{\citenamefont {Navarro}\ \emph {et~al.}(1997)\citenamefont {Navarro}, \citenamefont {Frenk},\ and\ \citenamefont {White}}]{Navarro:1996gj}%
  \BibitemOpen
  \bibfield  {author} {\bibinfo {author} {\bibfnamefont {J.~F.}\ \bibnamefont {Navarro}}, \bibinfo {author} {\bibfnamefont {C.~S.}\ \bibnamefont {Frenk}}, \ and\ \bibinfo {author} {\bibfnamefont {S.~D.~M.}\ \bibnamefont {White}},\ }\href {\doibase 10.1086/304888} {\bibfield  {journal} {\bibinfo  {journal} {Astrophys. J.}\ }\textbf {\bibinfo {volume} {490}},\ \bibinfo {pages} {493} (\bibinfo {year} {1997})},\ \Eprint {http://arxiv.org/abs/astro-ph/9611107} {arXiv:astro-ph/9611107} \BibitemShut {NoStop}%
\bibitem [{\citenamefont {Tinker}\ \emph {et~al.}(2008)\citenamefont {Tinker}, \citenamefont {Kravtsov}, \citenamefont {Klypin}, \citenamefont {Abazajian}, \citenamefont {Warren}, \citenamefont {Yepes}, \citenamefont {Gottlober},\ and\ \citenamefont {Holz}}]{Tinker:2008ff}%
  \BibitemOpen
  \bibfield  {author} {\bibinfo {author} {\bibfnamefont {J.~L.}\ \bibnamefont {Tinker}}, \bibinfo {author} {\bibfnamefont {A.~V.}\ \bibnamefont {Kravtsov}}, \bibinfo {author} {\bibfnamefont {A.}~\bibnamefont {Klypin}}, \bibinfo {author} {\bibfnamefont {K.}~\bibnamefont {Abazajian}}, \bibinfo {author} {\bibfnamefont {M.~S.}\ \bibnamefont {Warren}}, \bibinfo {author} {\bibfnamefont {G.}~\bibnamefont {Yepes}}, \bibinfo {author} {\bibfnamefont {S.}~\bibnamefont {Gottlober}}, \ and\ \bibinfo {author} {\bibfnamefont {D.~E.}\ \bibnamefont {Holz}},\ }\href {\doibase 10.1086/591439} {\bibfield  {journal} {\bibinfo  {journal} {Astrophys. J.}\ }\textbf {\bibinfo {volume} {688}},\ \bibinfo {pages} {709} (\bibinfo {year} {2008})},\ \Eprint {http://arxiv.org/abs/0803.2706} {arXiv:0803.2706 [astro-ph]} \BibitemShut {NoStop}%
\bibitem [{\citenamefont {Limber}(1954)}]{Limber:1954zz}%
  \BibitemOpen
  \bibfield  {author} {\bibinfo {author} {\bibfnamefont {D.~N.}\ \bibnamefont {Limber}},\ }\href {\doibase 10.1086/145870} {\bibfield  {journal} {\bibinfo  {journal} {Astrophys. J.}\ }\textbf {\bibinfo {volume} {119}},\ \bibinfo {pages} {655} (\bibinfo {year} {1954})}\BibitemShut {NoStop}%
\bibitem [{\citenamefont {Scoccimarro}\ \emph {et~al.}(2001)\citenamefont {Scoccimarro}, \citenamefont {Sheth}, \citenamefont {Hui},\ and\ \citenamefont {Jain}}]{Scoccimarro:2000gm}%
  \BibitemOpen
  \bibfield  {author} {\bibinfo {author} {\bibfnamefont {R.}~\bibnamefont {Scoccimarro}}, \bibinfo {author} {\bibfnamefont {R.~K.}\ \bibnamefont {Sheth}}, \bibinfo {author} {\bibfnamefont {L.}~\bibnamefont {Hui}}, \ and\ \bibinfo {author} {\bibfnamefont {B.}~\bibnamefont {Jain}},\ }\href {\doibase 10.1086/318261} {\bibfield  {journal} {\bibinfo  {journal} {Astrophys. J.}\ }\textbf {\bibinfo {volume} {546}},\ \bibinfo {pages} {20} (\bibinfo {year} {2001})},\ \Eprint {http://arxiv.org/abs/astro-ph/0006319} {arXiv:astro-ph/0006319} \BibitemShut {NoStop}%
\bibitem [{\citenamefont {Duffy}\ \emph {et~al.}(2008)\citenamefont {Duffy}, \citenamefont {Schaye}, \citenamefont {Kay},\ and\ \citenamefont {Dalla~Vecchia}}]{Duffy:2008pz}%
  \BibitemOpen
  \bibfield  {author} {\bibinfo {author} {\bibfnamefont {A.~R.}\ \bibnamefont {Duffy}}, \bibinfo {author} {\bibfnamefont {J.}~\bibnamefont {Schaye}}, \bibinfo {author} {\bibfnamefont {S.~T.}\ \bibnamefont {Kay}}, \ and\ \bibinfo {author} {\bibfnamefont {C.}~\bibnamefont {Dalla~Vecchia}},\ }\href {\doibase 10.1111/j.1745-3933.2008.00537.x} {\bibfield  {journal} {\bibinfo  {journal} {Mon. Not. Roy. Astron. Soc.}\ }\textbf {\bibinfo {volume} {390}},\ \bibinfo {pages} {L64} (\bibinfo {year} {2008})},\ \bibinfo {note} {[Erratum: Mon.Not.Roy.Astron.Soc. 415, L85 (2011)]},\ \Eprint {http://arxiv.org/abs/0804.2486} {arXiv:0804.2486 [astro-ph]} \BibitemShut {NoStop}%
\bibitem [{\citenamefont {Bhattacharya}\ \emph {et~al.}(2013)\citenamefont {Bhattacharya}, \citenamefont {Habib}, \citenamefont {Heitmann},\ and\ \citenamefont {Vikhlinin}}]{Bhattacharya:2011vr}%
  \BibitemOpen
  \bibfield  {author} {\bibinfo {author} {\bibfnamefont {S.}~\bibnamefont {Bhattacharya}}, \bibinfo {author} {\bibfnamefont {S.}~\bibnamefont {Habib}}, \bibinfo {author} {\bibfnamefont {K.}~\bibnamefont {Heitmann}}, \ and\ \bibinfo {author} {\bibfnamefont {A.}~\bibnamefont {Vikhlinin}},\ }\href {\doibase 10.1088/0004-637X/766/1/32} {\bibfield  {journal} {\bibinfo  {journal} {Astrophys. J.}\ }\textbf {\bibinfo {volume} {766}},\ \bibinfo {pages} {32} (\bibinfo {year} {2013})},\ \Eprint {http://arxiv.org/abs/1112.5479} {arXiv:1112.5479 [astro-ph.CO]} \BibitemShut {NoStop}%
\bibitem [{\citenamefont {Tinker}\ \emph {et~al.}(2010)\citenamefont {Tinker}, \citenamefont {Robertson}, \citenamefont {Kravtsov}, \citenamefont {Klypin}, \citenamefont {Warren}, \citenamefont {Yepes},\ and\ \citenamefont {Gottlober}}]{Tinker:2010my}%
  \BibitemOpen
  \bibfield  {author} {\bibinfo {author} {\bibfnamefont {J.~L.}\ \bibnamefont {Tinker}}, \bibinfo {author} {\bibfnamefont {B.~E.}\ \bibnamefont {Robertson}}, \bibinfo {author} {\bibfnamefont {A.~V.}\ \bibnamefont {Kravtsov}}, \bibinfo {author} {\bibfnamefont {A.}~\bibnamefont {Klypin}}, \bibinfo {author} {\bibfnamefont {M.~S.}\ \bibnamefont {Warren}}, \bibinfo {author} {\bibfnamefont {G.}~\bibnamefont {Yepes}}, \ and\ \bibinfo {author} {\bibfnamefont {S.}~\bibnamefont {Gottlober}},\ }\href {\doibase 10.1088/0004-637X/724/2/878} {\bibfield  {journal} {\bibinfo  {journal} {Astrophys. J.}\ }\textbf {\bibinfo {volume} {724}},\ \bibinfo {pages} {878} (\bibinfo {year} {2010})},\ \Eprint {http://arxiv.org/abs/1001.3162} {arXiv:1001.3162 [astro-ph.CO]} \BibitemShut {NoStop}%
\bibitem [{\citenamefont {Lewis}\ \emph {et~al.}(2000)\citenamefont {Lewis}, \citenamefont {Challinor},\ and\ \citenamefont {Lasenby}}]{Lewis:1999bs}%
  \BibitemOpen
  \bibfield  {author} {\bibinfo {author} {\bibfnamefont {A.}~\bibnamefont {Lewis}}, \bibinfo {author} {\bibfnamefont {A.}~\bibnamefont {Challinor}}, \ and\ \bibinfo {author} {\bibfnamefont {A.}~\bibnamefont {Lasenby}},\ }\href {\doibase 10.1086/309179} {\bibfield  {journal} {\bibinfo  {journal} {Astrophys. J.}\ }\textbf {\bibinfo {volume} {538}},\ \bibinfo {pages} {473} (\bibinfo {year} {2000})},\ \Eprint {http://arxiv.org/abs/astro-ph/9911177} {arXiv:astro-ph/9911177} \BibitemShut {NoStop}%
\bibitem [{\citenamefont {Coulton}\ \emph {et~al.}(2024)\citenamefont {Coulton} \emph {et~al.}}]{ACT:2024rue}%
  \BibitemOpen
  \bibfield  {author} {\bibinfo {author} {\bibfnamefont {W.~R.}\ \bibnamefont {Coulton}} \emph {et~al.} (\bibinfo {collaboration} {ACT}),\ }\href@noop {} {\  (\bibinfo {year} {2024})},\ \Eprint {http://arxiv.org/abs/2401.13033} {arXiv:2401.13033 [astro-ph.CO]} \BibitemShut {NoStop}%
\bibitem [{\citenamefont {Marsh}\ \emph {et~al.}(2022)\citenamefont {Marsh}, \citenamefont {Matthews}, \citenamefont {Reynolds},\ and\ \citenamefont {Carenza}}]{Marsh:2021ajy}%
  \BibitemOpen
  \bibfield  {author} {\bibinfo {author} {\bibfnamefont {M.~C.~D.}\ \bibnamefont {Marsh}}, \bibinfo {author} {\bibfnamefont {J.~H.}\ \bibnamefont {Matthews}}, \bibinfo {author} {\bibfnamefont {C.}~\bibnamefont {Reynolds}}, \ and\ \bibinfo {author} {\bibfnamefont {P.}~\bibnamefont {Carenza}},\ }\href {\doibase 10.1103/PhysRevD.105.016013} {\bibfield  {journal} {\bibinfo  {journal} {Phys. Rev. D}\ }\textbf {\bibinfo {volume} {105}},\ \bibinfo {pages} {016013} (\bibinfo {year} {2022})},\ \Eprint {http://arxiv.org/abs/2107.08040} {2107.08040} \BibitemShut {NoStop}%
\bibitem [{\citenamefont {Nelson}\ \emph {et~al.}(2023)\citenamefont {Nelson}, \citenamefont {Pillepich}, \citenamefont {Ayromlou}, \citenamefont {Lee}, \citenamefont {Lehle}, \citenamefont {Rohr},\ and\ \citenamefont {Truong}}]{Nelson:2023ilh}%
  \BibitemOpen
  \bibfield  {author} {\bibinfo {author} {\bibfnamefont {D.}~\bibnamefont {Nelson}}, \bibinfo {author} {\bibfnamefont {A.}~\bibnamefont {Pillepich}}, \bibinfo {author} {\bibfnamefont {M.}~\bibnamefont {Ayromlou}}, \bibinfo {author} {\bibfnamefont {W.}~\bibnamefont {Lee}}, \bibinfo {author} {\bibfnamefont {K.}~\bibnamefont {Lehle}}, \bibinfo {author} {\bibfnamefont {E.}~\bibnamefont {Rohr}}, \ and\ \bibinfo {author} {\bibfnamefont {N.}~\bibnamefont {Truong}},\ }\href@noop {} {\  (\bibinfo {year} {2023})},\ \Eprint {http://arxiv.org/abs/2311.06338} {arXiv:2311.06338 [astro-ph.GA]} \BibitemShut {NoStop}%
\bibitem [{\citenamefont {Battaglia}\ \emph {et~al.}(2012)\citenamefont {Battaglia}, \citenamefont {Bond}, \citenamefont {Pfrommer},\ and\ \citenamefont {Sievers}}]{Battaglia:2011cq}%
  \BibitemOpen
  \bibfield  {author} {\bibinfo {author} {\bibfnamefont {N.}~\bibnamefont {Battaglia}}, \bibinfo {author} {\bibfnamefont {J.~R.}\ \bibnamefont {Bond}}, \bibinfo {author} {\bibfnamefont {C.}~\bibnamefont {Pfrommer}}, \ and\ \bibinfo {author} {\bibfnamefont {J.~L.}\ \bibnamefont {Sievers}},\ }\href {\doibase 10.1088/0004-637X/758/2/75} {\bibfield  {journal} {\bibinfo  {journal} {Astrophys. J.}\ }\textbf {\bibinfo {volume} {758}},\ \bibinfo {pages} {75} (\bibinfo {year} {2012})},\ \Eprint {http://arxiv.org/abs/1109.3711} {arXiv:1109.3711 [astro-ph.CO]} \BibitemShut {NoStop}%
\bibitem [{\citenamefont {Lee}\ \emph {et~al.}(2022)\citenamefont {Lee}, \citenamefont {Coulton}, \citenamefont {Thiele},\ and\ \citenamefont {Ho}}]{Lee:2022tor}%
  \BibitemOpen
  \bibfield  {author} {\bibinfo {author} {\bibfnamefont {B.~K.~K.}\ \bibnamefont {Lee}}, \bibinfo {author} {\bibfnamefont {W.~R.}\ \bibnamefont {Coulton}}, \bibinfo {author} {\bibfnamefont {L.}~\bibnamefont {Thiele}}, \ and\ \bibinfo {author} {\bibfnamefont {S.}~\bibnamefont {Ho}},\ }\href {\doibase 10.1093/mnras/stac2602} {\bibfield  {journal} {\bibinfo  {journal} {Mon. Not. Roy. Astron. Soc.}\ }\textbf {\bibinfo {volume} {517}},\ \bibinfo {pages} {420} (\bibinfo {year} {2022})},\ \Eprint {http://arxiv.org/abs/2205.01710} {arXiv:2205.01710 [astro-ph.CO]} \BibitemShut {NoStop}%
\bibitem [{\citenamefont {{Pakmor}}\ \emph {et~al.}(2020)\citenamefont {{Pakmor}}, \citenamefont {{van de Voort}}, \citenamefont {{Bieri}}, \citenamefont {{G{\'o}mez}}, \citenamefont {{Grand}}, \citenamefont {{Guillet}}, \citenamefont {{Marinacci}}, \citenamefont {{Pfrommer}}, \citenamefont {{Simpson}},\ and\ \citenamefont {{Springel}}}]{2020MNRAS.498.3125P}%
  \BibitemOpen
  \bibfield  {author} {\bibinfo {author} {\bibfnamefont {R.}~\bibnamefont {{Pakmor}}}, \bibinfo {author} {\bibfnamefont {F.}~\bibnamefont {{van de Voort}}}, \bibinfo {author} {\bibfnamefont {R.}~\bibnamefont {{Bieri}}}, \bibinfo {author} {\bibfnamefont {F.~A.}\ \bibnamefont {{G{\'o}mez}}}, \bibinfo {author} {\bibfnamefont {R.~J.~J.}\ \bibnamefont {{Grand}}}, \bibinfo {author} {\bibfnamefont {T.}~\bibnamefont {{Guillet}}}, \bibinfo {author} {\bibfnamefont {F.}~\bibnamefont {{Marinacci}}}, \bibinfo {author} {\bibfnamefont {C.}~\bibnamefont {{Pfrommer}}}, \bibinfo {author} {\bibfnamefont {C.~M.}\ \bibnamefont {{Simpson}}}, \ and\ \bibinfo {author} {\bibfnamefont {V.}~\bibnamefont {{Springel}}},\ }\href {\doibase 10.1093/mnras/staa2530} {\bibfield  {journal} {\bibinfo  {journal} {\mnras}\ }\textbf {\bibinfo {volume} {498}},\ \bibinfo {pages} {3125} (\bibinfo {year} {2020})},\ \Eprint {http://arxiv.org/abs/1911.11163} {arXiv:1911.11163 [astro-ph.GA]} \BibitemShut {NoStop}%
\end{thebibliography}%
\bibliographystyle{apsrev4-1}

%%%%%%%%%%%%%%%%%%%%%%%%%%%%%%%%%%%%%%%%%%%%%%%%%%%%%%%%%%%%%%%%%%%%%%%%%%%%%%%%%%%%%%%%%%%%%%%%%%%%%%%%%%%%%%%%%%%%%
% Appendix
%%%%%%%%%%%%%%%%%%%%%%%%%%%%%%%%%%%%%%%%%%%%%%%%%%%%%%%%%%%%%%%%%%%%%%%%%%%%%%%%%%%%%%%%%%%%%%%%%%%%%%%%%%%%%%%%%%%%%
\clearpage
\pagebreak

\setcounter{secnumdepth}{2}
\setcounter{equation}{0}
\setcounter{figure}{0}
\setcounter{page}{1}

\renewcommand{\theequation}{S\arabic{equation}}
\renewcommand{\thefigure}{S\arabic{figure}}
\renewcommand{\theHfigure}{S\arabic{figure}}%
\renewcommand{\thetable}{S\arabic{table}}

\onecolumngrid
\clearpage
\pagebreak
\appendix

\onecolumngrid
\begin{center}
  \textbf{\large Supplemental Material}\\[.2cm]
  \vspace{0.05in}
  {Samuel Goldstein, Fiona McCarthy, Cristina Mondino, J.~Colin Hill, Junwu Huang, Matthew C.~Johnson}
\end{center}

\section{Mapmaking and measurement details}
\subsection{Axion-induced screening mapmaking}
\begin{figure}[!b]
\centering
\includegraphics[width=0.995\linewidth]{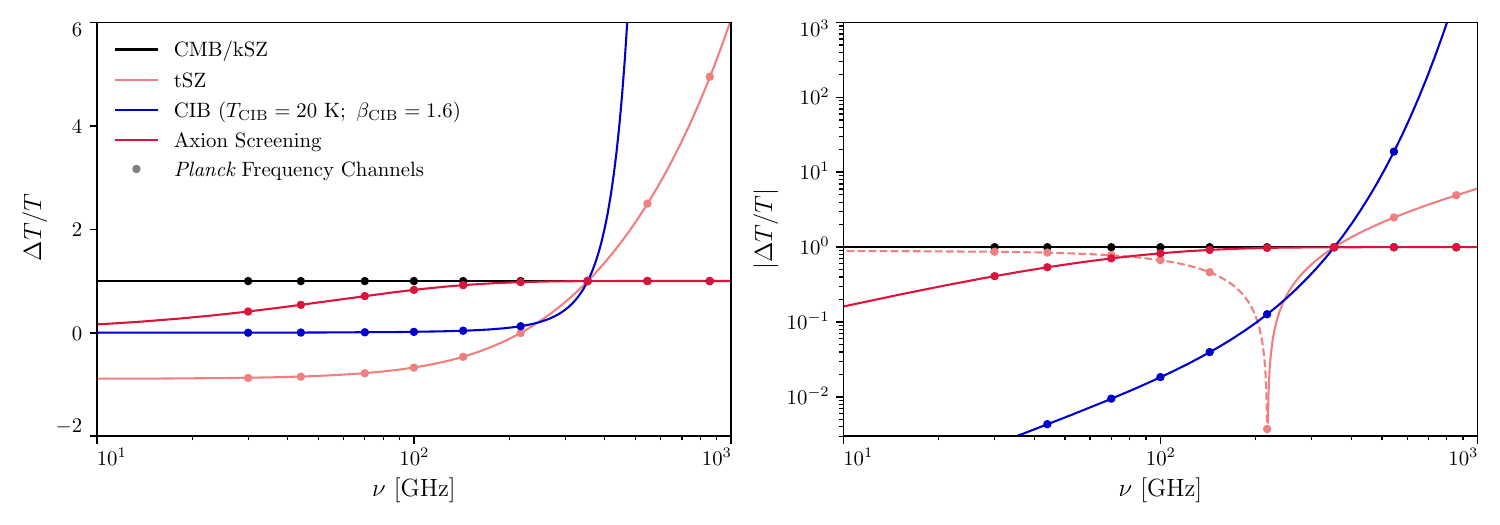}
\caption{Spectral energy distributions (SEDs) in CMB thermodynamic temperature units for the signals considered in this work. All SEDs are normalized to unity at $\nu=353$ GHz. For clarity, the right panel shows the same SEDs but with a logarithmically scaled $y$-axis. The solid points denote the \emph{Planck} frequency channels.}
\label{fig:SED}
\end{figure}

In this section, we describe our method for constructing the axion-induced patchy screening map. The resonant conversion of CMB photons into axions causes a spectral distortion in the CMB temperature with the following spectral energy distribution (SED) (see, \emph{e.g.},~\cite{Mondino:2024rif}):
\begin{equation}\label{eq:SED}
    \frac{\tilde{T}^a(x)}{T_{\rm CMB}}\propto x\left(\frac{1-e^{-x}}{x} \right).
\end{equation}
In Fig.~\ref{fig:SED}, we plot this SED alongside other relevant signals. The axion-induced screening signal (red) has a unique SED, which allows it to be distinguished from other components using multi-frequency CMB temperature measurements. At high frequencies, one might worry about separating the axion-induced screening signal from the CMB/kinematic SZ (kSZ, black) due to their similar SEDs. Fortunately, these contributions do not impact the \emph{cross-correlation} measurement analyzed in this work. Specifically, the cross-correlation $\langle T_{\rm CMB}\delta_g\rangle$ is effectively zero at all but the largest scales, where the integrated Sachs-Wolfe (ISW) effect becomes relevant~\cite{Sachs:1967er}.\footnote{At small scales, non-linear structure formation leads to a non-zero correlation between $T_{\rm CMB}$ and $\delta_g$~\cite{Rees:1968zza}. However, this effect is \emph{very} small and beyond the detection limits of current surveys~\cite{Ferraro:2022twg}.} The ISW contribution is negligible for the datasets and scales considered here~\cite{Krolewski:2021znk}. Similarly, $\langle T_{\rm kSZ}\delta_g\rangle$ is expected to be zero because the kSZ effect arises from line-of-sight peculiar velocities of galaxies and clusters, which are equally likely to be positive or negative~\cite{Dore:2003ex}. Consequently, any residual contributions from the CMB/kSZ in our axion maps will not bias our analysis, which is based on the cross-correlation between axion-induced screening and the projected galaxy distribution.

\begin{table*}[!t]
\begin{tabular}{|c|c|c|c|}
\hline
Harmonic scale $I$~&~$\ell_{\rm peak}$~& ~FWHM (degrees)~&~Frequencies included~\\\hline
    0 & 0 &  91.8& {30,44,70,100,143,217,353,545} GHz \\\hline
    1 & 100 & 35.6& {30,44,70,100,143,217,353,545} GHz\\\hline
    2 & 200 &25.2 & {30,44,70,100,143,217,353,545} GHz \\\hline
    3 & 300 &20.6&  {30,44,70,100,143,217,353,545} GHz\\\hline
    4 & 400 & 14.1 & {30,44,70,100,143,217,353,545} GHz\\\hline
    5 & 600 & 10.3 & {30,44,70,100,143,217,353,545} GHz\\\hline
    6 & 800 & 8.26 & {44,70,100,143,217,353,545} GHz\\\hline
    7 & 1000 & 6.31 & {70,100,143,217,353,545} GHz\\\hline
    8 & 1250 & 6.11 & {70,100,143,217,353,545} GHz\\\hline
    9 & 1400 & 4.74 & {70,100,143,217,353,545} GHz\\\hline
    10 & 1800 & 3.55 & {70,100,143,217,353,545} GHz \\\hline
    11 & 2200 & 1.34 & {143,217,353,545} GHz\\\hline
    12 & 4097 & 1.28 &{143,217,353,545} GHz \\\hline

\end{tabular}
\caption{Properties of the 13 cosine needlets used to construct the NILC maps of axion-induced screening. The table includes the angular scale at which each needlet filter peaks in harmonic space ($\ell_{\rm peak}$), the full-width half maximum (FWHM) of the real-space Gaussian kernel used for each filter scale, and the \emph{Planck} frequency channels that are used at each needlet scale.}\label{tab:realspace_filters}
\end{table*}

We construct maps of axion-induced screening from multi-frequency CMB temperature measurements using the internal linear combination (ILC) method~\cite{1992ApJ...396L...7B, WMAP:2003cmr, Tegmark:2003ve, Eriksen:2004jg}. The ILC method constructs the optimal, \emph{i.e.}, minimum-variance, linear combination of the observed frequency maps that preserves the SED of the signal of interest. When computing the optimal weights, it is important to choose an appropriate basis for the ILC. Here, we use needlet ILC (NILC)~\cite{Delabrouille:2008qd}, which computes the weights on a needlet frame. Needlets are spherical wavelets with compact support in both real and harmonic space~\cite{doi:10.1137/040614359}, allowing the weights to vary with both position and angular scale. Thus, NILC is particularly effective for CMB analyses, where Galactic contributions are often localized in real space and extragalactic contributions are better modeled in harmonic space. We use cosine needlets, which are defined by the harmonic-space needlet filter function $h^{(I)}_\ell:$
\begin{equation}
h^{(I)}_\ell= \begin{cases}
\cos\left(\frac{\pi}{2}\frac{\ell^I_{\rm peak}-\ell}{\ell^I_{\rm peak}-\ell^{I-1}_{\rm peak}}\right)&\ell^{I-1}_{\rm peak}\le\ell<\ell^{I}_{\rm peak}\\
\cos\left(\frac{\pi}{2}\frac{\ell-\ell^I_{\rm peak}}{\ell^{I+1}_{\rm peak}-\ell^I_{\rm peak}}\right)&\ell^{I}_{\rm peak}\le\ell<\ell^{I+1}_{\rm peak}\\
0 & {\rm otherwise},
\end{cases}
\end{equation}
where $I$ indexes the needlet scale. Following Ref.~\cite{McCarthy:2024ozh}, we use 13 cosine needlets with peak multipole values between $0$ and $4097.$  In addition to the harmonic-space filters, we use Gaussian real-space filters to compute the frequency-frequency covariance matrix.
The real-space filter scales are chosen to ensure that the ``ILC bias"~\cite{Delabrouille:2008qd}, which arises because the ILC weights are derived directly from the data (see Ref.~\cite{McCarthy:2023hpa} for a review), is less than 1\% in our final maps. Following Refs.~\cite{McCarthy:2023hpa, McCarthy:2023cwg, McCarthy:2024ozh}, we model the \emph{Planck} instrument beams~\cite{Planck:2015wtm, Planck:2015aiq} as Gaussians, with FWHM values given in Table I of Ref.~\cite{McCarthy:2023hpa}. After beam-deconvolving the \emph{Planck} maps, we reconvolve all maps to a common Gaussian beam with FWHM = 10 arcmin. Note that since the low-frequency \emph{Planck} maps have larger beams than the high-frequency maps, we exclude low-frequency channels when constructing the NILC map at the smallest needlet filter scales. Table~\ref{tab:realspace_filters} provides more details on the needlet filters used in this work and the \emph{Planck} maps included for each needlet filter scale. We construct the constrained NILC maps using \texttt{pyilc}~\cite{McCarthy:2023hpa}.\footnote{\href{https://github.com/jcolinhill/pyilc/}{https://github.com/jcolinhill/pyilc/}}  Note that we use the full \emph{Planck} passbands to compute the frequency responses for all components (\emph{i.e.}, we integrate each component SED over each frequency channel's passband), as described in Refs.~\cite{McCarthy:2023hpa,McCarthy:2024ozh}.

\begin{figure}[!t]
\centering
\includegraphics[width=0.995\linewidth]{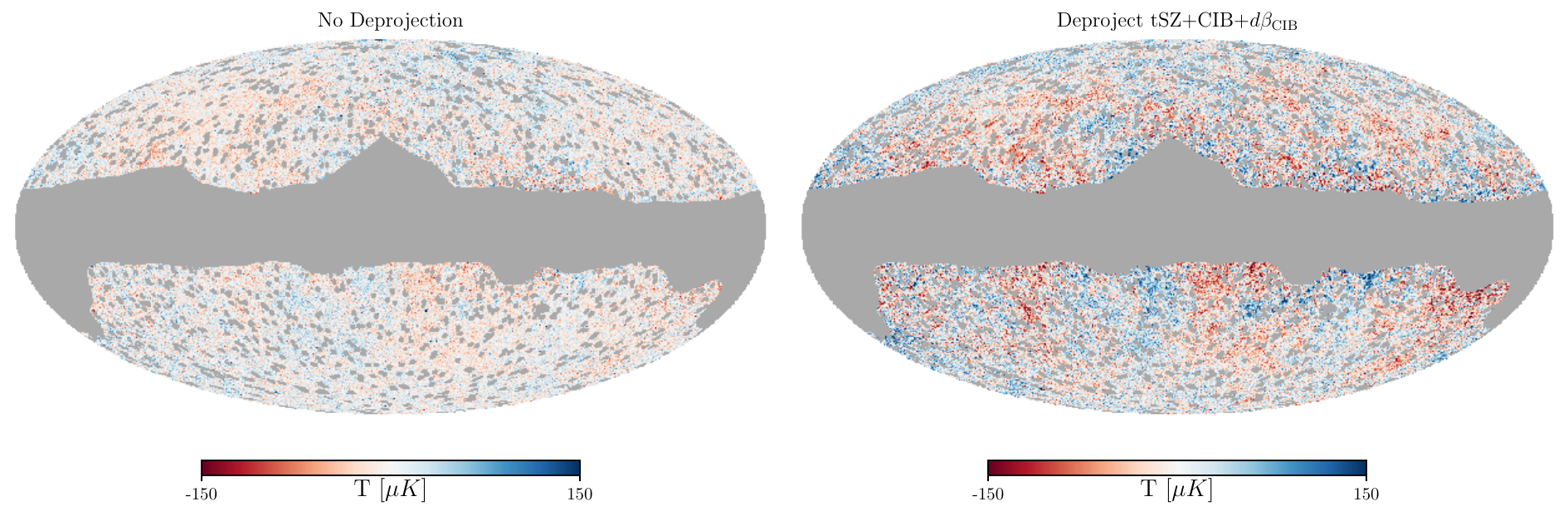}
\caption{Example maps of axion-induced screening constructed from \emph{Planck} CMB temperature maps. The maps are shown before (left) and after (right) deprojecting contributions from foregrounds that are correlated with the \emph{unWISE} galaxy sample. Note the visible increase in variance as a result of deprojection. The residual structures in the deprojected map are primarily due to the primary CMB, which we do not deproject, and to a lesser extent due to residual Galactic foregrounds. 
}
\label{fig:map_projections}
\end{figure}

To minimize contributions from contaminants that are correlated with the \emph{unWISE} galaxy distribution in our final axion map, we utilize the \emph{constrained} NILC method~\cite{Chen:2008gw, Remazeilles:2010hq}. Constrained ILC is designed to explicitly remove contaminants with known frequency dependence, albeit at the cost of increased noise in the final map. The primary contaminants expected to correlate with the \emph{unWISE} galaxies are the thermal Sunyaev-Zel'dovich (tSZ) effect~\cite{Sunyaev:1972eq, Sunyaev:1980nv} and the cosmic infrared background (CIB). While the tSZ effect can be deprojected exactly due to its known SED, the CIB is more challenging to deproject because it lacks a definitive SED. Following Refs.~\cite{McCarthy:2023hpa, McCarthy:2023cwg, McCarthy:2024ozh}, we approximate the CIB SED as a modified blackbody. For our fiducial analysis, we fix the CIB temperature $T_{\rm CIB}=20$ K and spectral index $\beta_{\rm CIB}=1.6$. To account for uncertainties in our model for the CIB SED, we also deproject the first derivative of the SED with respect to $\beta_{\rm CIB}$, as proposed in Ref.~\cite{Chluba:2017rtj}. Henceforth, we use the notation $d\beta_{\rm CIB}$ ($dT_{\rm CIB}$) to denote maps where the first derivative of the CIB SED with respect to  $\beta_{\rm CIB}$ ($T_{\rm CIB}$) has been deprojected. In Sec.~\ref{App:cross_correlation_measurement}, we demonstrate that this approach effectively removes any residual CIB contamination in our axion screening maps that is correlated with the \emph{unWISE} Blue galaxies.

Fig.~\ref{fig:map_projections} shows the NILC axion-induced screening map before (left) and after (right) deprojecting foreground contributions from the tSZ and CIB. Deprojecting the tSZ and CIB increases the variance of the final map, as expected. Notice that there is still significant structure in the deprojected map. This is largely due to the primary CMB, which, as discussed in more detail at the beginning of this section, we do not deproject in our analysis because it is uncorrelated with the \emph{unWISE} Blue galaxies over the range of scales used in our fiducial analysis. Additionally, the map shows evidence of residual Galactic contamination, particularly near the boundaries of the Galactic plane mask. In Sec.~\ref{App:cross_correlation_measurement}, we explicitly demonstrate that these residual contaminants do not bias the measured cross-correlation between the axion-induced screening map and the \emph{unWISE} Blue galaxies.

\begin{figure}[!t]
\centering
\includegraphics[width=0.5\linewidth]{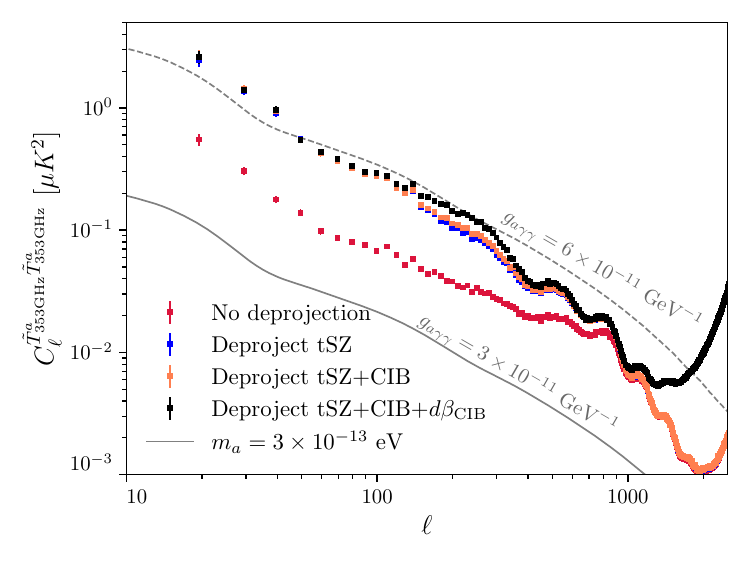}
\caption{Auto-power spectrum of the axion-induced CMB screening maps analyzed in this work. The different colors represent various choices of the deprojected components. The grey curves illustrate the expected signal for a $m_a=3\times 10^{-13}$ eV axion, assuming axion-photon couplings of $g_{a\gamma\gamma}=3\times 10^{-11}~{\rm GeV}^{-1}$ (solid) and $g_{a\gamma\gamma}=6\times 10^{-11}~{\rm GeV}^{-1}$ (dashed). The auto-power spectrum is significantly less sensitive to the axion-photon coupling than the cross-correlation analyzed in the \emph{Letter}.}
\label{fig:auto_spectrum}
\end{figure}

Fig.~\ref{fig:auto_spectrum} shows measurements of the auto-power spectrum of the axion-induced screening maps for various deprojection choices. Deprojecting foregrounds increases the variance of the final map. The acoustic oscillations of the primary CMB are clearly visible in the auto-power spectra because we do not deproject the primary CMB. For comparison, we include the theoretical prediction for $C_{\ell}^{\tilde{T}^a_{\rm 353 GHz}\tilde{T}^a_{\rm 353 GHz}}$ for a $m_a=3\times 10^{-13}~{\rm eV}$ axion, assuming two different axion-photon couplings.\footnote{Since the auto-spectrum includes contributions beyond the maximum redshift at which we measure the magnetic field profiles from IllustrisTNG ($z=1.5$), we compute $C_{\ell}^{\tilde{T}^a\tilde{T}^a}$ using the constant-$\beta$ magnetic field profile (see Sec.~\ref{sec:SM_Bfield_model}) assuming $\beta=100$.} Notice that the measured auto-power spectrum is significantly less sensitive to the axion-photon coupling than the cross-power spectrum presented in the \emph{Letter}, which constrains $g_{a\gamma\gamma}\lesssim 2\times 10^{-12}~{\rm GeV}^{-1}$ at the 95\% confidence limit for this axion mass. 

We note that Ref.~\cite{Mukherjee:2018zzg} constructed component-separated ILC maps of axion screening from \emph{Planck}-2015 temperature data. Using the auto-power spectrum of these maps, Ref.~\cite{Mukherjee:2018zzg} placed constraints on the axion-photon coupling by searching for non-resonant conversion of CMB photons into axions within the Galactic halo. In contrast, we use the cross-correlation of NILC axion screening maps with the projected \emph{unWISE} galaxy distribution to search for resonant conversion of CMB photons into axions in extragalactic halos. Note that resonant conversion has a different SED than the non-resonant case studied in Ref.~\cite{Mukherjee:2018zzg}, and thus our maps are not directly comparable. Finally, Ref.~\cite{Mukherjee:2018zzg} constrained $g_{a\gamma\gamma}\lesssim10^{-9}~{\rm GeV}^{-1}$ for $10^{-13}~{\rm eV}\lesssim m_a\lesssim 10^{-12}~{\rm eV}$, which is significantly weaker than the $g_{a\gamma\gamma}\lesssim10^{-12}~{\rm GeV}^{-1}$ bound derived in this work.

\begin{figure}[!t]
\centering
\includegraphics[width=0.995\linewidth]{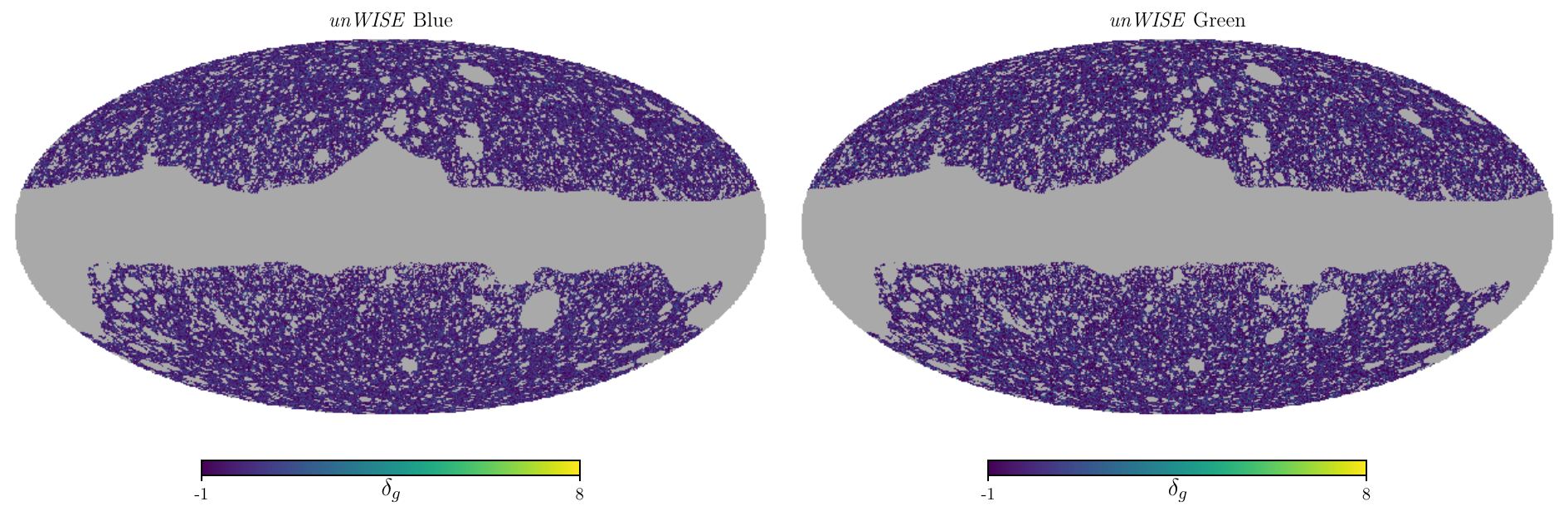}
\caption{Projected galaxy overdensity $\delta_g$ maps for the \emph{unWISE} Blue (left) and \emph{unWISE} Green (right) samples.
}
\label{fig:unwise_map_projections}
\end{figure}
\begin{figure}[!t]
\centering
\includegraphics[width=0.5\linewidth]{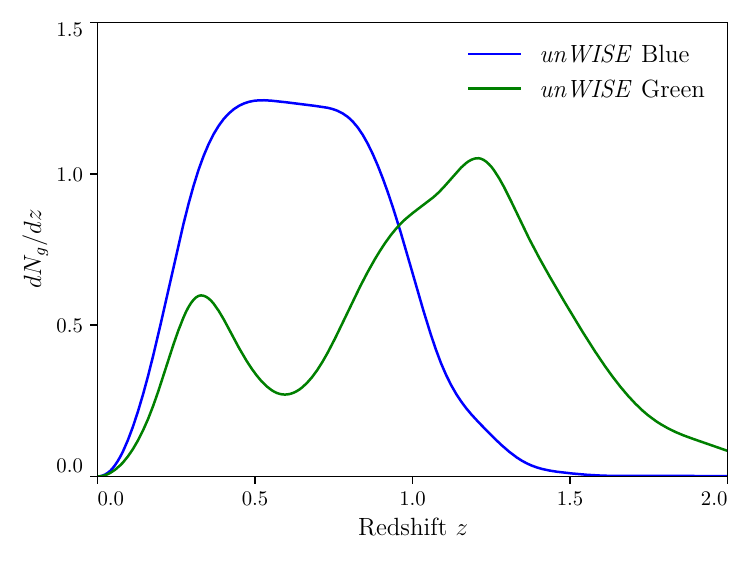}
\caption{Normalized galaxy redshift distributions for the \emph{unWISE} Blue and Green galaxy samples.
}
\label{fig:dN_dz_unwise}
\end{figure}
\subsection{The unWISE galaxy sample}

In this section, we review the main details of the \emph{unWISE} galaxy sample. For more details regarding \emph{unWISE}, see Refs.~\cite{Lang_2014, Meisner_2017, Meisner_2017a, Schlafly_2019, Krolewski:2019yrv}. The \emph{unWISE} galaxy sample consists of infrared-selected galaxies constructed from data collected by the \emph{Wide-field Infrared Survey Explorer} (\emph{WISE}) mission~\cite{2010AJ....140.1868W,2011ApJ...731...53M}. The \emph{unWISE} catalog contains over 500 million galaxies across the full sky. These galaxies are divided into three sub-samples: ``Blue," ``Green," and ``Red," with mean redshifts of $\bar{z} \approx$ 0.6, 1.1, and 1.5, respectively. In this work, we focus on the Blue sample due to its low redshift and high number density. The low-redshift Blue sample is the most relevant for our cross-correlation analysis because it is less correlated with the CIB and is easier to model in the halo model, where uncertainties in magnetic fields and electron distributions may complicate analyses of the higher-redshift galaxy samples. However, in what follows, we also analyze the cross-correlation of our axion screening maps with the \emph{unWISE} Green sample, for reference.

 Fig.~\ref{fig:unwise_map_projections} shows the projected galaxy overdensity $\delta_g$ for the \emph{unWISE} Blue and Green galaxy samples. Following Ref.~\cite{McCarthy:2024ozh}, we mask the \emph{unWISE} galaxy maps using the union of the \emph{unWISE} mask~\cite{Krolewski:2019yrv} and the \emph{Planck} CMB lensing mask, leaving 51\% of the sky unmasked.

In Fig.~\ref{fig:dN_dz_unwise}, we show the normalized redshift distribution of the \emph{unWISE} Blue and Green galaxy samples, which satisfy $\int \frac{dN_g}{dz}dz=1.$ The \emph{unWISE} Green sample extends to significantly higher redshifts and is therefore expected to be more strongly correlated with the CIB, which is predominantly sourced by high-redshift dusty star-forming galaxies, with a broad peak around $z \approx 2$~(e.g.,~\cite{Maniyar:2018xfk, Yan:2023okq}).

\begin{figure*}[!t]
\centering
\includegraphics[width=0.995\linewidth]{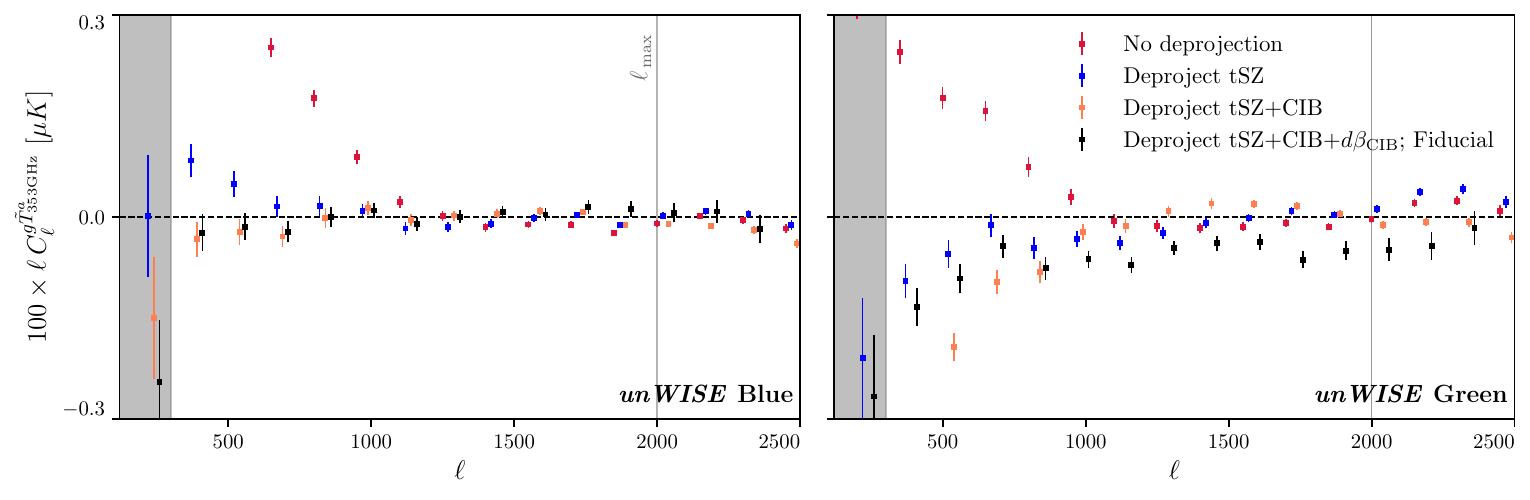}
\caption{Impact of changing the deprojected components on the measured cross-power spectrum of the axion screening map with \emph{unWISE} galaxies. Deprojecting the tSZ and CIB$+d\beta_{\rm CIB}$ components significantly impacts the measured cross-correlation. The cross-correlation of the fiducial axion screening map (black) with \emph{unWISE} Blue is consistent with zero for all multipoles between $300<\ell<2500$. The \emph{unWISE} Green cross-correlation is inconsistent with zero for all deprojection choices. The shaded grey region denotes multipoles that we exclude from our analysis due to residual extragalactic foreground contamination. The dashed grey line shows the maximum multipole included in our fiducial analysis. The points are horizontally offset for clarity.}
\label{fig:fig_change_deproj_components}
\end{figure*}
\begin{figure*}[!t]
\centering
\includegraphics[width=0.995\linewidth]{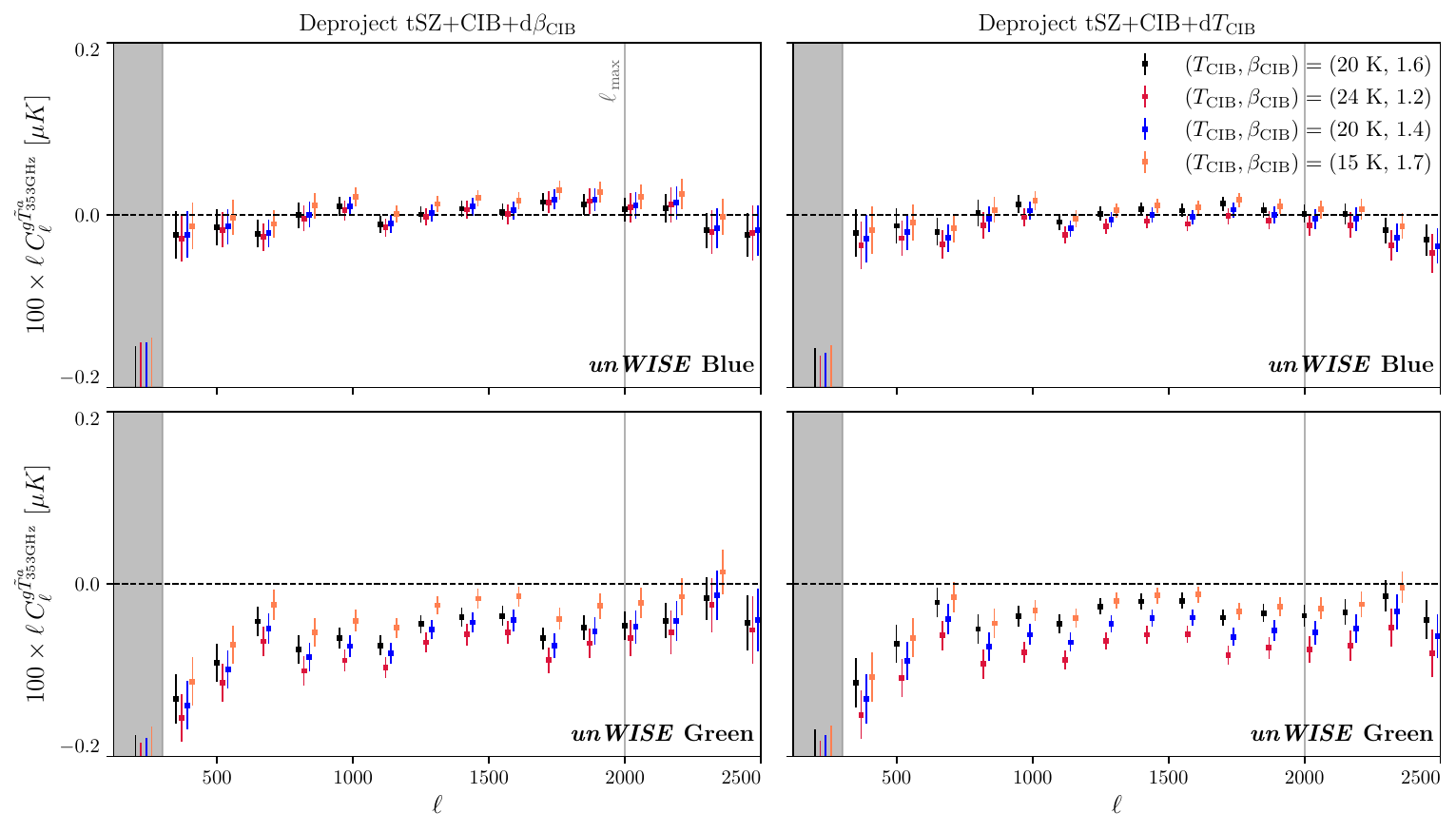}
\caption{Impact of changing the model for the CIB on the measured-cross correlation. The different colors indicate different choices of parameters for the CIB SED used in the constrained NILC axion screening map.  The left panels (right panels) show results for a constrained NILC with tSZ+CIB+d$\beta_{\rm CIB}$ deprojected (tSZ+CIB+d$T_{\rm CIB}$ deprojected). Whereas the \emph{unWISE} Blue measurements (top) are robust to changes in the CIB parameters, the \emph{unWISE} Green measurements fluctuate significantly. This demonstrates that the axion screening maps contain residual CIB contamination at high redshift and should not be used with the \emph{unWISE} Green galaxy catalog.}
\label{fig:fig_CIB_param_stability}
\end{figure*}
\subsection{Cross-correlation measurement}\label{App:cross_correlation_measurement}

In this section, we validate our cross-correlation measurements derived from the axion-induced screening maps. In Fig.~\ref{fig:fig_change_deproj_components}, we show the measured cross-power spectrum between the axion-induced screening map and the \emph{unWISE} galaxy samples for different deprojected components. For both galaxy samples, the cross-correlation changes considerably after deprojecting different combinations of the tSZ and CIB components, indicating that the undeprojected measurement is contaminated by these foregrounds. After deprojecting the tSZ, CIB, and $d\beta_{\rm CIB}$ components, we find no evidence for a significant cross-correlation between the axion-induced screening map and the \emph{unWISE} Blue galaxies for $\ell\gtrsim 300$. Notice that deprojecting the first derivative of the CIB with respect to $\beta_{\rm CIB}$ is crucial for removing CIB contributions at $\ell\gtrsim 1500$. For $\ell<300$, we detect a non-zero cross-correlation that, as described in more detail below, is likely due to residual extragalactic foreground contamination. Consequently, we exclude these scales from our analysis. The right panel of Fig.~\ref{fig:fig_change_deproj_components} shows the cross-correlation with the \emph{unWISE} Green galaxies for different deprojected components. The cross-correlation changes significantly after deprojecting different combinations of the tSZ and CIB fields. Moreover, the cross-correlation is inconsistent with zero after deprojecting the tSZ, CIB, and $d\beta_{\rm CIB}$ components. As described in more detail below, this is likely due to residual CIB contamination that correlates strongly with the high-redshift \emph{unWISE} Green galaxy sample.

\begin{figure*}[!t]
\centering
\includegraphics[width=0.995\linewidth]{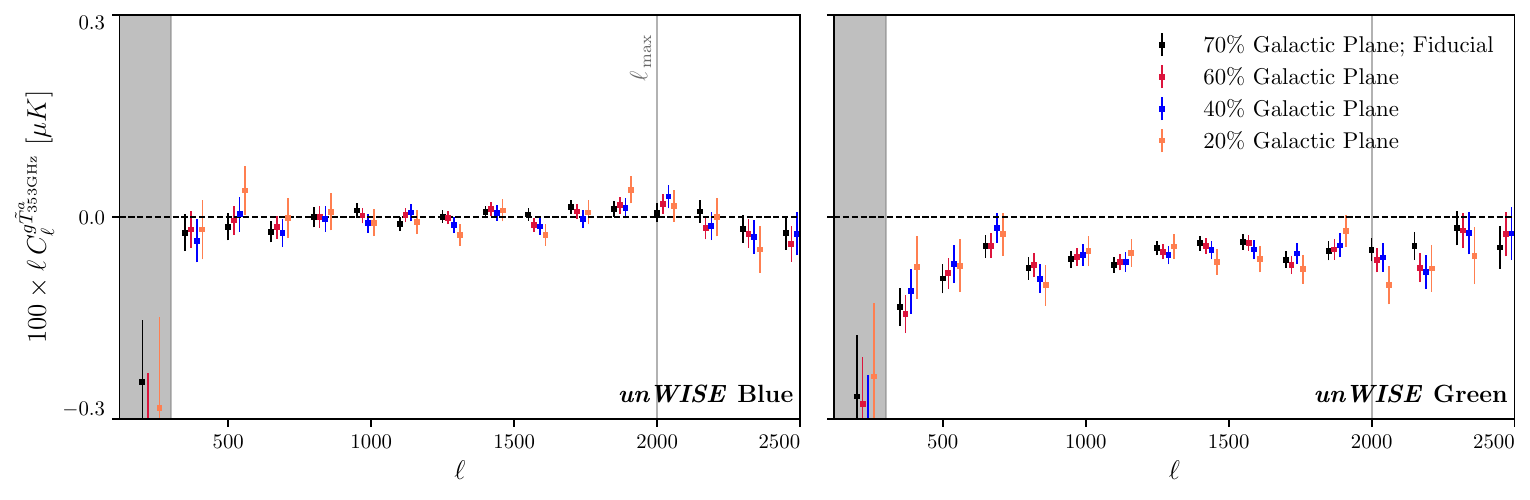}
\caption{Impact of changing the Galactic plane mask on the measured cross-power spectrum of the axion screening map with \emph{unWISE} galaxies. Our fiducial analysis (black) uses a mask that retains 70\% of the sky area. Using more conservative masks increases the variance of the cross-correlation, but does not significantly change the measurement.}
\label{fig:fig_change_gal_mask}
\end{figure*}
\begin{figure*}[!t]
\centering
\includegraphics[width=0.995\linewidth]{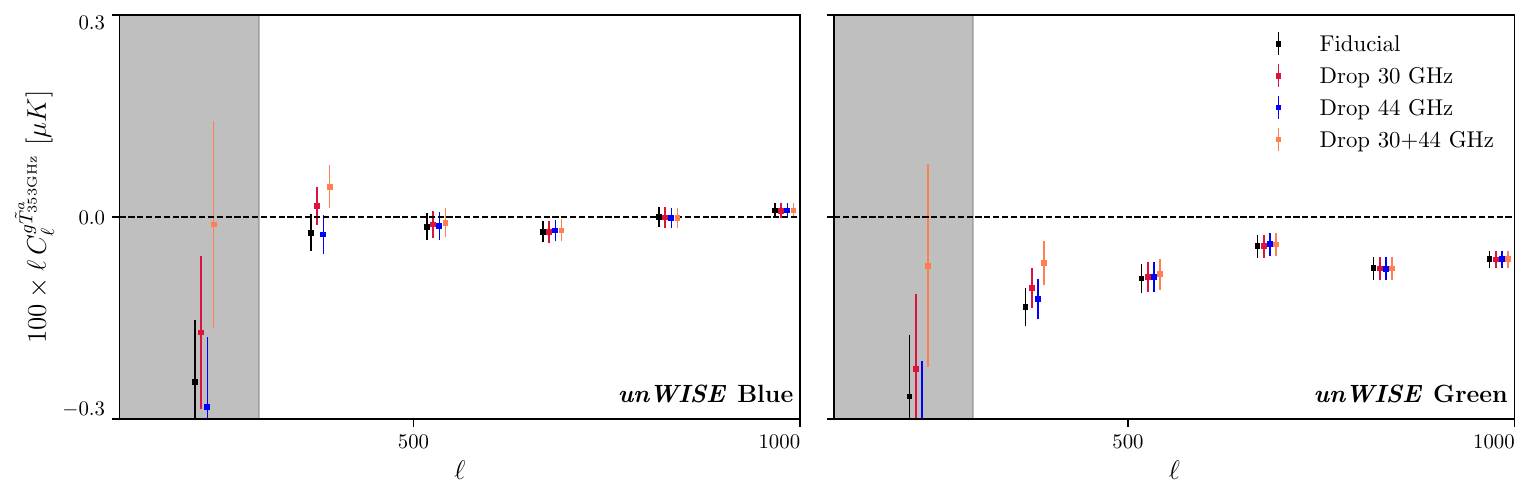}
\caption{Impact of not including the 30 GHz, 44 GHz, and both 30 and 44 GHz \emph{Planck} temperature maps on the measured cross-correlation. Note that the fiducial analysis (black) includes both the 30 and 44 GHz channels.}
\label{fig:fig_drop_channels}
\end{figure*}

To assess the robustness of our fiducial CIB deprojection, we show the measured cross-correlation for various CIB SEDs in Fig.~\ref{fig:fig_CIB_param_stability}. The different colors correspond to different choices of the parameters characterizing the modified blackbody CIB SED. The left (right) panels show analyses where we deproject the tSZ, CIB, and the first derivative of the CIB SED with respect to $\beta_{\rm CIB}$ ($T_{\rm CIB}$). Note that we do not have enough frequency channels to reliably deproject both $d\beta_{\rm CIB}$ and $dT_{\rm CIB}$ beyond the first six needlet scales, corresponding to $\ell\lesssim 1000$. The cross-correlation with \emph{unWISE} Blue remains consistent with zero for $\ell>300$ across all choices of CIB parameters, as well as when deprojecting $dT_{\rm CIB}$ instead of $d\beta_{\rm CIB}.$ This suggests that our fiducial analysis effectively removes any CIB contributions from the axion screening map that are correlated with \emph{unWISE} Blue, within the precision of the datasets used here. The bottom panel shows the cross-correlation with \emph{unWISE} Green for different models of the CIB SED. In this case, the measurement is highly sensitive to changes in the CIB parameters, and to the deprojection of $d\beta_{\rm CIB}$ versus $dT_{\rm CIB}$. Since the modified blackbody is only an approximation of the CIB SED and we do not \emph{a priori} know the values of $\beta_{\rm CIB}$ and $T_{\rm CIB}$, the strong dependence of $C_{\ell}^{g\tilde{T}^a}$ on these values, especially after deprojecting the first derivatives $d\beta_{\rm CIB}$ or $dT_{\rm CIB}$, suggests that our axion screening maps contain residual CIB contamination that is correlated with the \emph{unWISE} Green galaxies. Consequently, we exclude this galaxy sample from our analysis. 

Having validated the treatment of the tSZ and CIB contaminants in our fiducial analysis, we now assess the robustness of our measurements against potential Galactic contaminants. Fig.~\ref{fig:fig_change_gal_mask} shows the measured cross-power spectrum assuming different Galactic plane masks. While adopting a more conservative Galactic plane mask increases the variance of our measurement, the cross-power spectrum measurements are consistent across all scales regardless of the Galactic plane mask. Thus, we find no evidence that our cross-correlation measurement is biased by Galactic contamination, suggesting that any low-$\ell$ contamination in $C_{\ell}^{g\tilde{T}^a}$ could be of extragalactic origin.

Fig.~\ref{fig:fig_drop_channels} shows the measured cross-correlation after removing the low-frequency 30/44 GHz channels from the analysis. Excluding either the 30 GHz or the 44 GHz channels individually yields a non-zero cross-correlation at large scales that is consistent with the fiducial analysis. However, when both channels are excluded, the large-scale cross-correlation is mildly inconsistent ($\approx 2\sigma$) with the fiducial analysis, albeit with a significant increase in variance.

\begin{figure*}[!t]
\centering
\includegraphics[width=0.995\linewidth]{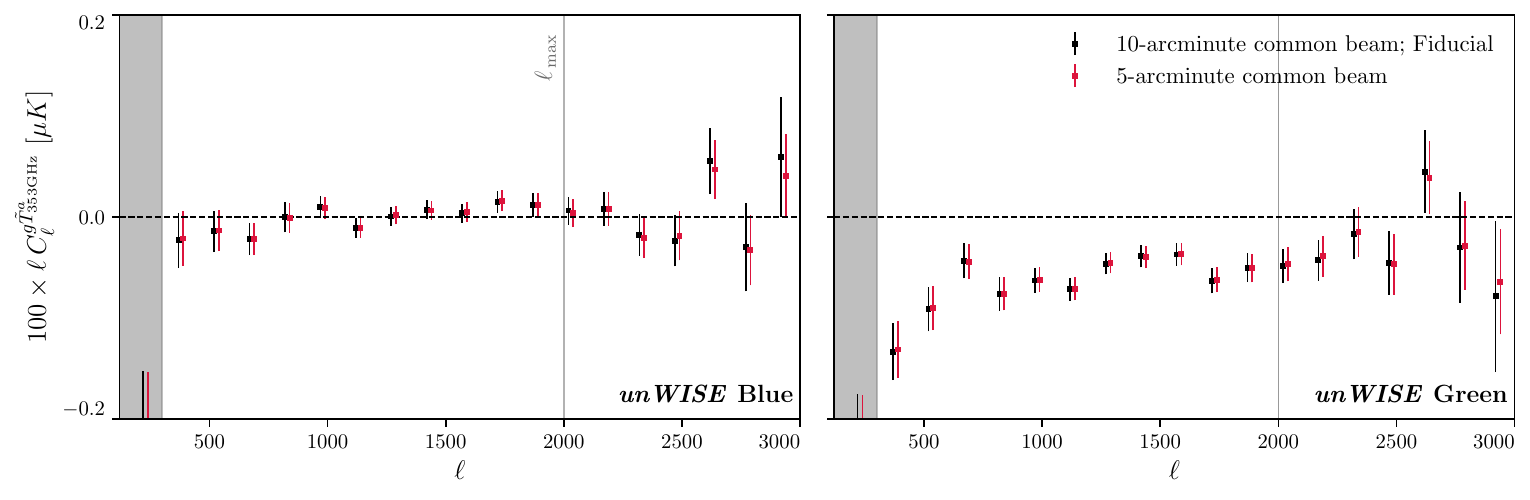}
\caption{Impact of varying the size of the common Gaussian beam used in constructing the NILC map on $C_{\ell}^{g \tilde{T}^a_{\rm 353 GHz}}$. The fiducial analysis (black) employs a FWHM = 10 arcmin beam. When using a 5-arcmin beam (red), the measurements are indistinguishable from the fiducial analysis up to $\ell \approx 2500$. For $\ell > 2500$, the cross-correlation measurements with the 5-arcminute beam have slightly lower variance than the fiducial measurement; however, these scales are excluded from our analysis. }
\label{fig:fig_change_beam}
\end{figure*}

Fig.~\ref{fig:fig_change_beam} shows the impact of varying the Gaussian beam size used in constructing the NILC map on the cross-correlation measurements. The size of the NILC beam can impact the final map because the NILC covariance is computed in real space, and is therefore sensitive to the smoothing of the small-scale modes~\cite{McCarthy:2023cwg}. In our fiducial analysis, all maps are convolved with a Gaussian beam with a 10 arcminute FWHM. Switching to a 5 arcminute beam does not alter our results, aside from a slight reduction in the variance of the measurement on small scales ($\ell\gtrsim 2500$), which are already excluded from our analysis. 

The tests performed in this section motivate the fiducial analysis choices used in the \emph{Letter}. Importantly, they demonstrate that the fiducial cross-correlation measurement is robust to residual Galactic and extragalactic foreground contamination, and is therefore suitable for analysis.

\section{Details of the theoretical model}
\subsection{Axion screening halo model}
In this section, we review the halo model calculation for the anisotropic screening of the CMB due to photon-axion conversion within LSS from Ref.~\cite{Mondino:2024rif}. Specifically, we focus on the model for the cross-spectrum, $C_{\ell}^{g\tilde{T}^a}$. For a more detailed treatment, we refer the reader to Ref.~\cite{Mondino:2024rif} (see also Refs.~\cite{Pirvu:2023lch, McCarthy:2024ozh} for a similar calculation for dark photons). The conversion of CMB photons into axions in a halo with mass $M$ at redshift $z$ leads to an anisotropic screening of the CMB.  The optical depth associated with this screening is given by~\cite{Pirvu:2023lch, Mondino:2024rif, McCarthy:2024ozh}\footnote{Note that this expression differs by a factor of $1/3$ from Eq. 3.9 of Ref.~\cite{Mondino:2024rif} because we include this factor, which arises from averaging over magnetic field orientations, in our definition of the conversion probability.} 
\begin{equation}\label{eq:screen_optical_depth}
        \tau^a_{\ell}(z,M)=\sqrt{\frac{4\pi}{2\ell+1}}N_{\rm res}(z,M|m_a)P^{\rm res}_{\gamma a}(z,M|\nu,m_a)\,u_{\ell }(z,M),
\end{equation}
where
\begin{equation}\label{eq:screening_angular_projection}
   u_{\ell }(z,M)\equiv \sqrt{\frac{2\ell+1}{4\pi}}\int d^2\hat{\Omega}\left[1-\frac{(\chi(z)\theta/R_{\rm res}(z,M|m_a))^2}{(1+z)^2} \right]^{-1/2}\mathcal{P}_\ell(\cos\theta)
\end{equation} 
encodes the angular dependence of the screening. In Eq.~\eqref{eq:screen_optical_depth} and Eq.~\eqref{eq:screening_angular_projection}, $\ell$ is the angular multipole, $\chi(z)$ is the comoving distance to redshift $z$, $N_{\rm res}(z,M|m_a)$ is the number of times the photon trajectory crosses the resonant radius within the halo, $P^{\rm res}_{\gamma a}(z,M|\nu,m_a)$ is the probability for a photon with frequency $\nu$ to resonantly convert to an axion with mass $m_a$ within the halo, $R_{\rm res}(z,M|m_a)$ is the radius at which a photon resonantly converts into an axion of mass $m_a$ within a halo of mass $M$ at redshift $z$, and $\mathcal{P}_\ell$ is the Legendre polynomial of degree $\ell.$ Following Ref.~\cite{Mondino:2024rif}, we set $N_{\rm res}=2$ if the resonance radius is smaller than the virial radius and $N_{\rm res}=0$ outside of the virial radius. That is, we only consider conversions that occur within the virial radius. In the Landau-Zener approximation, the resonant conversion probability is
\begin{equation}\label{eq:photon_axion_conv_probability_SM}
    P_{\gamma \rightarrow a}^{\rm res}(\bx(t_{\rm res}), \nu)\simeq \frac{2\pi^2\nu  g_{a\gamma\gamma}^2|{\bf{ B}}_\perp^{\rm res}|^2}{m_a^2}\left\vert \frac{d \ln m_{\gamma}^2(\bx)}{dt} \right\vert^{-1}_{t_{\rm res}},
\end{equation}
where ${\bf{B}}_\perp^{\rm res}$ is the component of the magnetic field transverse to the propagation direction of the photon evaluated at the resonance location, $t_{\rm res}$ is the time of resonance, and  $m_\gamma(\bx)\equiv \sqrt{4\pi\,\alpha \, n_e(\bx)/m_e}\approx 3.7\times 10^{-11}~{\rm eV}\,\sqrt{n_e/{\rm cm}^{-3}}$ is the photon plasma mass, where $\alpha$ is the fine-structure constant. Using Eq.~\eqref{eq:photon_axion_conv_probability} and assuming spherical symmetry, the resonant conversion probability can be written as
\begin{equation}
    P_{\gamma a}^{\rm res}(z, M|\nu,m_a)=2\pi^2\nu (1+z) g_{a\gamma\gamma}^2\frac{1}{3}|{\bf{B}}(R_{\rm res}|z,M)|^2\times\bigg|\frac{d m_{\gamma}^2(r|M,z)}{dr}\bigg|^{-1}_{R_{\rm res}},
\end{equation}
where $R_{\rm res}$ is the resonance radius (set by the electron number density profile) and $|{\bf{B}}(R_{\rm res}|z,M)|$ is the value of the magnetic field strength within a halo of mass $M$ at redshift $z$ evaluated at the resonance radius.

In thermodynamic temperature units, the screened CMB temperature fluctuation is related to the dark screening optical depth by
\begin{equation}
    \tilde{T}_\ell^a(\nu, \hatbn)=\left(\frac{1-e^{-x}}{x}\right)T_{\rm CMB}\,\tau_\ell^{a}(\nu, \hatbn),
\end{equation}
where $x\equiv 2\pi \nu/T_{\rm CMB}$ is the dimensionless frequency, $T_{\rm CMB}\approx 2.726$ K is the CMB monopole temperature, and the term in parentheses accounts for the conversion from specific intensity to CMB thermodynamic temperature units.

\begin{table*}[!t]
\begin{tabular}{|c|c|c|c|c|c|c|}
\hline
~&~$\alpha_s$~& ~$\sigma_{\ln M}$~&~$\log_{10}(M^{\rm HOD}_{\rm min}/M_{\odot})$~&~$\log_{10}(M_*/M_{\odot})$~&~$\lambda$~\\\hline
~\textbf{K22}; Fiducial~  & 1.304 &  0.687 & 11.97  &  12.87  & 1.087  \\\hline
K23  & 1.06 &  0.020 & 11.86 & 12.78 & 1.800  \\\hline

\end{tabular}
\caption{Halo occupation distribution (HOD) parameters used in this work. Our fiducial analysis uses the best-fit parameters from Ref.~\cite{Kusiak:2022xkt} (K22). We also consider an analysis using the best-fit parameters from Ref.~\cite{Kusiak:2023hrz} (K23).}\label{tab:HOD_parameters}
\end{table*}

To compute the angular power spectrum between the axion-induced screening, $\tilde{T}_{\ell}^a$, and the galaxy overdensity, $\delta_g,$ we need a model describing how galaxies populate halos. To this end, we use the Halo Occupation Distribution (HOD) model~\cite{Peacock:2000qk, Zheng:2007zg} (see also Refs.~\cite{Kusiak:2022xkt, Bolliet:2022pze, Kusiak:2023hrz} for specific details about the HOD used in this work). In the HOD framework, the galaxies are divided into ``centrals," which reside at the centers of the dark matter halos, and ``satellites," which are distributed throughout the halo according to a Navarro, Frenk, and White (NFW) profile~\cite{Navarro:1996gj}. The expected number of central and satellite galaxies as a function of halo mass, $M$, are given by 
\begin{equation}
N_c(M)=\frac{1}{2}+\frac{1}{2}{\rm erf}\left(\frac{\log_{10} M-\log_{10} M^{\rm HOD}_{\rm min}}{\sigma_{\ln M}}\right),  \ \ N_s(M)=N_c(M)\left(\frac{M}{M_*} \right)^{\alpha_s},
\end{equation}
respectively, where $M_*,$ $M^{\rm HOD}_{\rm min}$, $\sigma_{\ln M}$, and $\alpha_s$ are free parameters of the model that characterize a particular galaxy sample. The mean galaxy number density is then
\begin{equation}
\bar{n}_g(z)=\int dM \,n(z,M)\left[N_c(M)+N_s(M) \right].
\end{equation}
where $n(z,M)$ is the halo mass function. In this work, we use the halo mass function from Ref.~\cite{Tinker:2008ff}. The galaxy distribution is described by the galaxy kernel,
\begin{equation}
u_\ell^g(z,M)=\frac{H(z)}{\chi^2(z)}\frac{dN_g}{dz}\bar{n}_g^{-1}(z)\left[N_c(M)+N_s(M)u_\ell^{\rm NFW}(z,M) \right],
\end{equation}
where $H(z)$ is the Hubble parameter at redshift $z$, $\frac{dN_g}{dz}$ is the normalized redshift distribution of the galaxy sample, and $u_\ell^{\rm NFW}(z,M)$ is the Fourier transform of the truncated NFW profile evaluated at $k=(\ell+1/2)/\chi$.\footnote{We use the Limber approximation~\cite{Limber:1954zz}, which is accurate over the scales analyzed in this work.}  This is given by~\cite{Scoccimarro:2000gm}
\begin{align}\label{eq:uNFW_k}
    u^{\rm NFW}(k|z,M)=f_{\rm NFW}(\lambda c_{200c})\left[\cos(q)\left(\rm{Ci}(\bar{q})-{\rm Ci}(q)\right)+\sin(q)\left(\rm{Si}(\bar{q})-{\rm Si}(q)\right)-\frac{\sin(\bar{q}-q)}{\bar{q}} \right]
\end{align}
where $q=k\times \frac{r_{200c}}{c_{200c}}$, $\bar{q}=(1+\lambda c_{\rm 200c})q$, $c_{200c}$ is the concentration estimated from the concentration-mass relation in Ref.~\cite{Duffy:2008pz},\footnote{Note that this differs from the concentration-mass relation used in Refs.~\cite{Kusiak:2022xkt, Kusiak:2023hrz}, which use the relation from Ref.~\cite{Bhattacharya:2011vr}. We have checked that this has a negligible impact on our results.} and $r=\lambda R_{200c}$ is the truncation radius. The amplitude in Eq.~\eqref{eq:uNFW_k} is set by
\begin{equation}\label{eq:fNFW}
    f_{\rm NFW}(x)=\left(\ln(1+x)-\frac{x}{1+x} \right)^{-1}.
\end{equation}

In the halo model, the cross-spectrum, $C_{\ell}^{g\tilde{T}^a}$, is written as the sum of a 1-halo term and a 2-halo term:
\begin{equation}
   C_{\ell}^{g\tilde{T}^a}=C_{\ell}^{g\tilde{T}^a;\,1-{\rm halo}}+C_{\ell}^{g\tilde{T}^a;\,2-{\rm halo}}.
\end{equation}
The 1-halo term describes correlations within a single halo, and the 2-halo term describes correlations between two distinct halos. The 1-halo term is given by
\begin{equation}\label{eq:Cl_1halo}
    C_{\ell}^{g\tilde{T}^a;\,1-{\rm halo}}=\int\, dz\,\frac{\chi(z)^2}{H(z)} \int dM\, n(z,M)\,\tilde{T}^a_{\ell}(z,M)\,u_{\ell}^g(z,M).
\end{equation}
Similarly, the 2-halo term is 
\begin{equation}\label{eq:Cl_2halo}
    C_{\ell}^{g\tilde{T}^a,2-{\rm halo}}=\int\, dz\,\frac{\chi(z)^2}{H(z)}  \left[\prod_{i=1,2}\int\, dM_i \,n(z,M_i)\,b(z,M_i)\right] u_{\ell}^g(z,M_1)\tilde{T}^a_{\ell}(z,M_2) P^{\rm lin}_{mm}\left(\frac{\ell+1/2}{\chi(z)}, z \right),
\end{equation}
where $P^{\rm lin}_{mm}$ is the linear matter power spectrum and $b(z,M)$ is the linear halo bias, which, for the mass function used in this work, is given by Ref.~\cite{Tinker:2010my}.

\begin{figure*}[!t]
\centering
\includegraphics[width=0.995\linewidth]{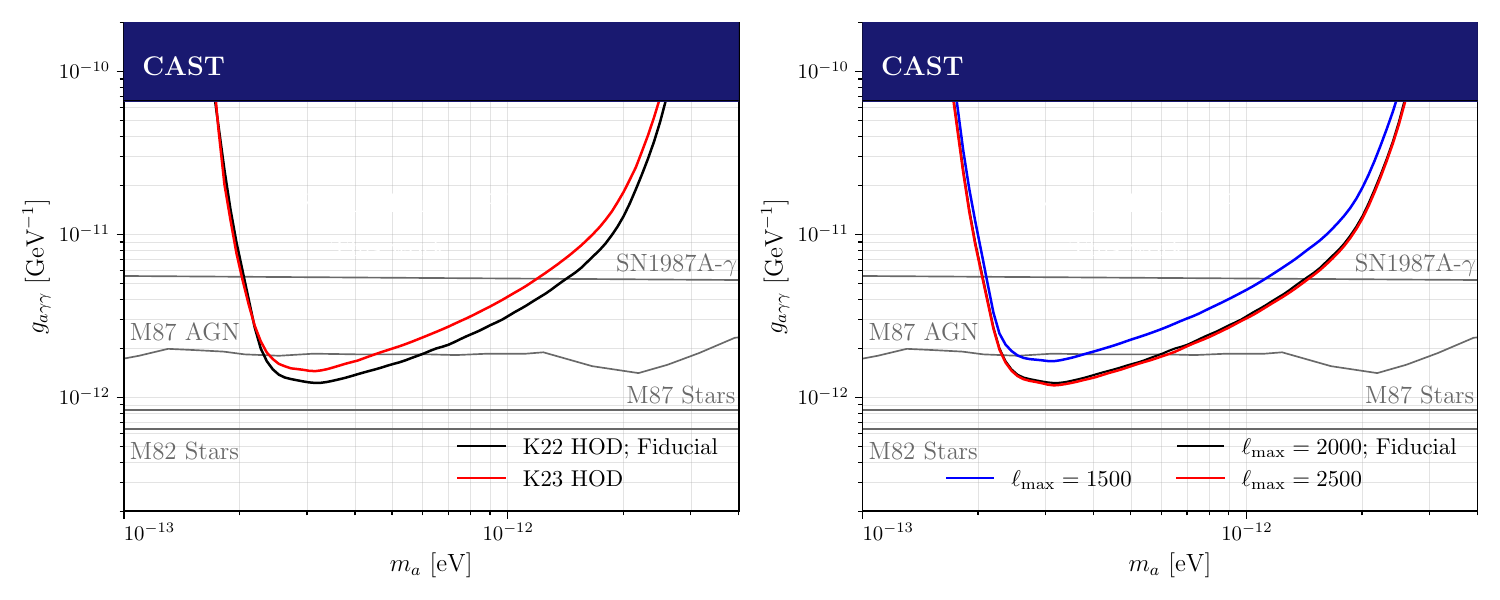}
\caption{\emph{Left}: Impact of varying the HOD parameters on the constraints on the axion-photon coupling. We compare the fiducial analysis (black), which uses the best-fit HOD model from Ref.~\cite{Kusiak:2022xkt}, with an analysis using the best-fit HOD parameters from Ref.~\cite{Kusiak:2023hrz} (red). \emph{Right}: Impact of varying the maximum multipole included in our analysis, $\ell_{\rm max}$, on the constraint on the axion-photon coupling. We find negligible improvement including modes beyond $\ell>2000.$}
\label{fig:varying_HOD}
\end{figure*}

We evaluate the integrals in Eq.~\eqref{eq:Cl_1halo} and Eq.~\eqref{eq:Cl_2halo} numerically using 50 redshift bins linearly spaced between $0.005<z<1.9$ and 100 mass bins logarithmically spaced between $10^{9}<M_{\rm vir}<5\times 10^{15}~M_\odot$. We have checked that these choices are sufficient for the \emph{unWISE} Blue HOD. We compute the matter power spectrum using \textsc{CAMB}~\cite{Lewis:1999bs}, and evaluate all halo-model-related quantities using \texttt{hmvec}.\footnote{\href{https://github.com/simonsobs/hmvec}{https://github.com/simonsobs/hmvec}}

The HOD model includes five free parameters: $\alpha_s$, $\sigma_{\ln M}$, $M^{\rm HOD}_{\rm min}$, $M_*$, and $\lambda$. For our fiducial analysis, we fix these parameters to their best-fit values for the \emph{unWISE} Blue galaxy sample from Ref.~\cite{Kusiak:2022xkt}. We refer to this model as the K22 HOD, and list the parameter values in Table~\ref{tab:HOD_parameters}. To assess the impact of uncertainties in the HOD on our constraints for the axion-photon coupling, we also consider the best-fit parameters derived in Ref.~\cite{Kusiak:2023hrz} for the same galaxy sample, referred to as the K23 HOD. The left panel of Fig.~\ref{fig:varying_HOD} compares the 95\% confidence limit on $g_{a\gamma\gamma}$ for these two sets of HOD parameters. For our fiducial analysis, we choose to use the K22 HOD because the K23 HOD is fit to smaller scales (up to $\ell=4000$), where the HOD framework is not expected to provide a good description of the data (formally, the PTE of the HOD fit is very low in K23). The K23 HOD was fit to provide an approximate description of the \emph{unWISE} Blue galaxy auto-spectrum at small scales (for kSZ analyses) and has several limitations, including an unphysically low value of $\sigma_{\ln M}$. Finally, the K22 HOD has a lower value of the truncation scale, $\lambda$, making the predictions less sensitive to the magnetic field profile in the outer regions of the halos. Despite the significant differences between the K22 and K23 HOD models, the constraints are broadly consistent between the two models, indicating that our fiducial analysis is robust to changes in the HOD.

The right panel of Fig.~\ref{fig:varying_HOD} shows how our constraints depend on the maximum multipole included in the analysis, $\ell_{\rm max}$. We find negligible improvements if we increase $\ell_{\rm max}$ from 2000 to 2500, thus we fix $\ell_{\rm max}=2000$ for the fiducial analysis. 

\subsection{Electron profile model}
\begin{figure*}[!t]
\centering
\includegraphics[width=0.995\linewidth]{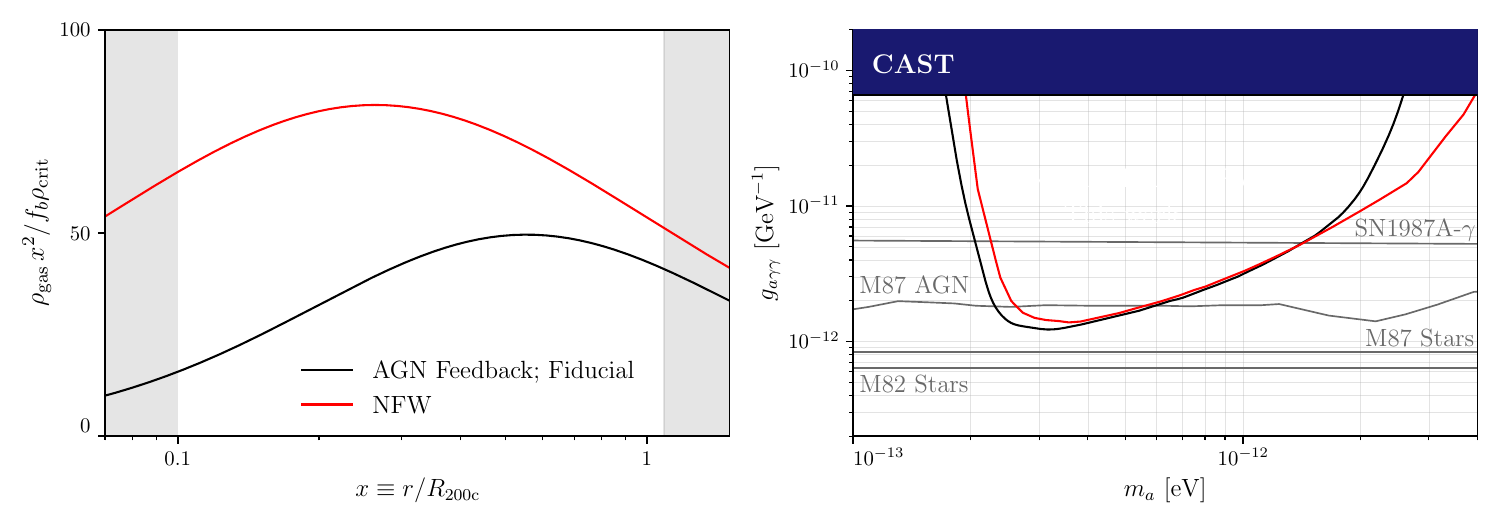}
\caption{\emph{Left}: Gas density profile models considered in this work. In our fiducial analysis, we use the AGN Feedback profile (black) from Ref.~\cite{Battaglia:2016xbi}. We also consider a variant of our analysis where the gas perfectly traces the dark matter, represented by an NFW profile (red)~\cite{Navarro:1996gj}. Profiles are shown for a $M_{\rm 200c}=3.3\times 10^{13}~M_\odot$ mass halo with $R_{ 200c}=0.6~{\rm Mpc}$ at redshift $z=0.55$. \emph{Right}: Impact of varying the gas profile model on the 95\% confidence interval for the axion-photon coupling.}
\label{fig:varying_electron_model}
\end{figure*}

The anisotropic screening of the CMB induced by photon-axion conversions depends on the distribution of the ionized gas (specifically free electrons) within halos. The distribution of gas within halos is affected by complex astrophysical feedback processes due to, \emph{e.g.}, Active Galactic Nuclei (AGN) and supernovae. These processes tend to push gas toward the outskirts of halos, thus flattening the electron profile. In our fiducial analysis, we assume that the gas follows the ``AGN Feedback" model calibrated from hydrodynamical simulations in~\cite{Battaglia:2016xbi}. Specifically, we model the gas density using a generalized NFW (gNFW) profile:\footnote{Assuming the plasma is fully ionized, the electron number density is related to the total gas density by $n_e(r)=\frac{1+X_H}{2 m_u}\rho_{\rm gas}(r),$ where $m_u$ is the atomic mass unit and $X_H\approx 0.76$ is the hydrogen mass fraction.}
\begin{equation}\label{eq:rho_gas_battaglia}
 \rho_{\rm gas}(r)=f_b\, \rho_{\rm crit.}(z)\, \rho_0
\,  \left(\frac{x}{x_c}\right) ^\gamma\left[1+\left(\frac{x}{x_c}\right)^\alpha \right]^{-\frac{\beta+\gamma}{\alpha}}
\end{equation}
where $f_b\equiv \Omega_b/\Omega_m$ is the baryon fraction , $\rho_{\rm crit.}$ is the critical density, $x\equiv r/R_{200c}$, and we fix $x_c=0.5$ and $\gamma=-0.2$. In Eq.~\eqref{eq:rho_gas_battaglia}, $\rho_0$, $\alpha$, and $\beta$ are mass- and redshift-dependent quantities parameterized by 
\begin{equation}
    X(M_{200c},z)=X_0\left( \frac{M_{200c}}{10^{14}~M_\odot}\right)^{X_M}(1+z)^{X_z} \text{ for } X\in \{\rho_0,\,\alpha,\,\beta\},
\end{equation}
where the best-fit values of $X_0,$ $X_M$, and $X_z$ are given in Table 2 of~\cite{Bolliet:2022pze}.

To assess how variations in the gas profile model can impact our constraint on $g_{a\gamma\gamma}$, we consider an alternate scenario where the gas directly traces the dark matter. In this case, the gas density is described by an NFW profile,
\begin{equation}
    \rho_{\rm gas}(r)=f_b\rho_s\frac{1}{\frac{r}{r_s}\left(1+\frac{r}{r_s} \right)^2},
\end{equation}
where $r_s=R_{200c}/c_{200c}$ and $\rho_s=\frac{M_{200c}}{4\pi r_s^3}f_{\rm NFW}(c_{200c})$ with $f_{\rm NFW}(x)$ given by Eq.~\eqref{eq:fNFW}. Although this scenario is observationally excluded~\cite{Amodeo:2020mmu,ACT:2024rue}, it provides a useful qualitative assessment of how variations in the electron profile can impact our constraint on $g_{a\gamma\gamma}$. The left panel of Fig.~\ref{fig:varying_electron_model} compares the radial gas density profile predicted by the AGN Feedback profile from Ref.~\cite{Battaglia:2016xbi} with the NFW profile for a $M_{200c}=3.3\times 10^{13}~M_\odot$ mass halo with $R_{200c}=0.6~{\rm Mpc}$ at redshift $z=0.55.$ Feedback processes redistribute gas towards the outskirts of the halo leading to a shallower electron number density profile.

The right panel of Fig.~\ref{fig:varying_electron_model} illustrates how our constraints change if we assume that the gas follows an NFW profile instead of the AGN Feedback profile. In general, our constraints are broadly consistent between the two profile choices. Assuming an NFW profile results in tighter constraints for more massive axions ($m_a\gtrsim 2\times 10^{-12}~{\rm eV}$), as these require higher electron number densities to satisfy the resonance condition. Conversely, assuming an NFW profile leads to slightly weaker constraints for low-mass axions ($m_a \lesssim 2\times 10^{-13}$ eV). This is because the NFW profile does not reach sufficiently low electron densities for resonant conversion to occur within the virial radius, for the halos described by the \emph{unWISE} Blue HOD. We also note that differences in the slopes of the NFW and AGN Feedback profiles impact the conversion probability, and thus the constraint on $g_{a\gamma\gamma}.$

These tests demonstrate that our constraints are robust to changes in the electron density model within the range of uncertainty probed by current cosmological surveys. In the future, it will be interesting to revisit these constraints with updated electron number density models that are calibrated using measurements of the kSZ effect~\cite{Amodeo:2020mmu, AtacamaCosmologyTelescope:2020wtv, Kusiak:2021hai} and fast radio bursts~\cite{Madhavacheril:2019buy, Prochaska_2019}.

\subsection{Magnetic field profile model}\label{sec:SM_Bfield_model}

\begin{figure}[!t]
\centering
\includegraphics[width=0.99\linewidth]{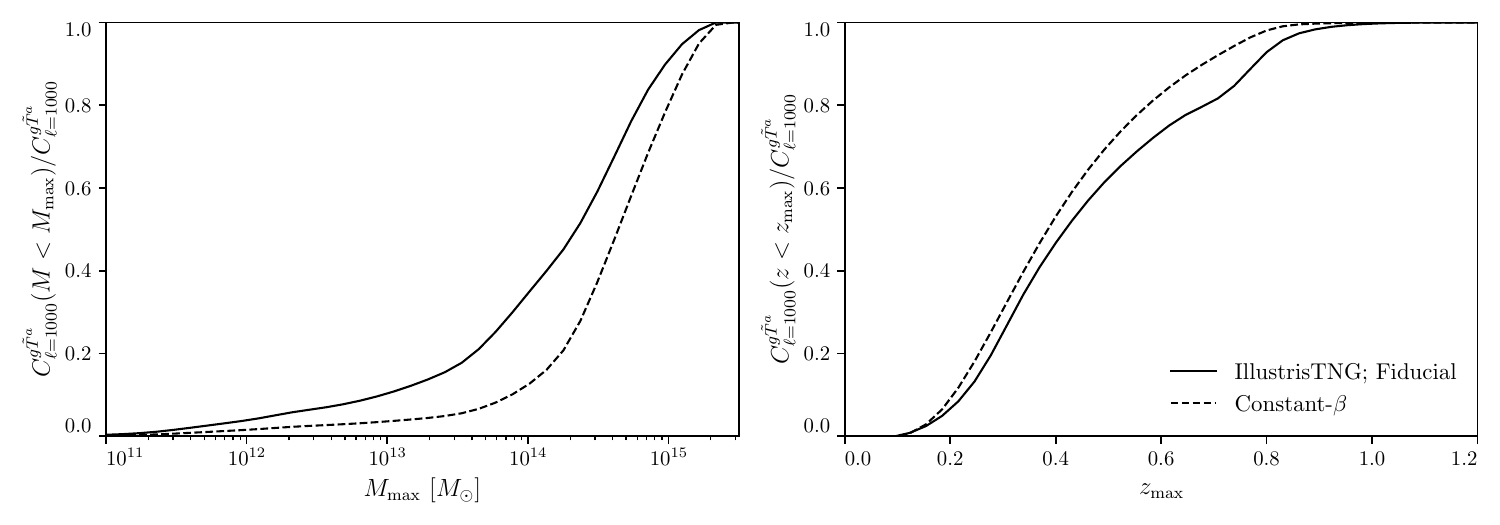}
\caption{\emph{Left}: Halo model prediction for the cumulative contribution to $C_{\ell}^{g\tilde{T}^a}$ as a function of halo mass. \emph{Right}: Halo model prediction for the cumulative contribution to $C_{\ell}^{g\tilde{T}^a}$ as a function of redshift. In both panels, solid lines represent the IllustrisTNG magnetic field profile, while dashed lines correspond to the constant-$\beta$ profile. All predictions are calculated at $\ell=1000$, for an axion mass of $m_a=3\times 10^{-13}~{\rm eV}$.}
\label{fig:fig_Cl_Mmax_zmax}
\end{figure}

The resonant axion-photon conversion probability (Eq.~\eqref{eq:photon_axion_conv_probability_SM}) depends on the value of the magnetic field at the resonance radius, and thus the halo magnetic field profile\footnote{{In the regime of photon energies and axion masses considered in this work, the conversion probability from Eq.~\eqref{eq:photon_axion_conv_probability_SM} does not depend on the magnetic field coherence length, as long as this is larger than at most roughly one hundred parsec. We refer the reader to Ref.~\cite{Mondino:2024rif}} (see Appendix A2) and~\cite{Marsh:2021ajy} for further discussions on the validity of this assumption.}. In our fiducial analysis, we estimate halo magnetic field profiles using the state-of-the-art IllustrisTNG cosmological magnetohydrodynamical simulations~\cite{Nelson:2018uso}. Specifically, we compute the magnetic field profiles from the TNG100-1 simulation and TNG300-1 simulations --- the highest-resolution runs of the IllustrisTNG (110.7 Mpc)$^3$ and (302.6 Mpc)$^3$ comoving volumes, respectively. Due to their large box size and high resolution, the IllustrisTNG simulations are well-suited for studying the magnetic field profiles in $10^{12}-10^{14.5}~M_\odot$ mass halos, which dominate the \emph{unWISE} Blue HOD~\cite{Kusiak:2022xkt}. It is crucial to have magnetic field profiles for the largest-mass halos in the \emph{unWISE} HOD, as these massive halos typically host the strongest magnetic fields. As a result, their contribution is significantly upweighted in the halo model prediction of the cross-correlation (see Fig.~\ref{fig:fig_Cl_Mmax_zmax}). 

We compute the spherically-averaged, mass-weighted mean radial magnetic field profiles in eight mass bins logarithmically spaced between $10^{11}~M_\odot \leq M_{200c}< 10^{15}~M_\odot$, at redshifts $z=0,~0.5,~1,$ and $1.5$, spanning the range of masses and redshifts in the \emph{unWISE} Blue HOD. Due to the limited number of halos with $M_{200c}>10^{14}~M_\odot$ in the TNG100-1 simulation volume, we use TNG300-1, which has a lower resolution (but larger volume) than TNG100-1, to compute the profiles for the two highest-mass bins.\footnote{Note that the TNG300-1 simulation does not have any halos between $10^{14.5}~M_\odot \leq M_{200c}<10^{15}~M_\odot$ at $z=1.5$. Therefore, we use the $z=1$ profiles for this mass bin. This has a negligible impact on our results, as the \emph{unWISE} Blue galaxies are predominately at redshifts $z<1$ (see Fig.~\ref{fig:dN_dz_unwise}). In future analyses, one could use the TNG-Cluster~\cite{Nelson:2023ilh} simulations to estimate these profiles.} To mitigate potential uncertainties in the simulation-based magnetic field profiles at low halo masses, we conservatively set the magnetic field to zero for all halos with $M_{200c}<10^{11}~M_\odot$. This has a negligible impact on our constraints except at the lowest axion masses we are sensitive to, $m_a\lesssim1.7\times 10^{-13}~{\rm eV}$, where the resonance condition requires a sufficiently low value of $n_e$. For each mass bin and redshift, we compute the magnetic field profile in 11 radial bins logarithmically spaced between $0.07\leq r/R_{\rm 200c} \leq 2.$ We interpolate the logarithm of the measured profiles as a function of the logarithm of the radial bin center using a fifth-degree univariate spline. Since the magnetic field profile depends only mildly on redshift, we use piecewise-constant interpolation over redshifts. For example, to estimate the magnetic field profile at redshift $z=0.7$, we use the measurements from the $z=0.5$ snapshot.

\begin{figure}[!t]
\centering
\includegraphics[width=0.99\linewidth]{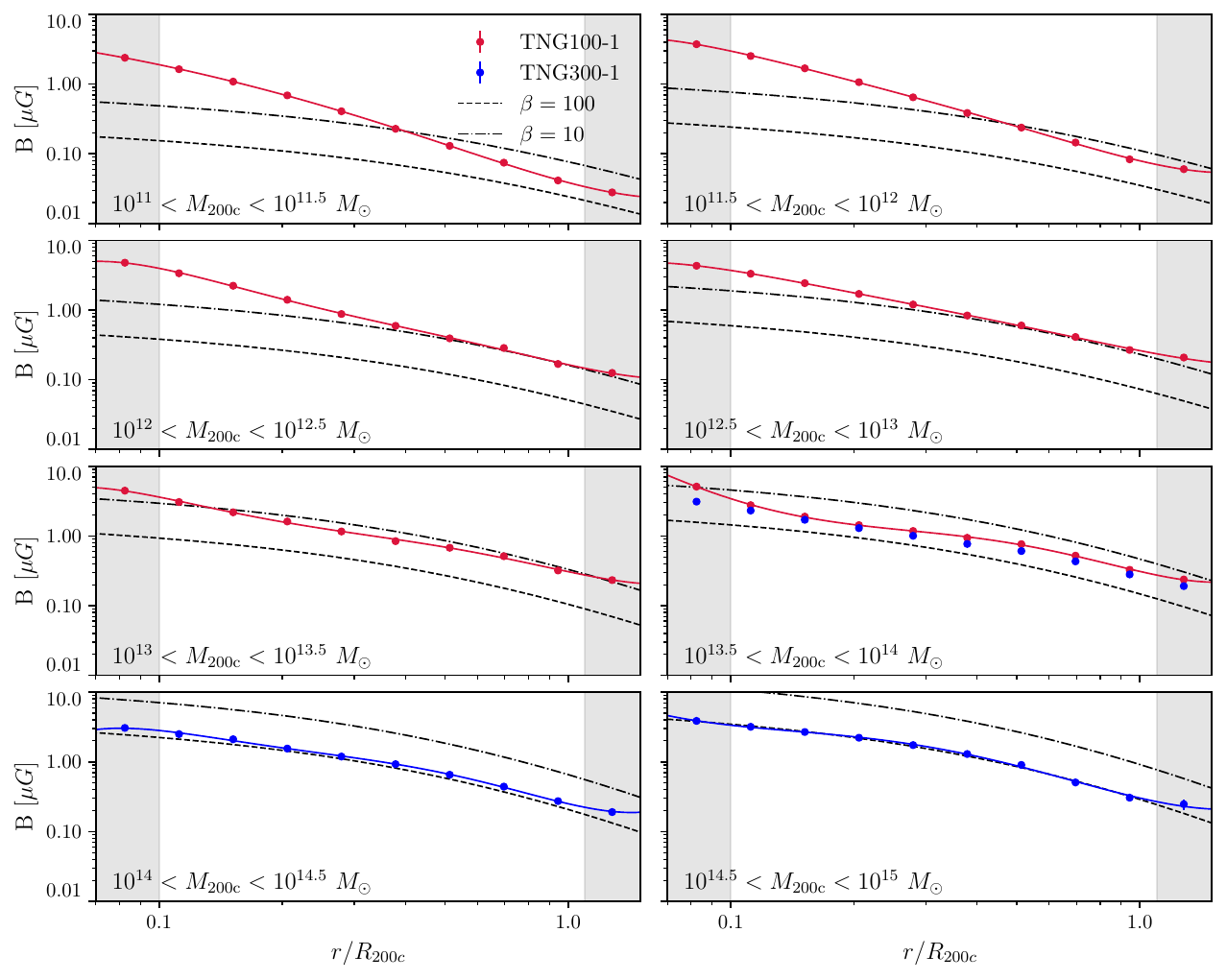}
\caption{Comparison of magnetic field profiles used in this work at redshift $z=0.5$. The red points show the mass-weighted mean magnetic field strength profiles measured from the TNG100-1 and TNG300-1 runs of the IllustrisTNG simulation, averaged over all halos within a given mass bin. The solid lines show the interpolated profiles used to compute the magnetic field strength at resonance in our fiducial halo model calculations. The black dashed lines show the constant-$\beta$ model assuming the B12 thermal pressure profile from Ref.~\cite{Battaglia:2011cq} for $\beta=10$ and $\beta=100$. The grey bands denote regions that are excluded from the halo model calculations.}
\label{fig:magnetic_field_profiles}
\end{figure}

As an alternative to the IllustrisTNG magnetic field model, we consider a constant-$\beta$ model, where $\beta \equiv P_{\rm th}(r)/P_{\rm mag}(r)$ is the ratio of the thermal to magnetic pressure. The constant-$\beta$ model has been used in previous high-energy astrophysics searches for photon-axion conversion (\emph{e.g.}, Ref.~\cite{Reynes:2021bpe}), and has been studied in the context of hydrodynamical simulations, including IllustrisTNG~\cite{Marinacci:2017wew}. In the constant-$\beta$ model, the magnetic field is given by 
\begin{equation}
   {\rm B}(r)=\sqrt{\frac{2 P_{\rm th}(r)}{\beta}}.
\end{equation}
We estimate the thermal pressure using the Battaglia (B12) pressure profile~\cite{Battaglia:2011cq}. The B12 profile models the pressure as a gNFW profile,
\begin{equation}
P_{\rm th}(r)=P_{200c} P_0\left(\frac{x}{x_c}\right)^\gamma\left(1+\left(\frac{x}{x_c}\right)^\alpha\right)^{-\beta_{\rm gNFW}},
\end{equation}
where $x\equiv r/R_{200c}$, and, following Ref.~\cite{Battaglia:2011cq}, we fix $\alpha=1$ and $\gamma=-0.3$. The amplitude of the pressure profile is
\begin{equation}
P_{200c}=\frac{G M_{200c} 200 \rho_{\rm crit}(z)f_b}{2R_{200c}},
\end{equation}
where $G$ is Newton's constant. Finally, $P_0$, $x_c$, and $\beta_{\rm gNFW}$ are mass- and redshift-dependent quantities parameterized by 
\begin{equation}
    X(M_{200c},z)=X_0\left( \frac{M_{200c}}{10^{14}~M_\odot}\right)^{X_M}(1+z)^{X_z} \text{ for } X\in \{P_0,\,x_c,\,\beta_{\rm gNFW}\},
\end{equation}
where $X_0, X_M$, and $X_z$ are derived from fitting hydrodynamical simulations (see Table 1 of Ref.~\cite{Battaglia:2011cq} for their values, and note that these are the same simulations from which our fiducial AGN Feedback gas profile is taken). Note that best-fit B12 pressure profile parameters were derived using halos with $M\gtrsim 5\times 10^{13}~M_\odot$. Therefore, we should not expect the B12 profile to fully model the pressure profile of the lower-mass halos included in our analysis~\cite{Lee:2022tor}.  Nevertheless, the B12 pressure profile combined with the constant-$\beta$ model provides a useful analytic alternative to cross-check the simulation-based analysis used in the main text.

In Fig.~\ref{fig:magnetic_field_profiles}, we compare the measured magnetic field profiles from IllustrisTNG with the constant-$\beta$ profiles, assuming the B12 pressure profile, for several values of $\beta.$ We show the profiles at $z=0.5$ in eight mass bins. The red (blue) markers show the mass-weighted mean profile estimated from the TNG100-1 (TNG300-1) simulations with error bars representing the standard deviation of the mean profile. In detail, we first estimate the covariance matrix for a single profile within a mass bin by computing the sample covariance of all halos in that bin. To approximate the error on the mean, we then rescale the sample covariance by $1/N_{\rm halo}$. The solid lines illustrate the smoothed profiles derived from spline interpolation, which we use in our fiducial analysis. For comparison, we show the profiles estimated from the TNG100-1 and TNG300-1 simulations for the $10^{13.5}<M_{200c}<10^{14}~M_\odot$ mass range. As discussed in Ref.~\cite{Marinacci:2017wew}, the magnetic field profile from the TNG300-1 simulations is lower than that from the TNG100-1 simulations due to the lower resolution of TNG300-1. This suggests that our constraints could improve with profiles from higher-resolution, large-volume simulations based on the IllustrisTNG physics model.

The black lines in Fig.~\ref{fig:magnetic_field_profiles} show the constant-$\beta$ profiles assuming $\beta=10$ and $\beta=100$, which roughly bracket the measured profiles from IllustrisTNG. The constant-$\beta$ profiles effectively capture the shape of the IllustrisTNG magnetic fields for massive halos ($M_{200c}\gtrsim 10^{14}~M_\odot$). Moreover, the value of $\beta$ needed to describe the IllustrisTNG magnetic field profiles tends to increase with halo mass. These findings are consistent with previous analyses of the magnetic field profiles in IllustrisTNG~\cite{Marinacci:2017wew} and \textsc{Auriga}~\cite{2020MNRAS.498.3125P,2024MNRAS.528.2308P}, although we emphasize that the constant-$\beta$ profiles used in this work were not computed from the IllustrisTNG thermal pressure profile. 

Fig.~\ref{fig:fig_Cl_Mmax_zmax} illustrates the mass and redshift dependence of the halo model prediction for $C_{\ell}^{g\tilde{T}^a}$. Specifically, the left panel of Fig.~\ref{fig:fig_Cl_Mmax_zmax} shows the cumulative contribution to $C_{\ell}^{g\tilde{T}^a}$ as a function of halo mass at $\ell=1000$, for an axion mass of $m_a=3\times 10^{-13}~{\rm eV}$. The solid line shows predictions from our fiducial analysis based on the IllustrisTNG magnetic field profiles, while the dashed line assumes the constant-$\beta$ magnetic field profile.\footnote{Note that the ratio that is shown is independent of both the value of $\beta$ and the axion-photon coupling $g_{a\gamma\gamma}$.} In both cases, the cross-correlation receives significant contributions from $M\gtrsim 10^{14}~M_\odot$ halos, particularly in the constant-$\beta$ model where the magnetic field profile depends more strongly on halo mass (see Fig.~\ref{fig:magnetic_field_profiles}). The right panel of Fig.~\ref{fig:fig_Cl_Mmax_zmax} shows the cumulative contribution to $C_{\ell}^{g\tilde{T}^a}$ as a function of redshift, assuming the same $\ell$ and axion mass as in the left panel.\footnote{The small feature in the fiducial analysis results from a discontinuity in our IllustrisTNG magnetic field model at $z=0.75$, which transitions from the $z=0.5$ to the $z=1.0$ profiles. } For both models, the cross-correlation is predominantly sourced by redshifts below $z\lesssim 1$, consistent with the \emph{unWISE} Blue $dN_g/dz$ (see Fig.~\ref{fig:dN_dz_unwise}). The fiducial model receives contributions from slightly higher redshifts than the constant-$\beta$ model, as the IllustrisTNG magnetic field profiles prefer a mild decrease in $\beta$ with redshift, for a fixed halo mass and assuming the B12 thermal pressure profile. These results could be useful for identifying the optimal LSS tracer for future axion searches using measurements of $C_{\ell}^{g\tilde{T}^a}$.

\begin{figure}[!t]
\centering
\includegraphics[width=0.6\linewidth]{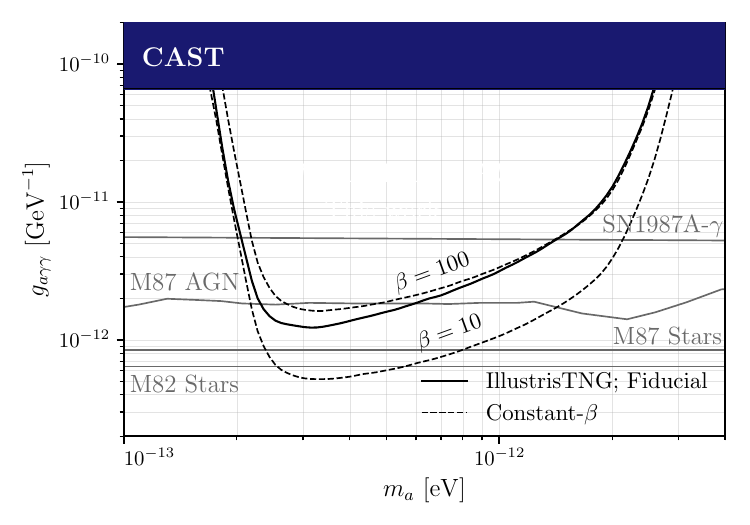}
\caption{Comparison of constraints on the axion-photon coupling from our fiducial analysis, which uses magnetic field profiles derived from the IllustrisTNG simulations, with analyses assuming a constant-$\beta$ magnetic field profile with $\beta=10$ and $\beta=100$.}\label{fig:constant_beta_constraints}
\end{figure}

Fig.~\ref{fig:constant_beta_constraints} compares our fiducial constraints on the axion-photon coupling with constraints derived using the constant-$\beta$ magnetic field profile, assuming a conservative value of $\beta=100$ and an optimistic value of $\beta=10$. Our fiducial constraints are broadly consistent with the $\beta=100$ analysis, assuming the B12 thermal pressure profile. At low axion masses, the fiducial constraints are marginally tighter than the $\beta=100$ analysis because conversions in this regime typically occur in lower-mass halos, where the IllustrisTNG profiles prefer $\beta<100.$ At higher axion masses, the fiducial analysis is consistent with the $\beta=100$ constraints, as resonant conversion in this regime occurs in the most massive halos in the \emph{unWISE} HOD, where the IllustrisTNG profiles are consistent with our constant-$\beta=100$ profiles. 

Ultimately, one of the central challenges to astrophysical axion searches is modeling the complex magnetic fields in astrophysical environments. In this work, we model the spherically-averaged halo magnetic field strength using the IllustrisTNG simulations --- arguably the most advanced cosmological magnetohydrodynamical simulations that span the mass and redshift ranges analyzed in this work. Our fiducial constraints are broadly consistent with an alternative analysis using a constant-$\beta=100$ magnetic field profile with a simulation-based model for the thermal pressure profile. In the future, it will be interesting to revisit these constraints with updated magnetic field models informed by new observations and simulations.

%
%TC:endignore

\end{document}